\documentclass[a4paper,12pt]{book}
\usepackage{times}
\usepackage{amssymb,amsmath,amsfonts,amsthm}
\usepackage{hyperref}
\usepackage{graphicx}
\def\be{\begin{equation}}
\def\ee{\end{equation}}
\def\bea{\begin{eqnarray}}
\def\eea{\end{eqnarray}}
\def\vec#1{\mbox{\boldmath$#1$}}
\def\hsp5{\hspace{5mm}}
\def\lb{\label}
\def\bi{\bibitem}
\def\ct{\cite}
\def\ptl{\partial}

\def\case#1/#2{\textstyle\frac{#1}{#2}}
\newcommand{\leftout}[1]{}
\begin{document}

\vspace*{1cm}
\begin{center}
{\Huge\sc
										An Introduction to \\
                    Business Mathematics
\par}
\par\vfill\vfill
                           Lecture notes for
                           the Bachelor degree programmes \\ 
							IB/IMC/IMA/ITM/IEVM/ACM/IEM/IMM \\
							at Karlshochschule International University
\par\vfill
							Module
\par\vfill
							0.1.1 IMQM: Introduction to Management and its
							Quantitative Methods
\par\vfill\vfill\vfill

                    {\large\sc Henk van Elst}

\par\vfill

                 August 30, 2015
\par\vfill\vfill

                 Fakult\"at I: Betriebswirtschaft und Management\\
                 Karlshochschule\\
                 International University\\
                 Karlstra\ss e 36--38\\
                 76133 Karlsruhe\\
                 Germany
\par\vfill
                 E--mail: {\tt hvanelst@karlshochschule.de}
\par\vfill\vfill
                 E--Print: 
\href{http://arxiv.org/abs/1509.04333}{arXiv:1509.04333v2 
[q-fin.GN]}
\par\vfill\vfill\vfill
									\copyright\ 2009--2015 Karlshochschule
									International University
									and Henk van Elst
                    
\end{center}
\vspace*{1cm}
\sloppy
      \renewcommand{\thepage}{}                 
\addcontentsline{toc}{chapter}{Abstract}
\chapter*{}
\vspace{-8ex}
\section*{Abstract}
{\small These lecture notes provide a self-contained introduction 
to the mathematical methods required in a Bachelor degree 
programme in Business, Economics, or Management. In particular, 
the topics covered comprise real-valued vector and matrix algebra, 
systems of linear algebraic equations, Leontief's stationary 
input--output matrix model, linear programming, elementary 
financial mathematics, as well as differential and integral 
calculus of real-valued functions of one real variable. A special 
focus is set on applications in quantitative economical modelling.}

\vspace{10mm}
\noindent
\underline{Cite as:} 
\href{http://arxiv.org/abs/1509.04333}{arXiv:1509.04333v2
[q-fin.GN]}
\vfill

\medskip
\noindent
These lecture notes were typeset in \LaTeXe.

      \newpage \thispagestyle{empty}
      \setcounter{page}{1}
\tableofcontents
      \newpage \thispagestyle{empty}
      \cleardoublepage \pagenumbering{arabic}
\addcontentsline{toc}{chapter}{Qualification objectives of the module (excerpt)}
\chapter*{Qualification objectives of the module (excerpt)}
The qualification objectives shall be reached by an integrative
approach.

\medskip
\noindent
A broad instructive range is aspired. The students shall acquire
a 360 degree orientation concerning the task- and
personnel-related tasks and roles of a manager and supporting
tools and methods and be able to describe the coherence in an
integrative way. The knowledge concerning the tasks and the
understanding of methods and tools shall be strengthened by a
constructivist approach based on case studies and exercises.

\medskip
\noindent
Students who have successfully participated in this module
will be able to
\begin{itemize}
\item \ldots,
\item solve problems in Linear Algebra and Analysis and
apply such mathematical methods to quantitative problems
in management.
\item apply and challenge the knowledge critically on
current issues and selected case studies.
\end{itemize}
%

\addcontentsline{toc}{chapter}{Introduction}
\chapter*{Introduction}
These lecture notes contain the entire material of the 
quantitative methods part of the first semester module {\bf 0.1.1 
IMQM: Introduction to Management and its Quantitative Methods} at 
Karlshochschule International University. The aim is to provide a 
selection of tried-and-tested mathematical tools that proved 
efficient in actual practical problems of {\bf Economics} and {\bf 
Management}. These tools constitute the foundation for a 
systematic treatment of the typical kinds of quantitative problems 
one is confronted with in a Bachelor degree programme. 
Nevertheless, they provide a sufficient amount of points of 
contact with a quantitatively oriented subsequent Master degree 
programme in {\bf Economics}, {\bf Management}, or the {\bf Social 
Sciences}.

\medskip
\noindent
The prerequisites for a proper understanding of these lecture 
notes are modest, as they do not go much beyond the basic A-levels 
standards in {\bf Mathematics}. Besides the four fundamental 
arithmetical operations of addition, subtraction, multiplication 
and division of real numbers, you should be familiar, e.g., with 
manipulating fractions, dealing with powers of real numbers, the 
binomial formulae, determining the point of intersection for two 
straight lines in the Euclidian plane, solving a quadratic 
algebraic equation, and the rules of differentiation of 
real-valued functions of one variable.

\medskip
\noindent
It might be useful for the reader to have available a modern 
{\bf graphic display calculator (GDC)} for dealing with some of 
the calculations that necessarily arise along the way, when 
confronted with specific quantitative problems. Some current 
models used in public schools and in undergraduate studies are, 
amongst others,
\begin{itemize}
\item Texas Instruments \emph{TI--84 plus},
\item Casio \emph{CFX--9850GB PLUS}.
\end{itemize}
However, the reader is strongly encouraged to think about 
resorting, as an alternative, to a {\bf spreadsheet programme} 
such as EXCEL or OpenOffice to handle the calculations one 
encounters in one's quantitative work.

\medskip
\noindent
The central theme of these lecture notes is the acquisition and 
application of a number of effective mathematical methods in a 
business oriented environment. In particular, we hereby focus on 
{\bf quantitative processes} of the sort
\[
\text{INPUT} \rightarrow \text{OUTPUT} \ ,
\]
for which different kinds of {\bf functional relationships} 
between some numerical {\bf INPUT quantities} and some numerical 
{\bf OUTPUT quantities} are being considered. Of special interest 
in this context will be {\bf ratios} of the structure
\[
\frac{\text{OUTPUT}}{\text{INPUT}} \ .
\]
In this respect, it is a general objective in {\bf Economics} to 
look for ways to optimise the value of such ratios (in favour of 
some {\bf economic agent}), either by seeking to increase the 
OUTPUT when the INPUT is confined to be fixed, or by seeking to 
decrease the INPUT when the OUTPUT is confined to be fixed. 
Consequently, most of the subsequent considerations in these 
lecture notes will therefore deal with issues of {\bf 
optimisation} of given {\bf functional relationships} between some 
{\bf variables}, which manifest themselves either in {\bf 
minimisation} or in {\bf maximisation} procedures.

\medskip
\noindent
The structure of these lecture notes is the following. Part~I 
presents selected mathematical methods from {\bf Linear Algebra}, 
which are discussed in Chs.~\ref{ch1} to \ref{ch5}. Applications 
of these methods focus on the quantitative aspects of flows 
of goods in simple economic models, as well as on problems in 
linear programming. In Part~II, which is limited to 
Ch.~\ref{ch6}, we turn to discuss elementary aspects of {\bf 
Financial Mathematics}. Fundamental principles of {\bf Analysis}, 
comprising differential and integral calculus for real-valued 
functions of one real variable, and their application to 
quantitative economic problems, are reviewed in Part~III; this
extends across Chs.~\ref{ch7} and \ref{ch8}.

\medskip
\noindent
We emphasise the fact that there are \emph{no} explicit examples 
nor exercises included in these lecture notes. These are reserved 
exclusively for the lectures given throughout term time.

\medskip
\noindent
Recommended textbooks accompanying the lectures are the works by 
Asano (2013)~\ct{asa2013}, Dowling (2009)~\ct{dow2009}, Dow\-ling 
(1990)~\ct{dow1990}, Bauer \emph{et al} (2008)~\ct{bauetal2008},
Bosch (2003)~\ct{bos2003}, and H\"ulsmann \emph{et al} 
(2005)~\ct{hueetal2005}. Some standard references of {\bf Applied 
Mathematics} are, e.g., Bronstein \emph{et al} 
(2005)~\ct{broetal2005} and Arens \emph{et al} 
(2008)~\ct{areetal2008}. Should the reader feel inspired by the 
aesthetics, beauty, ellegance and efficiency of the mathematical 
methods presented, and, hence, would like to know more about their 
background and relevance, as well as being introduced to further 
mathematical techniques of interest, she/he is recommended to take 
a look at the brilliant books by Penrose (2004)~\ct{pen2004}, 
Singh (1997)~\ct{sin1997}, Gleick(1987)~\ct{gle1987} and Smith 
(2007)~\ct{smi2007}. Note that most of the textbooks and 
monographs mentioned in this Introduction are available from 
the library at Karlshochschule International University.

\medskip
\noindent
Finally, we draw the reader's attention to the fact that the
*.pdf version of these lecture notes contains interactive features 
such as fully hyperlinked references to original publications at 
the websites \href{http://dx.doi.org}{{\tt dx.doi.org}} and 
\href{http://www.jstor.org}{{\tt jstor.org}}, as well 
as active links to biographical information on scientists that 
have been influential in the historical development of {\bf 
Mathematics}, hosted by the websites 
\href{http://www-history.mcs.st-and.ac.uk/}{The MacTutor History 
of Mathematics archive ({\tt www-history.mcs.st-and.ac.uk})} and 
\href{http://en.wikipedia.org/wiki/Main_Page}{{\tt 
en.wikipedia.org}}.

\chapter[Vector algebra in Euclidian space ${\mathbb R}^{n}$
]{Vector algebra in Euclidian space ${\mathbb R}^{n}$}
\lb{ch1}
Let us begin our elementary considerations of {\bf vector algebra} 
with the introduction of a special class of mathematical objects. 
These will be useful at a later stage, when we turn to formulate 
certain problems of a quantitative nature in a compact and elegant 
way. Besides introducing these mathematical objects, we also need 
to define which kinds of mathematical operations they can be 
subjected to, and what computational rules we have to take care of.

\section[Basic concepts]%
{Basic concepts}
\lb{sec:vekeinf}
Given be a set $V$ of mathematical objects $\vec{a}$ which, for now, we want to consider merely as a collection of $n$ arbitrary real numbers $a_{1}$, \ldots, $a_{i}$, \ldots, $a_{n}$. In explicit terms,
\be
\lb{vecsp}
V =  \left\{
\vec{a} = \left(
\begin{array}{c}
a_{1} \\ \vdots \\ a_{i} \\ \vdots \\ a_{n}
\end{array}\right)
\left|\, a_{i} \in {\mathbb R},\hspace{1mm}
i=1, \dots, n\right.\right\} \ .
\ee
Formally the $n$ real numbers considered can either be assembled in an ordered pattern as a column or a row. We define

\medskip
\noindent
\underline{\bf Def.:}
Real-valued {\bf column vector} with $n$ 
components
\be
\fbox{$\displaystyle
\vec{a} := \left(
\begin{array}{c}
a_{1} \\ \vdots \\ a_{i} \\ \vdots \\ a_{n}
\end{array}
\right) \ ,
\hspace{10mm}
a_{i} \in {\mathbb R},\hspace{3mm}
i=1, \dots, n \ ,
$}
\ee
Notation: $\vec{a} \in \mathbb{R}^{n \times 1}$,

\medskip
\noindent
and

\medskip
\noindent
\underline{\bf Def.:}
Real-valued {\bf row vector} with $n$ 
components
\be
\fbox{$\displaystyle
\vec{a}^{T} := \left(
a_{1}, \dots, a_{i}, \dots, a_{n}\right) \ ,
\hspace{10mm}
a_{i} \in {\mathbb R},\hspace{3mm}
i=1, \dots, n \ ,
$}
\ee
Notation: $\vec{a}^{T} \in \mathbb{R}^{1 \times n}$.

\medskip
\noindent
Correspondingly, we define the $n$-component objects
\be
\vec{0} := \left(
\begin{array}{c}
0 \\ \vdots \\ 0 \\ \vdots \\ 0
\end{array}
\right)
\qquad\text{and}\qquad
\vec{0}^{T} := \left(
0, \dots, 0, \dots, 0\right)
\ee
to constitute related {\bf zero vectors}.

\medskip
\noindent
Next we define for like objects in the set $V$, i.e., either for 
$n$-component column vectors or for $n$-component row vectors, two 
simple computational operations. These are

\medskip
\noindent
\underline{\bf Def.:}
{\bf Addition} of vectors
\be
\lb{vekadd}
\fbox{$\displaystyle
\vec{a} + \vec{b}
:= \left(
\begin{array}{c}
a_{1} \\ \vdots \\ a_{i} \\ \vdots \\ a_{n}
\end{array}
\right)
+ \left(
\begin{array}{c}
b_{1} \\ \vdots \\ b_{i} \\ \vdots \\ b_{n}
\end{array}
\right)
= \left(
\begin{array}{c}
a_{1}+b_{1} \\ \vdots \\ a_{i}+b_{i} \\ \vdots \\
a_{n}+b_{n}
\end{array}
\right) \ ,
\hspace{10mm}
a_{i}, b_{i} \in {\mathbb R} \ ,
$}
\ee

\medskip
\noindent
and

\medskip
\noindent
\underline{\bf Def.:}
{\bf Rescaling} of vectors
\be
\lb{vekskal}
\fbox{$\displaystyle
\lambda\vec{a}
:= \left(
\begin{array}{c}
\lambda a_{1} \\ \vdots \\ \lambda a_{i} \\ \vdots \\
\lambda a_{n}
\end{array}
\right) \ ,
\hspace{10mm}
\lambda, a_{i} \in {\mathbb R} \ .
$}
\ee

\medskip
\noindent
The rescaling of a vector $\vec{a}$ with an arbitrary non-zero real number  $\lambda$ has the following effects:
\begin{itemize}
	\item $|\lambda| > 1$ --- stretching of the length of $\vec{a}$
	\item $0 < |\lambda| < 1$ --- shrinking of the length of 
	$\vec{a}$
	\item $\lambda < 0$ --- directional reversal of $\vec{a}$.
\end{itemize}
The notion of the length of a vector $\vec{a}$ will be made precise shortly.

\medskip
\noindent
The addition and the rescaling of $n$-component vectors satisfy the following addition and multiplication laws:

\medskip
\noindent
{\bf Computational rules for addition and rescaling of vectors}

\noindent
For vectors $\vec{a}, \vec{b}, \vec{c} \in {\mathbb R}^{n}$:
\begin{enumerate}
	\item $\vec{a}+\vec{b} = \vec{b}+\vec{a}$
	\hfill ({\bf commutative addition})
	\item $\vec{a}+(\vec{b}+\vec{c}) = (\vec{a}+\vec{b})+\vec{c}$
	\hfill ({\bf associative addition})
	\item $\vec{a}+\vec{0}=\vec{a}$ \hfill ({\bf addition identity element})
	\item For every $\vec{a}, \vec{b} \in {\mathbb R}^{n}$, there exists exactly one  $\vec{x} \in {\mathbb R}^{n}$ such that $\vec{a}+\vec{x}=\vec{b}$
	
{} \hfill ({\bf invertibility of addition})
	\item $(\lambda\mu)\vec{a}=\lambda(\mu\vec{a})$ with $\lambda
	\in {\mathbb R}$ \hfill ({\bf associative rescaling})
	\item $1\vec{a}=\vec{a}$ \hfill ({\bf rescaling identity element})
	\item $\lambda(\vec{a}+\vec{b}) = \lambda\vec{a}+\lambda\vec{b}$;
	
	$(\lambda+\mu)\vec{a} = \lambda\vec{a}+\mu\vec{a}$ with $\lambda,
	\mu	\in {\mathbb R}$ \hfill ({\bf distributive rescaling}).
\end{enumerate}

\medskip
\noindent
In conclusion of this section, we remark that every set of 
mathematical objects $V$ constructed in line with 
Eq.~(\ref{vecsp}), with an addition and a rescaling defined 
according to Eqs.~(\ref{vekadd}) and~(\ref{vekskal}), and 
satisfying the laws stated above, constitutes a {\bf linear vector 
space over Euclidian space} ${\mathbb R}^{n}$.\footnote{This is 
named after the ancient greek mathematician
\href{http://www-groups.dcs.st-and.ac.uk/~history/Biographies/Euclid.html}{Euclid of Alexandria (about 325~BC--265~BC)}.}

\section[Dimension and basis of ${\mathbb R}^{n}$]%
{Dimension and basis of ${\mathbb R}^{n}$}
\lb{sec:vekdim}
Let there be given $m$ $n$-component vectors\footnote{A slightly s
horter notation for $n$-component column vectors $\vec{a} \in 
\mathbb{R}^{n\times 1}$ is given by $\vec{a} \in \mathbb{R}^{n}$; 
likewise $\vec{a}^{T} \in \mathbb{R}^{n}$ for $n$-component row 
vectors $\vec{a}^{T} \in \mathbb{R}^{1\times n}$.} $\vec{a}_{1}, 
\ldots, \vec{a}_{i}, \ldots, \vec{a}_{m} \in {\mathbb R}^{n}$, as 
well as $m$ real numbers $\lambda_{1}, \ldots, \lambda_{i}, 
\ldots, \lambda_{m} \in {\mathbb R}$. The new $n$-component vector 
$\vec{b}$ resulting from the addition of arbitrarily rescaled 
versions of these $m$ vectors according to
\be
\lb{linkomb}
\fbox{$\displaystyle
\vec{b} = \lambda_{1}\vec{a}_{1}+\ldots+\lambda_{i}\vec{a}_{i}
+\ldots+\lambda_{m}\vec{a}_{m}
=: \sum_{i=1}^{m}\lambda_{i}\vec{a}_{i} \in {\mathbb R}^{n}
$}
\ee
is referred to as a {\bf linear combination} 
of the $m$ vectors $\vec{a}_{i},\,i=1, \ldots, m$.

\medskip
\noindent
\underline{\bf Def.:} A set of $m$ vectors $\vec{a}_{1}, \ldots,
\vec{a}_{i}, \ldots, \vec{a}_{m} \in {\mathbb R}^{n}$ is called 
{\bf linearly independent} when the condition
\be
\lb{linunabh}
\boldsymbol{0} \stackrel{!}{=} \lambda_{1}\vec{a}_{1}+\ldots
+\lambda_{i}\vec{a}_{i}+\ldots+\lambda_{m}\vec{a}_{m}
= \sum_{i=1}^{m}\lambda_{i}\vec{a}_{i} \ ,
\ee
i.e., the problem of forming the {\bf zero vector} $\boldsymbol{0} 
\in {\mathbb R}^{n}$ from a linear combination of the $m$ vectors 
$\vec{a}_{1}, \ldots, \vec{a}_{i}, \ldots, \vec{a}_{m} \in 
{\mathbb R}^{n}$, can \emph{only} be solved trivially, namely by 
$0=\lambda_{1}=\ldots=\lambda_{i}=\ldots=\lambda_{m}$. When, 
however, this condition can be solved non-trivially, with some 
$\lambda_{i} \neq 0$, then the set of $m$ vectors $\vec{a}_{1},
\ldots, \vec{a}_{i}, \ldots,\vec{a}_{m} \in {\mathbb R}^{n}$ is 
called {\bf linearly dependent}.

\medskip
\noindent
In Euclidian space ${\mathbb R}^{n}$, there is a maximum number 
$n$ (!) of vectors which can be linearly independent. This maximum 
number is referred to as the {\bf dimension of Euclidian 
space} ${\mathbb R}^{n}$. Every set of $n$ linearly independent 
vectors in Euclidian space ${\mathbb R}^{n}$ constitutes a 
possible {\bf basis of Euclidian space} ${\mathbb R}^{n}$.
If the set $\{\vec{a}_{1}, \ldots, \vec{a}_{i}, 
\ldots,\vec{a}_{n}\}$ constitutes a basis of ${\mathbb R}^{n}$, 
then every other vector $\vec{b} \in {\mathbb R}^{n}$ can be 
expressed in terms of these basis vectors by
\be
\vec{b} = \beta_{1}\vec{a}_{1}+\ldots+\beta_{i}\vec{a}_{i}
+\ldots+\beta_{n}\vec{a}_{n}
= \sum_{i=1}^{n}\beta_{i}\vec{a}_{i} \ .
\ee
The rescaling factors $\beta_{i}\in{\mathbb R}$ of the 
$\vec{a}_{i}\in{\mathbb R}^{n}$ are called the {\bf components of 
vector} $\vec{b}$ {\bf with respect to the basis}
$\{\vec{a}_{1}, \ldots, \vec{a}_{i}, \ldots, \vec{a}_{n}\}$.

\vspace{5mm}
\noindent
\underline{\bf Remark:} The $n$ {\bf unit vectors}
\be
\lb{kanbasis}
\vec{e}_{1} := \left(
\begin{array}{c}
1 \\ 0 \\ \vdots \\ 0
\end{array}
\right) \ , \hspace{5mm}
\vec{e}_{2} := \left(
\begin{array}{c}
0 \\ 1 \\ \vdots \\ 0
\end{array}
\right) \ , \hspace{5mm}
\dots \ , \hspace{5mm}
\vec{e}_{n} := \left(
\begin{array}{c}
0 \\ 0 \\ \vdots \\ 1
\end{array}
\right) \ , \hspace{5mm}
\ee
constitute the so-called {\bf canonical basis of Euclidian 
space} ${\mathbb R}^{n}$. With respect to this basis, all vectors 
$\vec{b} \in \mathbb{R}^{n}$ can be represented as a linear combinationen
\be
\vec{b} = \left(
\begin{array}{c}
b_{1} \\ b_{2} \\ \vdots \\ b_{n}
\end{array}
\right)
= b_{1}\vec{e}_{1}+b_{2}\vec{e}_{2}+\dots
+b_{n}\vec{e}_{n}
= \sum_{i=1}^{n}b_{i}\vec{e}_{i} \ .
\ee
%

\section[Euclidian scalar product]%
{Euclidian scalar product}
\lb{sec:vekskal}
Finally, to conclude this section, we introduce a third mathematical operation defined for vectors on $\mathbb{R}^{n}$.

\medskip
\noindent
\underline{\bf Def.:} For an $n$-component row vector $\vec{a}^{T} 
\in \mathbb{R}^{1 \times n}$ and an $n$-component column vector 
$\vec{b} \in \mathbb{R}^{n \times 1}$, the {\bf Euclidian scalar 
product}
\be
\lb{skalprod}
\fbox{$\displaystyle
\vec{a}^{T}\cdot\vec{b}
:= \left(
a_{1}, \ldots, a_{i}, \ldots a_{n}\right)
\left(
\begin{array}{c}
b_{1} \\ \vdots \\ b_{i} \\ \vdots \\ b_{n}
\end{array}
\right)
= a_{1}b_{1}+\ldots+a_{i}b_{i}\ldots+a_{n}b_{n}
=: \sum_{i=1}^{n}a_{i}b_{i}
$}
\ee
defines a mapping $f: \mathbb{R}^{1 \times n} \times \mathbb{R}^{n 
\times 1} \rightarrow \mathbb{R}$ from the product set of 
$n$-component row and column vectors to the set of real numbers. 
Note that, in contrast to the addition and the rescaling of 
$n$-component vectors, the outcome of forming a Euclidian scalar 
product between two $n$-component vectors is a \emph{single real 
number}.

\medskip
\noindent
In the context of the Euclidian scalar product, two non-zero 
vectors $\vec{a}, \vec{b} \in {\mathbb R}^{n}$ (wit $\vec{a}
\neq \vec{0} \neq \vec{b}$) are referred to as {\bf mutually 
orthogonal} when they exhibit the property that $0 = 
\vec{a}^{T}\cdot\vec{b} = \vec{b}^{T}\cdot\vec{a}$.

\medskip
\noindent
{\bf Computational rules for Euclidian scalar product of 
vectors}

\noindent
For vectors $\vec{a}, \vec{b}, \vec{c} \in {\mathbb R}^{n}$:

\begin{enumerate}
	\item $(\vec{a}+\vec{b})^{T}\cdot\vec{c}
	= \vec{a}^{T}\cdot\vec{c}+\vec{b}^{T}\cdot\vec{c}$
	\hfill ({\bf distributive scalar product})
	\item $\vec{a}^{T}\cdot\vec{b} = \vec{b}^{T}\cdot\vec{a}$
	\hfill ({\bf commutative scalar product})
	\item $(\lambda\vec{a}^{T})\cdot\vec{b} =
	\lambda(\vec{a}^{T}\cdot\vec{b})$ with $\lambda \in {\mathbb R}$
	\hfill ({\bf homogeneous scalar product})
	\item $\vec{a}^{T}\cdot\vec{a} > 0$ for all $\vec{a}
	\neq \vec{0}$ \hfill ({\bf positive definite scalar product}).
\end{enumerate}

\medskip
\noindent
Now we turn to introduce the notion of the length of an 
$n$-component vector.

\medskip
\noindent
\underline{\bf Def.:} The {\bf length} of a vector $\vec{a} \in 
\mathbb{R}^{n}$ is defined via the Euclidian scalar product as
\be
\fbox{$\displaystyle
|\vec{a}| :=
\sqrt{\vec{a}^{T}\cdot\vec{a}}
= \sqrt{a_{1}^{2}+\ldots+a_{i}^{2}+\ldots+a_{n}^{2}}
=: \sqrt{\sum_{i=1}^{n}a_{i}^{2}} \ .
$}
\ee

\medskip
\noindent
Technically one refers to the non-negative real number $|\vec{a}|$ 
as the {\bf absolute value} or the {\bf Euclidian norm} of the 
vector $\vec{a}\in \mathbb{R}^{n}$. The length of $\vec{a} \in
\mathbb{R}^{n}$ has the following properties: 
\begin{itemize}
	\item $|\vec{a}| \geq 0$, and $|\vec{a}|=0 \Leftrightarrow
	\vec{a}=\vec{0}$;
	\item $|\lambda\vec{a}| = |\lambda||\vec{a}|$ for $\lambda \in
	{\mathbb R}$;
	\item $|\vec{a}+\vec{b}| \leq |\vec{a}|+|\vec{b}|$
	\hfill ({\bf triangle inequality}).
\end{itemize}

\medskip
\noindent
Every non-zero vector $\vec{a} \in \mathbb{R}^{n}$, i.e., 
$|\vec{a}|>0$, can be rescaled by the reciprocal of its length. 
This procedure defines the

\medskip
\noindent
\underline{\bf Def.:}
{\bf Normalisation} of a vector $\vec{a} \in
\mathbb{R}^{n}$;
\be
\hat{\vec{a}} := \frac{\vec{a}}{|\vec{a}|}
\qquad\Rightarrow\qquad
|\hat{\vec{a}}| = 1 \ .
\ee

\medskip
\noindent
By this method one generates a vector of length $1$, i.e., a {\bf 
unit vector} $\hat{\vec{a}}$. To denote unit vectors we will 
employ the ``hat'' symbol.

\medskip
\noindent
Lastly, also by means of the Euclidian scalar product, we introduce the angle enclosed between two non-zero vectors.

\medskip
\noindent
\underline{\bf Def.:}
{\bf Angle} enclosed between $\vec{a},\vec{b} \neq \vec{0}
\in {\mathbb R}^{n}$ 
\be
\fbox{$\displaystyle
\cos[\varphi(\vec{a},\vec{b})]
= \frac{\vec{a}^{T}}{|\vec{a}|}\cdot
\frac{\vec{b}}{|\vec{b}|}
= \hat{\vec{a}}{}^{T}\cdot\hat{\vec{b}}
\hspace{5mm} \Rightarrow \hspace{5mm}
\varphi(\vec{a},\vec{b})
= \cos^{-1}(\hat{\vec{a}}{}^{T}\cdot\hat{\vec{b}}) \ .
$}
\ee
\underline{\bf Remark:} The inverse cosine function\footnote{The 
notion of on inverse function will be discussed later in 
Ch.~\ref{ch7}.} $\cos^{-1}(\ldots)$ is available on every standard 
GDC or spreadsheet.


\chapter[Matrices]{Matrices}
\lb{ch2}
In this chapter, we introduce a second class of mathematical 
objects that are more general than vectors. For these objects, we 
will also define certain mathematical operations, and a set of 
computational rules that apply in this context.

\section[Matrices as linear mappings]%
{Matrices as linear mappings}
\lb{sec:matlinab}
Consider given a collection of $m \times n$ arbitrary real numbers
$a_{11}$, $a_{12}$ \ldots, $a_{ij}$, \ldots,
$a_{mn}$, which we arrange systematically in a particular kind of 
array.

\medskip
\noindent
\underline{\bf Def.:} A real-valued $\boldsymbol{(m\times n)}${\bf -matrix} is formally defined to constitute an array of real 
numbers according to
\be
\mathbf{A}
:= \left(\begin{array}{cccccc}
   	a_{11} & a_{12} & \ldots & a_{1j} & \ldots & a_{1n} \\
   	a_{21} & a_{22} & \ldots & a_{2j} & \ldots & a_{2n} \\
    \vdots & \vdots & \ddots & \vdots & \ddots & \vdots \\
    a_{i1} & a_{i2} & \ldots & a_{ij} & \ldots & a_{in} \\
    \vdots & \vdots & \ddots & \vdots & \ddots & \vdots \\
    a_{m1} & a_{m2} & \ldots & a_{mj} & \ldots & a_{mn}
	\end{array}\right) \ ,
\ee
where $a_{ij} \in {\mathbb R}$,	$i=1, \dots, m; j=1, \dots, n$.

\noindent
Notation: $\mathbf{A} \in \mathbb{R}^{m \times n}$.

\medskip
\noindent
Characteristic features of this array of real numbers are:
\begin{itemize}
	\item $m$ denotes the number of {\bf rows} of $\mathbf{A}$, $n$ 
	the number of {\bf columns} of $\mathbf{A}$.
	
	\item $a_{ij}$ represents the {\bf elements} of $\mathbf{A}$;
	$a_{ij}$ is located at the point of intersection of the $i$th 
	row and the $j$th column of $\mathbf{A}$.
	
	\item elements of the $i$th row constitute the {\bf row vector}
	$\left(a_{i1}, a_{i2}, \ldots, a_{ij}, \ldots, a_{in}\right)$,
	elements of the $j$th column the {\bf column vector}
	$\left(\begin{array}{c} a_{1j} \\ a_{2j} \\ \vdots \\ a_{ij} \\
	\vdots \\ a_{mj}
	\end{array}\right)$.
\end{itemize}
Formally column vectors need to be viewed as $(n\times 
1)$-matrices, row vectors as $(1\times n)$-matrices. An 
$\boldsymbol{(m \times n)}${\bf -zero matrix}, denoted by
$\mathbf{0}$, has all its elements equal to zero, i.e.,
\be
\mathbf{0}
:= \left(\begin{array}{cccc}
   	0 & 0 & \ldots & 0 \\
   	0 & 0 & \ldots & 0 \\
    \vdots & \vdots & \ddots & \vdots \\
    0 & 0 & \ldots & 0
	\end{array}\right) \ .
\ee

\medskip
\noindent
Matrices which have an \emph{equal} number of rows and columns, 
i.e. $m=n$, are referred to as {\bf quadratic matrices}. In 
particular, the $\boldsymbol{(n \times n)}${\bf -unit matrix} (or 
identity matrix)
\be
\lb{einmatr}
\mathbf{1} := \left(\begin{array}{cccccc}
   	1 & 0 & \ldots & 0 & \ldots & 0 \\
   	0 & 1 & \ldots & 0 & \ldots & 0 \\
    \vdots & \vdots & \ddots & \vdots & \ddots & \vdots \\
    0 & 0 & \ldots & 1 & \ldots & 0 \\
    \vdots & \vdots & \ddots & \vdots & \ddots & \vdots \\
    0 & 0 & \ldots & 0 & \ldots & 1
	\end{array}\right)
\ee
holds a special status in the family of $(n\times n)$-matrices.

\medskip
\noindent
Now we make explicit in what sense we will comprehend $(m\times 
n)$-matrices as mathematical objects.

\medskip
\noindent
\underline{\bf Def.:} A real-valued matrix $\mathbf{A} \in
\mathbb{R}^{m \times n}$ defines by the computational operation
\bea
\lb{matabb}
\mathbf{A}\vec{x}
& := & \left(\begin{array}{cccccc}
   	a_{11} & a_{12} & \ldots & a_{1j} & \ldots & a_{1n} \\
   	a_{21} & a_{22} & \ldots & a_{2j} & \ldots & a_{2n} \\
    \vdots & \vdots & \ddots & \vdots & \ddots & \vdots \\
    a_{i1} & a_{i2} & \ldots & a_{ij} & \ldots & a_{in} \\
    \vdots & \vdots & \ddots & \vdots & \ddots & \vdots \\
    a_{m1} & a_{m2} & \ldots & a_{mj} & \ldots & a_{mn}
	\end{array}\right)
	\left(\begin{array}{c}
x_{1} \\ x_{2} \\ \vdots \\ x_{j} \\ \vdots \\ x_{n}
\end{array}\right) \nonumber \\
& := & \left(\begin{array}{c}
a_{11}x_{1}+a_{12}x_{2}+\ldots+a_{1j}x_{j}+\ldots+a_{1n}x_{n} \\
a_{21}x_{1}+a_{22}x_{2}+\ldots+a_{2j}x_{j}+\ldots+a_{2n}x_{n} \\
\vdots \\
a_{i1}x_{1}+a_{i2}x_{2}+\ldots+a_{ij}x_{j}+\ldots+a_{in}x_{n} \\
\vdots \\
a_{m1}x_{1}+a_{m2}x_{2}+\ldots+a_{mj}x_{j}+\ldots+a_{mn}x_{n}
\end{array}\right)
=: \left(
\begin{array}{c}
y_{1} \\
y_{2} \\
\vdots \\
y_{i} \\
\vdots \\
y_{m}
\end{array}
\right)
= \vec{y}
\eea
a {\bf mapping} $\mathbf{A}: \mathbb{R}^{n \times 1} \rightarrow 
\mathbb{R}^{m \times 1}$, i.e. a mapping from the set of 
real-valued $n$-component column vectors (here: $\vec{x}$) to the 
set of real-valued $m$-component column vectors (here: $\vec{y}$).

\medskip
\noindent
In loose analogy to the photographic process, $\vec{x}$ can be 
viewed as representing an ``object,'' $\mathbf{A}$ a ``camera,'' 
and $\vec{y}$ the resultant ``image.''

\medskip
\noindent
Since for real-valued vectors $\vec{x}_{1},\vec{x}_{2} \in 
\mathbb{R}^{n \times 1}$ and real numbers $\lambda \in 
\mathbb{R}$, mappings defined by real-valued matrices $\mathbf{A} 
\in \mathbb{R}^{m \times n}$ exhibit the two special properties
\be
\lb{lin}
\fbox{$\displaystyle\begin{array}{c}
\mathbf{A}(\vec{x}_{1}+\vec{x}_{2})
= (\mathbf{A}\vec{x}_{1})+(\mathbf{A}\vec{x}_{2}) \\[5mm]
\mathbf{A}(\lambda\vec{x}_{1})
= \lambda(\mathbf{A}\vec{x}_{1}) \ ,
\end{array}
$}
\ee
they constitute {\bf linear mappings}.\footnote{It is important to 
note at this point that many advanced mathematical models designed 
to describe quantitative aspects of some natural and economic 
phenomena do \emph{not} satisfy the conditions (\ref{lin}), as 
they employ \emph{non-linear mappings} for this purpose. However, 
in such contexts, linear mappings often provide useful first 
approximations.}

\medskip
\noindent
We now turn to discuss the most important mathematical operations 
defined for $(m \times n)$-matrices, as well as the computational 
rules that obtain.

\section[Basic concepts]%
{Basic concepts}
\lb{sec:matrech}

\noindent
\underline{\bf Def.:}
{\bf Transpose} of a matrix

\noindent
For $\mathbf{A} \in \mathbb{R}^{m \times n}$, we define the 
process of transposing $\mathbf{A}$ by
\be
\fbox{$\displaystyle
\mathbf{A}^{T}\!: \quad
a_{ij}^{T} := a_{ji} \ ,
$}
\ee
where $i=1,\ldots,m$ und $j=1,\ldots,n$. Note that it holds that
$\mathbf{A}^{T} \in \mathbb{R}^{n \times m}$.

\medskip
\noindent
When transposing an $(m \times n)$-matrix, one simply has to 
exchange the matrix' rows with its columns (and vice versa): the 
elements of the first row become the elements of the first column, 
etc. It follows that, in particular,
\be
(\mathbf{A}^{T})^{T} = \mathbf{A}
\ee
applies.

\medskip
\noindent
Two special cases may occur for quadratic matrices (where $m=n$):
\begin{itemize}
	\item When $\mathbf{A}^{T}=\mathbf{A}$, one refers to 
	$\mathbf{A}$ as a {\bf symmetric matrix}.
	\item When $\mathbf{A}^{T}=-\mathbf{A}$, one refers to 
	$\mathbf{A}$ as an {\bf antisymmetric matrix}.
\end{itemize}

\medskip
\noindent
\underline{\bf Def.:}
{\bf Addition} of matrices

\noindent
For $\mathbf{A}, \mathbf{B} \in \mathbb{R}^{m \times n}$, the sum 
is given by
\be
\fbox{$
\displaystyle
\mathbf{A} + \mathbf{B} =: \mathbf{C}\!: \quad
a_{ij} + b_{ij} =: c_{ij} \ ,
$}
\ee
with $i=1,\ldots,m$ and $j=1,\ldots,n$.

\medskip
\noindent
Note that an addition can be performed meaningfully only for 
matrices of the \emph{same format}.

\pagebreak
\medskip
\noindent
\underline{\bf Def.:}
{\bf Rescaling} of matrices

\noindent
For $\mathbf{A} \in \mathbb{R}^{m \times n}$
and $\lambda \in \mathbb{R}\backslash \{0\}$, let
\be
\fbox{$\displaystyle
\lambda\mathbf{A} =: \mathbf{C}\!: \quad
\lambda a_{ij} =: c_{ij} \ ,
$}
\ee
where $i=1,\ldots,m$ and $j=1,\ldots,n$.

\medskip
\noindent
When rescaling a matrix, all its elements simply have to be 
multiplied by the same non-zero real number~$\lambda$.

\medskip
\noindent
{\bf Computational rules for addition and rescaling of matrices}

\noindent
For matrices $\mathbf{A}, \mathbf{B}, \mathbf{C}
\in \mathbb{R}^{m \times n}$:

\begin{enumerate}
	\item $\mathbf{A}+\mathbf{B} = \mathbf{B}+\mathbf{A}$
	\hfill ({\bf commutative addition})
	\item $\mathbf{A}+(\mathbf{B}+\mathbf{C})
	= (\mathbf{A}+\mathbf{B})+\mathbf{C}$
	\hfill ({\bf associative addition})
	\item $\mathbf{A}+\mathbf{0} = \mathbf{A}$
	\hfill ({\bf addition identity element})
	\item For every $\mathbf{A}$ and $\mathbf{B}$, there exists 
	exactly one $\mathbf{Z}$ such that 
	$\mathbf{A}+\mathbf{Z}=\mathbf{B}$.
	
	\hfill ({\bf invertibility of addition})
	\item $(\lambda\mu)\mathbf{A}=\lambda(\mu\mathbf{A})$
	with $\lambda,\mu	\in \mathbb{R} \backslash \{0\}$ 
	\hfill ({\bf associative rescaling})
	\item $1\mathbf{A}=\mathbf{A}$
	\hfill ({\bf rescaling identity element})
	\item $\lambda(\mathbf{A}+\mathbf{B})
	= \lambda\mathbf{A}+\lambda\mathbf{B}$;
	
	$(\lambda+\mu)\mathbf{A} = \lambda\mathbf{A}+\mu\mathbf{A}$
	with $\lambda, \mu	\in \mathbb{R} \backslash \{0\}$
	\hfill ({\bf distributive rescaling})
	
	\item $(\mathbf{A}+\mathbf{B})^{T} = \mathbf{A}^{T} + 
	\mathbf{B}^{T}$ \hfill ({\bf transposition rule 1})
	
	\item $(\lambda\mathbf{A})^{T} = \lambda\mathbf{A}^{T}$ with 
	$\lambda \in \mathbb{R} \backslash \{0\}$. \hfill ({\bf 
	transposition rule 2})
\end{enumerate}

\medskip
\noindent
Next we introduce a particularly useful mathematical operation for 
matrices.

\section[Matrix multiplication]%
{Matrix multiplication}
\lb{sec:matmult}

\noindent
\underline{\bf Def.:}
For a real-valued $(m \times n)$-matrix $\mathbf{A}$ and a 
real-valued $(n \times r)$-matrix $\mathbf{B}$, a {\bf matrix 
multiplication} is defined by
\be
\fbox{$
\displaystyle\begin{array}{c}
\mathbf{A}\mathbf{B} =: \mathbf{C} \\[5mm]
a_{i1}b_{1j}+\ldots+a_{ik}b_{kj}+\ldots
+a_{in}b_{nj}
=: \sum_{k=1}^{n}a_{ik}b_{kj} =: c_{ij} \ ,
\end{array}
$}
\ee
with $i=1,\ldots,m$ and $j=1,\ldots,r$, thus yielding as an 
outcome a real-valued $(m \times r)$-matrix $\mathbf{C}$.

\medskip
\noindent
The element of $\mathbf{C}$ at the intersection of the $i$th row 
and the $j$th column is determined by the computational rule
\be
c_{ij} = \text{Euclidian scalar product of $i$th row vector of
$\mathbf{A}$ and $j$th column vector of $\mathbf{B}$} \ .
\ee
It is important to realise that the definition of a matrix 
multiplication just provided depends in an essential way on the 
fact that \emph{matrix $\mathbf{A}$ on the left in the product 
needs to have as many (!) columns as matrix $\mathbf{B}$ on the 
right rows}. Otherwise, a matrix multiplication \emph{cannot} be 
defined in a meaningful way.

\medskip
\noindent
\underline{\bf GDC:} For matrices ${\tt [A]}$ and ${\tt [B]}$ 
edited beforehand, of matching formats, their matrix 
multiplication can be evaluated in mode {\tt MATRIX} $\rightarrow$ 
{\tt NAMES} by ${\tt [A]*[B]}$.

\vspace{5mm}
\noindent
{\bf Computational rules for matrix multiplication}

\noindent
For $\mathbf{A}, \mathbf{B}, \mathbf{C}$ real-valued matrices of 
correspondingly matching formats we have:

\begin{enumerate}
	\item $\mathbf{A}\mathbf{B} = \mathbf{0}$ is possible with $\mathbf{A}\neq\mathbf{0},\mathbf{B}\neq\mathbf{0}$.
	\hfill ({\bf zero divisor})
	\item $\mathbf{A}(\mathbf{B}\mathbf{C})
	= (\mathbf{A}\mathbf{B})\mathbf{C}$
	\hfill ({\bf associative matrix multiplication})
	\item $\mathbf{A}
	\underbrace{\mathbf{1}}_{\in \mathbb{R}^{n \times n}}
	=\underbrace{\mathbf{1}}_{\in \mathbb{R}^{m \times m}}
	\mathbf{A}=\mathbf{A}$
	\hfill ({\bf multiplicative identity element})
	\item $(\mathbf{A}+\mathbf{B})\mathbf{C}
	= \mathbf{A}\mathbf{C}+\mathbf{B}\mathbf{C}$
	
	$\mathbf{C}(\mathbf{A}+\mathbf{B})
	= \mathbf{C}\mathbf{A}+\mathbf{C}\mathbf{B}$
	\hfill ({\bf distributive matrix multiplication})
	\item $\mathbf{A}(\lambda\mathbf{B})
	=(\lambda\mathbf{A})\mathbf{B}
	=\lambda(\mathbf{A}\mathbf{B})$
	with $\lambda \in \mathbb{R}$ 
	\hfill ({\bf homogeneous matrix multiplication})
	\item $(\mathbf{A}\mathbf{B})^{T}
	=\mathbf{B}^{T}\mathbf{A}^{T}$
	\hfill ({\bf transposition rule}).
\end{enumerate}


\chapter[Systems of linear algebraic equations]{Systems of linear 
algebraic equations}
\lb{ch3}

\vspace{10mm}
\noindent
In this chapter, we turn to address a particular field of 
application of the notions of matrices and vectors, or of linear 
mappings in general.

\section[Basic concepts]%
{Basic concepts}
\lb{sec:lgsgrund}
Let us begin with a system of $m \in \mathbb{N}$ \emph{linear} 
algebraic equations, wherein every single equation can be 
understood to constitute a {\bf constraint} on the range of values 
of $n \in \mathbb{N}$ variables $x_{1}, 
\ldots, x_{n} \in \mathbb{R}$. The objective is to determine all 
possible values of $x_{1}, \ldots, x_{n} \in \mathbb{R}$ which 
satisfy these constraints simultaneously. Problems of this kind, 
namely {\bf systems of linear algebraic equations}, are often 
represented in the form

\medskip
\noindent
$\bullet$ Representation~1:\\[-7mm]
\bea
a_{11}x_{1}+\ldots+a_{1j}x_{j}+\ldots+a_{1n}x_{n} & = & b_{1} 
\nonumber \\
 & \vdots & \nonumber \\
a_{i1}x_{1}+\ldots+a_{ij}x_{j}+\ldots+a_{in}x_{n} & = & b_{i} \\
 & \vdots & \nonumber \\
a_{m1}x_{1}+\ldots+a_{mj}x_{j}+\ldots+a_{mn}x_{n} & = & b_{m}
\nonumber \ .
\eea
Depending on how the natural numbers $m$ and $n$ relate to one 
another, systems of linear algebraic equations can be classified 
as follows:\\[-7mm]
\begin{itemize}
\item $m < n$: fewer equations than variables; the linear system 
is {\bf under-determined},\\[-7mm]

\item $m = n$: same number of equations as variables;
the linear system is {\bf well-determined},\\[-7mm]

\item $m > n$: more equations than variables; the linear system is 
{\bf over-determined}.\\[-7mm]
\end{itemize}
A more compact representation of a linear system of format
$(m \times n)$ is given by

\medskip
\noindent
$\bullet$ Representation~2:
\be
\fbox{$\displaystyle
\mathbf{A}\vec{x} =
\left(\begin{array}{ccccc}
a_{11} & \ldots & a_{1j} & \ldots & a_{1n} \\
\vdots & \ddots & \vdots & \ddots & \vdots \\
a_{i1} & \ldots & a_{ij} & \ldots & a_{in} \\
\vdots & \ddots & \vdots & \ddots & \vdots \\
a_{m1} & \ldots & a_{mj} & \ldots & a_{mn}
\end{array}\right)
\left(\begin{array}{c}
x_{1} \\
\vdots \\
x_{j} \\
\vdots \\
x_{n}
\end{array}\right)
= \left(
\begin{array}{c}
b_{1} \\
\vdots \\
b_{i} \\
\vdots \\
b_{m}
\end{array}\right)
= \vec{b} \ .
$}
\ee
The mathematical objects employed in this variant of a linear 
system are as follows: $\mathbf{A}$ takes the central role of the 
{\bf coefficient matrix} of the linear system, of format $(m 
\times n)$, $\vec{x}$ is its {\bf variable vector}, of format $(n 
\times 1)$, and, lastly, $\vec{b}$ is its {\bf image vector}, of 
format $(m \times 1)$.

\medskip
\noindent
When dealing with systems of linear algebraic equations in the 
form of Representation 2, i.e. $\mathbf{A}\vec{x} = \vec{b}$, the 
main question to be answered is:

\medskip
\noindent
\underline{\bf Question:}
For given {\bf coefficient matrix} $\mathbf{A}$ and {\bf image 
vector} $\vec{b}$, can we find a {\bf variable vector} $\vec{x}$ 
that $\mathbf{A}$ maps onto $\vec{b}$?

\medskip
\noindent
In a sense this describes the inversion of the photographic 
process we had previously referred to: we \emph{have} given the 
camera and we already \emph{know} the image, but we have yet to 
find a matching object. Remarkably, to address this issue, we can 
fall back on a simple algorithmic method due to the German 
mathematician and astronomer 
\href{http://www-groups.dcs.st-and.ac.uk/~history/Biographies/Gauss.html}{Carl Friedrich Gau\ss\ (1777--1855)}.

\section[Gau\ss ian elimination]%
{Gau\ss ian elimination}
\lb{sec:lgsgauss}
The algorithmic solution technique developed by Gau\ss\ is based 
on the insight that the solution set of a {\bf linear system} of 
$m$ algebraic equations for $n$ real-valued variables, i.e.
\be
\fbox{$\displaystyle
\mathbf{A}\vec{x} = \vec{b} \ ,
$}
\ee
remains unchanged under the following algebraic {\bf equivalence 
transformations} of the linear system:
\begin{enumerate}
\item changing the order amongst the equations,
\item multiplication of any equation by a non-zero real number $c \neq 0$,
\item addition of a multiple of one equation to another equation,
\item changing the order amongst the equations.
\end{enumerate}
Specifically, this implies that we may manipulate a given linear 
system by means of these four different kinds of equivalence 
transformations without ever changing its identity. In concrete 
cases, however, one should not apply these equivalence 
transformations at random but rather follow a target oriented 
strategy. This is what Gau\ss ian elimination can provide.

\medskip
\noindent
\underline{\bf Target:} To cast the {\bf augmented coefficient 
matrix} $(\mathbf{A}|\vec{b})$, i.e., the array
\be
\begin{array}{ccccc|c}
a_{11} & \ldots & a_{1j} & \ldots & a_{1n} & b_{1} \\
\vdots & \ddots & \vdots & \ddots & \vdots & \vdots \\
a_{i1} & \ldots & a_{ij} & \ldots & a_{in} & b_{i} \\
\vdots & \ddots & \vdots & \ddots & \vdots & \vdots \\
a_{m1} & \ldots & a_{mj} & \ldots & a_{mn} & b_{m}
\end{array} \ ,
\ee
when possible, into {\bf upper triangular form}
\be
\begin{array}{ccccc|c}
1 & \ldots & \tilde{a}_{1j} & \ldots & \tilde{a}_{1n} & \tilde{b}_{1} \\
\vdots & \ddots & \vdots & \ddots & \vdots & \vdots \\
0 & \ldots & \tilde{a}_{ij} & \ldots & \tilde{a}_{in} & \tilde{b}_{i} \\
\vdots & \ddots & \vdots & \ddots & \vdots & \vdots \\
0 & \ldots & 0 & \ldots & \tilde{a}_{mn} & \tilde{b}_{m}
\end{array} \ ,
\ee
by means of the four kinds of equivalence transformations such 
that the resultant simpler final linear system may easily be 
solved using {\bf backward substitution}.

\medskip
\noindent
Three exclusive cases of possible {\bf solution content} for a 
given system of linear algebraic equations do exist. The linear 
system may possess either
\begin{enumerate}
\item \emph{no solution} at all, or

\item a \emph{unique solution}, or

\item \emph{multiple solutions}.
\end{enumerate}
\underline {\bf Remark:} Linear systems that are under-determined, 
i.e., when $m < n$, can \emph{never} be solved uniquely due to the 
fact that in such a case there not exist enough equations to 
constrain the values of \emph{all} of the $n$ variables.

\medskip
\noindent
\underline {\bf GDC:} For a stored augmented coefficient matrix 
${\tt [A]}$ of format $(m \times n+1)$, associated with a given  
$(m \times n)$ linear system, select mode {\tt MATRIX} 
$\rightarrow$ {\tt MATH} and then call the function ${\tt 
rref([A])}$. It is possible that backward substitution needs to be 
employed to obtain the final solution.

\medskip
\noindent
For completeness, we want to turn briefly to the issue of 
solvability of a system of linear algebraic equations. To this 
end, we need to introduce the notion of the rank of a matrix.

\section[Rank of a matrix]%
{Rank of a matrix}
\lb{sec:lgsrang}

\noindent
\underline{\bf Def.:}
A real-valued matrix $\mathbf{A} \in \mathbb{R}^{m \times n}$ 
possesses the {\bf rank}
\be
\fbox{$\displaystyle
\text{rank}(\mathbf{A}) = r \ , \qquad
r \leq \text{min}\{m,n\}
$}
\ee
if and only if $r$ is the {\bf maximum number} of row resp.~column 
vectors of $\mathbf{A}$ which are linearly independent. Clearly, 
$r$ can only be as large as the smaller of the numbers $m$ and $n$ 
that determine the format of $\mathbf{A}$.

\medskip
\noindent
For {\bf quadratic matrices} $\mathbf{A} \in
\mathbb{R}^{n \times n}$, there is available a more elegant 
measure to determine its rank. This (in the present case 
real-valued) measure is referred to as the {\bf determinant} of 
matrix $\mathbf{A}$, $\det(\mathbf{A})$, and is defined as follows.

\medskip
\noindent
\underline{\bf Def.:}
\begin{itemize}
\item[(i)]~When $\mathbf{A} \in \mathbb{R}^{2 \times 2}$,
its {\bf determinant} is given by
\be
\det(\mathbf{A})
:= \left|\begin{array}{cc}
   	a_{11} & a_{12} \\
   	a_{21} & a_{22}
	\end{array}\right|
:= a_{11}a_{22} - a_{12}a_{21} \ ,
\ee
i.e. the difference between the products of $\mathbf{A}$'s 
on-diagonal elements and $\mathbf{A}$'s off-diagonal elements.

\item[(ii)]~When $\mathbf{A} \in \mathbb{R}^{3 \times 3}$, the 
definition of $\mathbf{A}$'s {\bf determinant} is more complex. In 
that case it is given by
\bea
\det(\mathbf{A})
& := & \left|\begin{array}{ccc}
   	a_{11} & a_{12} & a_{13} \\
   	a_{21} & a_{22} & a_{23} \\
   	a_{31} & a_{32} & a_{33}
	\end{array}\right| \nonumber \\
& := & a_{11}(a_{22}a_{33}-a_{32}a_{23})
+ a_{21}(a_{32}a_{13}-a_{12}a_{33})
+ a_{31}(a_{12}a_{23}-a_{22}a_{13}) \ .
\eea
Observe, term by term, the cyclic permutation of the first index 
of the elements $a_{ij}$ according to the rule $1 \rightarrow 2 
\rightarrow 3 \rightarrow 1$.

\item[(iii)]~Finally, for the (slightly involved) definition of 
the {\bf determinant} of a higher-dimensional matrix $\mathbf{A} 
\in \mathbb{R}^{n \times n}$, please refer to the literature; e.g. 
Bronstein \emph{et al} (2005)~\ct[p~267]{broetal2005}.
\end{itemize}

\medskip
\noindent
To determine the rank of a given quadtratic matrix $\mathbf{A}
\in \mathbb{R}^{n \times n}$, one now installs the following 
criteria: $\text{rank}(\mathbf{A}) = r = n$, if $\det(\mathbf{A}) 
\neq 0$, and $\text{rank}(\mathbf{A}) = r < n$, if 
$\det(\mathbf{A}) = 0$. In the first case, $\mathbf{A}$ is 
referred to as {\bf regular}, in the second as {\bf singular}. For 
quadratic matrices $\mathbf{A}$ that are singular, 
$\text{rank}(\mathbf{A}) = r$ (with $r < n$) is given by the 
number $r$ of rows (or columns) of the largest possible non-zero 
subdeterminant of $\mathbf{A}$.

\medskip
\noindent
\underline {\bf GDC:} For a stored quadratic matrix ${\tt [A]}$, 
select mode {\tt MATRIX} $\rightarrow$ {\tt MATH} and obtain its 
determinant by calling the function ${\tt det([A])}$.

\section[Solution criteria]%
{Criteria for solving systems of linear algebraic equations}
\lb{sec:lgsloes}
Making use of the concept of the {\bf rank} of a real-valued 
matrix $\mathbf{A} \in \mathbb{R}^{m \times n}$, we can now 
summarise the solution content of a specific system of linear 
algebraic equations of format $(m \times n)$ in a table. For given 
linear system
$$
\mathbf{A}\vec{x} = \vec{b} \ ,
$$
with coefficient matrix $\mathbf{A} \in \mathbb{R}^{m \times n}$, 
variable vector $\vec{x} \in \mathbb{R}^{n \times 1}$ and image 
vector $\vec{b} \in \mathbb{R}^{m \times 1}$,
there exist(s)
\begin{center}
	\begin{tabular}[h]{c|c|c}
		\hline\hline
		 & & \\
		 & $\vec{b} \neq \vec{0}$ &
		 $\vec{b} = \vec{0}$ \\
		 & & \\
		\hline
		 & & \\
		1. $\text{rank}(\mathbf{A}) \neq \text{rank}(\mathbf{A}|\vec{b})$ &
		no solution & -------- \\
		 & & \\
		\hline
		 & & \\
		2. $\text{rank}(\mathbf{A}) = \text{rank}(\mathbf{A}|\vec{b}) = r$ &
		& \\
		 & & \\
		(a) \quad $r = n$ & a unique & $\vec{x} = \vec{0}$ \\
		 & solution & \\
		 & & \\
		(b) \quad $r < n$ & multiple & multiple \\
		 & solutions: & solutions: \\
		 & $n-r$ free &  $n-r$ free \\
		 & parameters & parameters \\
		 & & \\
		\hline\hline
	 \end{tabular}
\end{center}

\noindent
$(\mathbf{A}|\vec{b})$ here denotes the augmented coefficient 
matrix.

\medskip
\noindent
Next we discuss a particularly useful property of \emph{regular} 
quadratic matrices.

\section[Inverse of a regular $(n\times n)$-matrix]%
{Inverse of a regular $(n\times n)$-matrix}
\lb{sec:lgsinv}

\noindent
\underline{\bf Def.:}
Let a real-valued quadratic matrix $\mathbf{A} \in
\mathbb{R}^{n \times n}$ be {\bf regular}, i.e., $\det(\mathbf{A}) 
\in \mathbb{R}\backslash\{0\}$. Then there exists an {\bf inverse 
matrix} $\mathbf{A}^{-1}$ to $\mathbf{A}$ defined by the 
characterising properties
\be
\fbox{$\displaystyle
\mathbf{A}^{-1}\mathbf{A} = \mathbf{A}\mathbf{A}^{-1}
= \mathbf{1} \ .
$}
\ee
Here $\mathbf{1}$ denotes the $\boldsymbol{(n \times n)}${\bf -unit matrix} [cf.\ Eq.~(\ref{einmatr})].

\medskip
\noindent
When a computational device is not at hand, the inverse matrix 
$\mathbf{A}^{-1}$ of a regular quadratic matrix $\mathbf{A}$ can 
be obtained by solving the matrix-valued linear system
\be
\mathbf{A}\mathbf{X} \stackrel{!}{=} \mathbf{1}
\ee
for the unknown matrix $\mathbf{X}$ by means of {\bf simultaneous 
Gau\ss ian elimination}.

\medskip
\noindent
\underline{\bf GDC:} For a stored quadratic matrix ${\tt [A]}$, 
its inverse matrix can be simply obtained as ${\tt [A]}^{-1}$, 
where the $x^{-1}$ function key needs to be used.

\pagebreak
\medskip
\noindent
{\bf Computational rules for the inverse operation}

\nopagebreak
\noindent
For $\mathbf{A},\mathbf{B} \in \mathbb{R}^{n \times n}$,
with $\det(\mathbf{A}) \neq 0 \neq \det(\mathbf{B})$, it holds that

\begin{enumerate}
\item $(\mathbf{A}^{-1})^{-1} = \mathbf{A}$
\item $(\mathbf{A}\mathbf{B})^{-1}
= \mathbf{B}^{-1}\mathbf{A}^{-1}$
\item $(\mathbf{A}^{T})^{-1} = (\mathbf{A}^{-1})^{T}$
\item $\displaystyle (\lambda\mathbf{A})^{-1}
= \frac{1}{\lambda}\,\mathbf{A}^{-1}$.
\end{enumerate}

\medskip
\noindent
The special interest in applications in the concept of {\bf 
inverse matrices} arises for the following reason. Consider given 
a well-determined linear system
\[
\mathbf{A}\vec{x}=\vec{b} \ ,
\]
with \emph{regular} quadratic coefficient matrix $\mathbf{A} 
\in \mathbb{R}^{n \times n}$, i.e., $\det(\mathbf{A}) \neq 0$. 
Then, for $\mathbf{A}$, there exists an inverse matrix 
$\mathbf{A}^{-1}$. Matrix-multiplying both sides of the equation 
above \emph{from the left~(!)} by the inverse $\mathbf{A}^{-1}$, 
results in
\be
\underbrace{\mathbf{A}^{-1}(\mathbf{A}\vec{x})
= (\mathbf{A}^{-1}\mathbf{A})\vec{x}
= \mathbf{1}\vec{x}
= \vec{x}}_{\text{left-hand side}}
= \underbrace{\mathbf{A}^{-1}\vec{b}}_{\text{right-hand side}} \ .
\ee
In this case, the \emph{unique solution~(!)} $\vec{x} = 
\mathbf{A}^{-1}\vec{b}$ of the linear system arises simply from 
matrix multiplication of the image vector $\vec{b}$ by the inverse 
matrix of $\mathbf{A}$. (Of course, it might actually require a 
bit of computational work to determine $\mathbf{A}^{-1}$.)

\section[Outlook]{Outlook}
\lb{sec:lgsausblick}
There are a number of exciting advanced topics in {\bf Linear 
Algebra}. Amongst them one finds the concept of the characteristic 
{\bf eigenvalues} and associated {\bf eigenvectors} of {\bf 
quadratic matrices}, which has particularly high relevance in 
practical applications. The question to be answered here is the 
following: for given real-valued quadratic matrix $\mathbf{A} \in 
\mathbb{R}^{n \times n}$, do there exist real numbers $\lambda_{n} 
\in \mathbb{R}$ and real-valued vectors $\vec{v}_{n} \in 
\mathbb{R}^{n \times 1}$ which satisfy the condition
\be
\lb{eigenveq1}
\mathbf{A}\vec{v}_{n} \stackrel{!}{=} \lambda_{n}\vec{v}_{n} \ ?
\ee
Put differently: for which vectors $\vec{v}_{n} \in \mathbb{R}^{n 
\times 1}$ does their mapping by a quadratic matrix $\mathbf{A} 
\in \mathbb{R}^{n \times n}$ amount to simple rescalings by real 
numbers $\lambda_{n} \in \mathbb{R}$?

\medskip
\noindent
By re-arranging, Eq.~(\ref{eigenveq1}) can be recast into the form
\be
\lb{eigenveq2}
\boldsymbol{0} \stackrel{!}{=}
\left(\mathbf{A}-\lambda_{n}\boldsymbol{1}\right)\vec{v}_{n} \ ,
\ee
with $\boldsymbol{1}$ an $(n \times n)$-unit matrix
[cf.\ Eq.~(\ref{einmatr})] and $\boldsymbol{0}$ an $n$-component 
zero vector. This condition corresponds to a homogeneous system of 
linear algebraic equations of format $(n \times n)$. Non-trivial 
solutions $\vec{v}_{n} \neq \boldsymbol{0}$ to this system exist 
provided that the so-called {\bf characteristic equation}
\be
0 \stackrel{!}{=} 
\det\left(\mathbf{A}-\lambda_{n}\boldsymbol{1}\right) \ ,
\ee
a polynomial of degree $n$ (cf. Sec.~\ref{subsec:polynomials}), 
allows for real-valued roots $\lambda_{n} \in \mathbb{R}$. Note 
that \emph{symmetric} quadratic matrices (cf. 
Sec.~\ref{sec:matrech}) possess exclusively real-valued 
eigenvalues $\lambda_{n}$. When these eigenvalues turn out to be 
all \emph{different}, then the associated 
eigenvectors~$\vec{v}_{n}$ prove to be mutually orthogonal.

\medskip
\noindent
Knowledge of the spectrum of {\bf eigenvalues} $\lambda_{n} \in 
\mathbb{R}$ and associated {\bf eigenvectors} $\vec{v}_{n} \in 
\mathbb{R}^{n \times 1}$ of a real-valued matrix
$\mathbf{A} \in \mathbb{R}^{n \times n}$ provides the basis of a 
transformation of $\mathbf{A}$ to its {\bf diagonal 
form}~$\mathbf{A}_{\lambda_{n}}$, thus yielding a diagonal matrix 
which features the eigenvalues $\lambda_{n}$ as its on-diagonal 
elements; cf. Leon (2009)~\ct{leo2009}.

\medskip
\noindent
Amongst other examples, the concept of eigenvalues and 
eigenvectors of quadratic real-valued matrices plays a special 
role in {\bf Statistics}, in the context of exploratory {\bf 
principal component analyses} of multivariate data sets, where the 
objective is to identify dominant intrinsic structures; cf. Hair 
\emph{et al} (2010)~\ct[Ch.~3]{haietal2010} and 
Ref.~\ct[App.~A]{hve2015}.


\chapter[Leontief's 
input--output matrix model]{Leontief's stationary
input--output matrix model}
\lb{ch4}

\vspace{10mm}
\noindent
We now turn to discuss some specific applications of {\bf Linear 
Algebra} in economic theory. To begin with, let us consider 
quantitative aspects of the exchange of goods between a certain 
number of {\bf economic agents}. We here aim at a simplified 
abstract description of real economic processes.

\section[General considerations]%
{General considerations}
\lb{sec:leongrund}
The quantitative model to be described in the following is due to 
the Russian economist 
\href{http://en.wikipedia.org/wiki/Leontief}{Wassily Wassilyovich 
Leontief (1905--1999)}, cf. Leontief (1936)~\ct{leo1936}, for 
which, besides other important contributions, he was awarded the 
1973 
\href{http://www.nobelprize.org/nobel_prizes/economics/laureates/1973/}{Sveriges Riksbank Prize in Economic Sciences in Memory of 
Alfred Nobel}. 

\medskip
\noindent
Suppose given an economic system consisting of $n \in 
\mathbb{N}$~{\bf interdependent economic agents} exchanging 
between them the goods they produce. For simplicity we want to 
\emph{assume} that every one of these {\bf economic agents} 
represents the production of a \emph{single} good only. Presently 
we intend to monitor the flow of goods in this simple economic 
system during a specified {\bf reference period of time}. The 
total numbers of the $n$~goods leaving the production sector of 
this model constitute the {\bf OUTPUT quantities}. The {\bf INPUT 
quantities} to the production sector are twofold. On the one hand, 
there are {\bf exogenous} INPUT quantities which we take to be 
given by $m \in \mathbb{N}$ different kinds of external {\bf 
resources} needed in differing proportions to produce the 
$n$~goods. On the other hand, due to their mutual interdependence, 
some of the {\bf economic agents} require {\bf goods made by their 
neighbours} to be able to produce their own goods; these 
constitute the {\bf endogenous} INPUT quantities of the system. 
Likewise, the production sector's total OUTPUT during the chosen 
reference period of the $n$~goods can be viewed to flow through 
one of \emph{two} separate channels: (i)~the {\bf exogenous} 
channel linking the production sector to {\bf external consumers} 
representing an open market, and (ii)~the {\bf endogenous} channel 
linking the {\bf economic agents} to their {\bf neighbours} (thus 
respresenting their interdependencies). It is supposed that 
momentum is injected into the economic system, triggering the flow 
of goods between the different actors, by the prospect of {\bf 
increasing the value} of the INPUT quantities, in line with the 
notion of the economic {\bf value chain}.

\medskip
\noindent
Leontief's model is based on the following three elementary

\medskip
\noindent
{\bf Assumptions}:
\begin{enumerate}
\item For all goods involved the functional relationship between 
INPUT and OUTPUT quantities be of a {\bf linear nature} [cf.\ 
Eq.~(\ref{lin})].

\item The proportions of ``INPUT quantities to OUTPUT quantities'' be {\bf constant} over the reference period of time considered; the flows of goods are thus considered to be {\bf stationary}.

\item {\bf Economic equilibrium} obtains during the reference period of time: the numbers of goods then supplied equal the numbers of goods then demanded.
\end{enumerate}
The mathematical formulation of Leontief's quantitative
model employs the following

\medskip
\noindent
{\bf Vector- and matrix-valued quantities}:
%
\begin{enumerate}

\item $\vec{q}$ --- {\bf total output vector}
$\in \mathbb{R}^{n \times 1}$, components $q_{i} \geq 
0~\text{units}$
\hfill (dim: $\text{units}$)

\item $\vec{y}$ --- {\bf final demand vector}
$\in \mathbb{R}^{n \times 1}$, components $y_{i} \geq 
0~\text{units}$
\hfill (dim: $\text{units}$)

\item $\mathbf{P}$ --- {\bf input--output matrix}
$\in \mathbb{R}^{n \times n}$, components $P_{ij} \geq 0$
\hfill (dim: 1)

\item $(\mathbf{1}-\mathbf{P})$ --- {\bf technology matrix}
$\in \mathbb{R}^{n \times n}$, regular, hence, invertible
\hfill (dim: 1)

\item $(\mathbf{1}-\mathbf{P})^{-1}$ --- {\bf total demand matrix}
$\in \mathbb{R}^{n \times n}$
\hfill (dim: 1)

\item $\vec{v}$ --- {\bf resource vector}
$\in \mathbb{R}^{m \times 1}$, components $v_{i} \geq 
0~\text{units}$
\hfill (dim: $\text{units}$)

\item $\mathbf{R}$ --- {\bf resource consumption matrix}
$\in \mathbb{R}^{m \times n}$, components $R_{ij} \geq 0$,
\hfill (dim: 1)

\end{enumerate}
where $\mathbf{1}$ denotes the {\bf $\boldsymbol{(n \times 
n)}$-unit matrix} [cf.\ Eq.~(\ref{einmatr})]. Note that the 
components of all the vectors involved, as well as of the 
input--output matrix and of the resource consumption matrix, can 
assume \emph{non-negative values (!)} only.

\section[Input--output matrix and resource consumption matrix]%
{Input--output matrix and resource consumption matrix}
\lb{sec:inoutmat}
We now turn to provide the definition of the two central matrix-valued quantities in Leontief's model. We will also highlight their main characteristic features.

\subsection{Input--output matrix}
Suppose the {\bf reference period of time} has ended for the 
economic system in question, i.e. the stationary {\bf flows of 
goods} have stopped eventually. We now want to take stock of the 
{\bf numbers of goods} that have been delivered by each of the 
$n$~{\bf economic agents} in the system. Say that during the 
reference period considered, agent~$1$ delivered of their good the 
number $n_{11}$ to themselves, the number $n_{12}$ to agent~$2$, 
the number $n_{13}$ to agent~$3$, and so on, and, lastly, the 
number $n_{1n}$ to agent~$n$. The number delivered by agent~$1$ to 
external consumers shall be denoted by $y_{1}$. Since in this 
model a good produced \emph{cannot} all of a sudden disappear 
again, and since by Assumption~3 above the number of goods 
supplied must be equal to the number of goods demanded, we find 
that for the total output of agent~$1$ it holds that 
$q_{1}:=n_{11} + \ldots + n_{1j} + \ldots + n_{1n} + y_{1}$. 
Analogous relations hold for the total output $q_{2}$, $q_{3}$, 
\ldots, $q_{n}$ of each of the remaining $n-1$ agents. We thus 
obtain the intermediate result
\begin{eqnarray}
q_{1} & = & n_{11} + \ldots + n_{1j} + \ldots + n_{1n} + y_{1} > 0
\\
 & \vdots & \nonumber \\
q_{i} & = & n_{i1} + \ldots + n_{ij} + \ldots + n_{in} + y_{i} > 0
\\
 & \vdots & \nonumber \\
q_{n} & = & n_{n1} + \ldots + n_{nj} + \ldots + n_{nn} + y_{n} > 0
\ .
\end{eqnarray}
This simple system of {\bf balance equations} can be summarised in terms of a standard {\bf input--output table} as follows:
\begin{center}
    \begin{tabular}[h]{c|ccccc|c|c}
    \hline\hline
    [Values in $\text{units}$] & agent~$1$ & $\cdots$ & agent~$j$ & $\cdots$ & agent~$n$ & external consumers & $\Sigma$: total output  \\
    \hline
    agent~$1$ & $n_{11}$ & $\ldots$ & $n_{1j}$ & $\ldots$ & $n_{1n}$ & $y_{1}$ & $q_{1}$ \\
    $\vdots$ & $\vdots$ & $\ddots$ & $\vdots$ & $\ddots$ & $\vdots$ & $\vdots$ & $\vdots$ \\
    agent~$i$ & $n_{i1}$ & \ldots & $n_{ij}$ & \ldots & $n_{in}$ & $y_{i}$ & $q_{i}$ \\
    $\vdots$ & $\vdots$ & $\ddots$ & $\vdots$ & $\ddots$ & $\vdots$ & $\vdots$ & $\vdots$ \\
    agent~$n$ & $n_{n1}$ & \ldots & $n_{nj}$ & \ldots & $n_{nn}$ & $y_{n}$ & $q_{n}$ \\
    \hline\hline
    \end{tabular}
\end{center}
The first column of this table lists all the $n$ different {\bf 
sources of flows of goods} (or suppliers of goods), while its 
first row shows the $n+1$ different {\bf sinks of flows of goods} 
(or consumers of goods). The last column contains the total output 
of each of the $n$ agents in the {\bf reference period of time}.

\medskip
\noindent
Next we compute for each of the $n$ agents the respective values 
of the \emph{non-negative ratios}
\be
P_{ij} := \frac{\text{INPUT\ (in\ units)\ of\ agent\ $i$\ for
\ agent $j$\ (during\ reference\ period)}}{
\text{OUTPUT\ (in\ units)\ of\ agent\ $j$\ (during\ reference
\ period)}} \ ,
\ee
or, employing a compact and economical index 
notation,\footnote{Note that the normalisation quantities in these 
ratios $P_{ij}$ are given by the total output $q_{j}$ of the 
receiving agent~$j$ and \emph{not} by the total output $q_{i}$ of 
the supplying agent~$i$. In the latter case the $P_{ij}$ would 
represent percentages of the total output $q_{i}$.}
\be
\fbox{$\displaystyle
P_{ij} := \frac{n_{ij}}{q_{j}} \ ,
$}
\ee
with $i,j = 1, \ldots, n$. These $n \times n = n^{2}$ different 
ratios may be naturally viewed as the elements of a quadratic 
matrix $\mathbf{P}$ of format $(n \times n)$. In general, this 
matrix is given by
\be
\fbox{$\displaystyle
\mathbf{P} =
\left(\begin{array}{ccccc}
\frac{n_{11}}{n_{11}+\ldots+n_{1j}+\ldots+n_{1n}+y_{1}} &
\ldots &
\frac{n_{1j}}{n_{j1}+\ldots+n_{jj}+\ldots+n_{jn}+y_{j}} &
\ldots &
\frac{n_{1n}}{n_{n1}+\ldots+n_{nj}+\ldots+n_{nn}+y_{n}} \\
\vdots & \ddots & \vdots & \ddots & \vdots \\
\frac{n_{i1}}{n_{11}+\ldots+n_{1j}+\ldots+n_{1n}+y_{1}} &
\ldots &
\frac{n_{ij}}{n_{j1}+\ldots+n_{jj}+\ldots+n_{jn}+y_{j}} &
\ldots &
\frac{n_{in}}{n_{n1}+\ldots+n_{nj}+\ldots+n_{nn}+y_{n}} \\
\vdots & \ddots & \vdots & \ddots & \vdots \\
\frac{n_{n1}}{n_{11}+\ldots+n_{1j}+\ldots+n_{1n}+y_{1}} &
\ldots &
\frac{n_{nj}}{n_{j1}+\ldots+n_{jj}+\ldots+n_{jn}+y_{j}} &
\ldots &
\frac{n_{nn}}{n_{n1}+\ldots+n_{nj}+\ldots+n_{nn}+y_{n}}
\end{array}\right) \ ,
$}
\ee
and is referred to as Leontief's {\bf input--output matrix} of the 
stationary economic system under investigation.

\medskip
\noindent
For the very simple case with just $n=3$ producing agents, the 
input--output matrix reduces to
\[
\displaystyle
\mathbf{P} =
\left(\begin{array}{ccc}
\frac{n_{11}}{n_{11}+n_{12}+n_{13}+y_{1}} &
\frac{n_{12}}{n_{21}+n_{22}+n_{23}+y_{2}} &
\frac{n_{13}}{n_{31}+n_{32}+n_{33}+y_{3}} \\
\frac{n_{21}}{n_{11}+n_{12}+n_{13}+y_{1}} &
\frac{n_{22}}{n_{21}+n_{22}+n_{23}+y_{2}} &
\frac{n_{23}}{n_{31}+n_{32}+n_{33}+y_{3}} \\
\frac{n_{31}}{n_{11}+n_{12}+n_{13}+y_{1}} &
\frac{n_{32}}{n_{21}+n_{22}+n_{23}+y_{2}} &
\frac{n_{33}}{n_{31}+n_{32}+n_{33}+y_{3}}
\end{array}\right) \ .
\]
It is important to realise that for an actual economic system the 
input--output matrix $\mathbf{P}$ can be determined only once 
\emph{the reference period of time chosen has come to an end}.

\medskip
\noindent
The utility of Leontief's stationary input--output matrix model is 
in its application for the purpose of {\bf forecasting}. This is 
done on the basis of an {\bf extrapolation}, namely by 
\emph{assuming} that an input--output matrix 
$\mathbf{P}_{\text{reference\ period}}$ obtained from data taken 
during a specific reference period also is valid (to an 
acceptable degree of accuracy) during a subsequent period, i.e.,
\be
\fbox{$\displaystyle
\mathbf{P}_{\text{subsequent\ period}}
\approx
\mathbf{P}_{\text{reference period}} \ ,
$}
\ee
or, in component form,
\be
\left.P_{ij}\right|_{\text{subsequent\ period}}
= \left.\frac{n_{ij}}{q_{j}}\right|_{\text{subsequent\ period}}
\approx
\left.P_{ij}\right|_{\text{reference\ period}}
= \left.\frac{n_{ij}}{q_{j}}\right|_{\text{reference\ period}} \ .
\ee
In this way it becomes possible to compute for a given (idealised) 
economic system approximate numbers of {\bf INPUT quantities} 
required during a near future production period from the known 
numbers of {\bf OUTPUT quantities} of the most recent production 
period. Long-term empirical experience has shown that this method 
generally leads to useful results to a reasonable approximation. 
All of these  calculations are grounded on linear relationships 
describing the quantitative aspects of stationary flows of goods, 
as we will soon elucidate.

\subsection{Resource consumption matrix}
The second matrix-valued quantity central to Leontief's stationary 
model is the {\bf resource consumption matrix} $\mathbf{R}$. This 
may be interpreted as providing a recipe for the amounts of the 
$m$~different kinds of external resources (the exogenous {\bf 
INPUT quantities}) that are needed in the production of the 
$n$~goods (the {\bf OUTPUT quantities}). Its elements are defined 
as the ratios
\be
\fbox{$\displaystyle
R_{ij} := \text{amounts\ (in\ units)\ required\ of\ resource\ $i$
\ for\ the\ production\ of\ one\ unit\ of\ good\ $j$} \ ,
$}
\ee
with $i=1,\ldots,m$ und $j=1,\ldots,n$. The rows of matrix 
$\mathbf{R}$ thus contain information concerning the 
$m$~resources, the columns information concerning the $n$~goods. 
Note that in general the $(m \times n)$ {\bf resource consumption 
matrix}~$\mathbf{R}$ is \emph{not (!)} a quadratic matrix and, 
therefore, in general \emph{not} invertible.

\section[Stationary linear flows of goods]%
{Stationary linear flows of goods}
\lb{sec:qstroeme}
\subsection{Flows of goods: endogenous INPUT to total OUTPUT}
We now turn to a quantitative description of the stationary {\bf 
flows of goods} that are associated with the {\bf total 
output}~$\vec{q}$ during a specific period of time considered. 
According to Leontief's Assumption~1, there exists a \emph{linear} 
functional relationship between the endogenous vector-valued {\bf 
INPUT quantity} $\vec{q}-\vec{y}$ and the vector-valued {\bf 
OUTPUT quantity}~$\vec{q}$. This may be represented in terms of a 
matrix-valued relationship as
\be
\lb{strom1}
\fbox{$\displaystyle
\vec{q}-\vec{y} = \mathbf{P}\vec{q}
\quad \Leftrightarrow \quad
q_{i}-y_{i} = \sum_{j=1}^{n}P_{ij}q_{j} \ ,
$}
\ee
with $i = 1, \ldots, n$, in which the {\bf input--output 
matrix}~$\mathbf{P}$ takes the role of mediating a mapping between 
either of these vector-valued quantities. According to 
Assumption~2, the elements of the {\bf input--output 
matrix}~$\mathbf{P}$ remain \emph{constant} for the period of time 
considered, i.e. the corresponding flows of goods are assumed to 
be {\bf stationary}.

\medskip
\noindent
Relation~(\ref{strom1}) may also be motivated from an alternative 
perspective that takes the {\bf physical sciences} as a guidline. 
Namely, the total numbers~$\vec{q}$ of the $n$~goods produced 
during the period of time considered which, by Assumption~3, are 
equal to the numbers supplied of the $n$~goods satisfy a {\bf  
conservation law}: ``whatever has been produced of the $n$ goods 
during the period of time considered \emph{cannot} get lost in 
this period.'' In quantitative terms this simple relationship may 
be cast into the form
\[
\underbrace{\vec{q}}_{\text{total output}}
=\underbrace{\vec{y}}_{\text{final demand (exogenous)}}
+\underbrace{\mathbf{P}\vec{q}}_{\text{deliveries to production sector (endogenous)}} \ .
\]

\medskip
\noindent
For computational purposes this central stationary flow of goods 
relation~(\ref{strom1}) may be rearranged as is convenient. In 
this context it is helpful to make use of the matrix identity  
$\vec{q}=\mathbf{1}\vec{q}$, where $\mathbf{1}$ denotes the
{\bf $\boldsymbol{(n \times n)}$-unit matrix} [cf.\ 
Eq.~(\ref{einmatr})].

\medskip
\noindent
{\bf Examples:}
\begin{itemize}
\item[(i)]~given/known: $\mathbf{P}$, $\vec{q}$

\medskip
\noindent
Then it applies that
\be
\lb{strom12}
\vec{y} = (\mathbf{1}-\mathbf{P})\vec{q}
\quad \Leftrightarrow \quad
y_{i} = \sum_{j=1}^{n}(\delta_{ij}-P_{ij})q_{j} \ ,
\ee
with $i = 1, \ldots, n$; $(\mathbf{1}-\mathbf{P})$ represents the 
invertible  {\bf technology matrix} of the economic system 
regarded.

\item[(ii)]~given/known: $\mathbf{P}$, $\vec{y}$

\medskip
\noindent
Then it holds that
\be
\lb{strom13}
\vec{q} = (\mathbf{1}-\mathbf{P})^{-1}\vec{y}
\quad \Leftrightarrow \quad
q_{i} = \sum_{j=1}^{n}(\delta_{ij}-P_{ij})^{-1}y_{j} \ ,
\ee
with $i = 1, \ldots, n$; $(\mathbf{1}-\mathbf{P})^{-1}$ here 
denotes the {\bf total demand matrix}, i.e., the inverse of the 
technology matrix.
\end{itemize}
%

\subsection{Flows of goods: exogenous INPUT to total OUTPUT}
Likewise, by Assumption~1, a \emph{linear} functional relationship 
is supposed to exist between the exogenous vector-valued {\bf 
INPUT quantity}~$\vec{v}$ and the vector-valued {\bf OUTPUT 
quantity}~$\vec{q}$. In matrix language this can be expressed by

\be
\lb{strom2}
\fbox{$\displaystyle
\vec{v} = \mathbf{R}\vec{q}
\quad\Leftrightarrow\quad
v_{i} = \sum_{j=1}^{n}R_{ij}q_{j} \ ,
$}
\ee
with $i = 1, \ldots, m$. By Assumption~2, the elements of the {\bf 
resource consumption matrix}~$\mathbf{R}$ remain \emph{constant} 
during the period of time considered, i.e., the corresponding 
resource flows are supposed to be {\bf stationary}.

\medskip
\noindent
By combination of Eqs.~(\ref{strom2}) and~(\ref{strom13}), it is 
possible to compute the numbers~$\vec{v}$ of resources required 
(during the period of time considered) for the production of the 
$n$~goods for given final demand~$\vec{y}$. It applies that
\be
\lb{strom22}
\vec{v} = \mathbf{R}\vec{q}
= \mathbf{R}(\mathbf{1}-\mathbf{P})^{-1}\vec{y}
\quad\Leftrightarrow\quad
v_{i} = \sum_{j=1}^{n}\sum_{k=1}^{n}R_{ij}
(\delta_{jk}-P_{jk})^{-1}y_{k} \ ,
\ee
with $i = 1, \ldots, m$.

\medskip
\noindent
\underline{\bf GDC:} For problems with $n \leq 5$, and known matrices $\mathbf{P}$ and $\mathbf{R}$, Eqs.~(\ref{strom12}), (\ref{strom13}) and (\ref{strom22}) can be immediately used to calculate the quantities $\vec{q}$ from given quantities $\vec{y}$, or vice versa.

\section[Outlook]%
{Outlook}
\lb{sec:geldstroeme}
Leontief's input--output matrix model may be extended in a 
straightforward fashion to include more advanced considerations of 
{\bf economic theory}. Supposing a closed though not necessarily 
stationary economic system~$G$ comprising $n$ interdependent {\bf 
economic agents} producing $n$ different goods, one may assign 
{\bf monetary values} to the {\bf INPUT quantity}~$\vec{v}$ as 
well as to the {\bf OUTPUT quantities} $\vec{q}$ and $\vec{y}$ of 
the system. Besides the numbers of goods produced and the 
associated flows of goods one may monitor with respect to~$G$ for 
a given period of time, one can in addition analyse in time and 
space the {\bf amount of money} coupled to the different goods, 
and the corresponding {\bf flows of money}. However, contrary to 
the number of goods, in general there does \emph{not} exist a 
{\bf conservation law} for the amount of money with respect 
to~$G$. This may render the analysis of flows of money more 
difficult, because, in the sense of an {\bf increase in value}, 
\emph{money can either be generated inside~$G$ during the period 
of time considered or it can likewise be annihilated}; it is 
\emph{not} just limited to either flowing into respectively 
flowing out of $G$. Central to considerations of this kind is a 
{\bf balance equation} for the amount of money contained in $G$ 
during a given period of time, which is an \emph{additive} 
quantity. Such balance equations constitute familiar tools in {\bf 
Physics} (cf. Herrmann (2003)~\ct[p~7ff]{her2003}). Its structure 
in the present case is given by\footnote{Here the symbols CU and 
TU denote ``currency units'' and ``time units,'' respectively.} 
\[
\left(\begin{array}{c}
\text{{\bf rate\ of\ change\ in\ time}} \\
\text{of\ the\ {\bf amount of money}}\\
\text{in}\ G\ \text{[in CU/TU]}
\end{array}\right)
= \left(\begin{array}{c}
\text{{\bf flux of money}} \\
\text{into}\ G\ \text{[in CU/TU]}
\end{array}\right)
+ \left(\begin{array}{c}
\text{{\bf rate of generation of money}} \\
\text{in}\ G\ \text{[in CU/TU]}
\end{array}\right) \ .
\]
Note that, with respect to $G$, both fluxes of money and rates of 
generation of money can in principle possess either sign, positive 
or negative. To deal with these quantitative issues properly, one 
requires the technical tools of the {\bf differential and integral 
calculus} which we will discuss at an elementary level in 
Chs.~\ref{ch7} and~\ref{ch8}. We make contact here with the 
interdisciplinary science of {\bf Econophysics} (cf., e.g., 
Bouchaud and Potters (2003) \ct{boupot2003}), a very interesting 
and challenging subject which, however, is beyond the scope of 
these lecture notes.

\medskip
\noindent
Leontief's input--output matrix model, and its possible extension 
as outlined here, provide the quantitative basis for 
considerations of economical ratios of the kind
\[
\frac{\text{OUTPUT [in~$\text{units}$]}}{\text{INPUT 
[in~$\text{units}$]}} \ ,
\]
as mentioned in the Introduction. In addition, \emph{dimensionless}
(scale-invariant) ratios of the form
\[
\frac{\text{REVENUE [in~$\text{CU}$]}}{\text{COSTS 
[in~$\text{CU}$]}} \ ,
\]
referred to as {\bf economic efficiency}, can be computed for and 
compared between different economic systems and their underlying 
production sectors. In Ch.~\ref{ch7} we will briefly reconsider 
this issue.


\chapter[Linear programming]{Linear programming}
\lb{ch5}

\vspace{10mm}
\noindent
On the backdrop of the {\bf economic principle}, we discuss in 
this chapter a special class of quantitative problems that 
frequently arise in specific practical applications in {\bf 
Business} and {\bf Management}. Generally one distinguishes 
between two variants of the {\bf economic principle}: either 
(i)~to draw maximum utility from limited resources, or (ii)~to 
reach a specific target with minimum effort (costs). With regard 
to the ratio $(\text{OUTPUT})/(\text{INPUT})$ put into focus in 
the Introduction, the issue is to find an {\bf optimal value} for 
this ratio under given {\bf boundary conditions}. This aim can be 
realised either (i)~by increasing the (positive) value of the 
numerator for fixed (positive) value of the denominator, or 
(ii)~by decreasing the (positive) value of the denominator for 
fixed (positive) value of the numerator. The class of quantitative 
problems to be looked at in some detail in this chapter typically 
relate to boundary conditions according to case~(i).

\section[Exposition of a quantitative problem]{Exposition of a
quantitative problem}
\lb{sec:lopeinf}
To be maximised is a (non-negative) real-valued quantity~$z$, 
which depends in a \emph{linear functional fashion} on a fixed 
number of ~$n$ (non-negative) real-valued variables $x_{1}, 
\ldots, x_{n}$. We suppose that the $n$~variables $x_{1}, \ldots, 
x_{n}$ in turn are constrained by a fixed number~$m$ of algebraic 
conditions, which also are assumed to depend on $x_{1}, \ldots, 
x_{n}$ in a \emph{linear fashion}. These $m$~constraints, or 
restrictions, shall have the character of imposing upper limits on 
$m$~different kinds of resources.

\medskip
\noindent
\underline{\bf Def.:}
Consider a matrix~$\mathbf{A} \in
\mathbb{R}^{m \times n}$, a vector~$\vec{b} \in
\mathbb{R}^{m \times 1}$, two vectors $\vec{c},\vec{x} \in
\mathbb{R}^{n \times 1}$, and a constant $d \in \mathbb{R}$. A quantitative problem of the form
\be
\fbox{$\displaystyle
\text{max}\left\{z=\vec{c}^{T}\cdot\vec{x}
+d\left|\mathbf{A}\vec{x}
\leq \vec{b}, \vec{x} \geq \vec{0}\right.\right\} \ ,
$}
\ee
or, expressed in terms of a component notation,
\begin{eqnarray}
\lb{lops1}
\text{max}\ z(x_{1},\ldots,x_{n})
= c_{1}x_{1} + \ldots + c_{n}x_{n} + d & & \\
a_{11}x_{1}+\ldots+a_{1n}x_{n} & \leq & b_{1} \\
 & \vdots & \nonumber \\
a_{m1}x_{1}+\ldots+a_{mn}x_{n} & \leq & b_{m} \\
x_{1} & \geq & 0 \\
 & \vdots & \nonumber \\
\lb{lopsm1}
x_{n} & \geq & 0 \ ,
\end{eqnarray}
is referred to as a  {\bf standard maximum problem of 
linear programming} with $n$~real-valued variables. The different 
quantities and relations appearing in this definition are 
called

\begin{itemize}

\item $z(x_{1},\ldots,x_{n})$ --- {\bf linear objective function}, the dependent variable,

\item $x_{1},\ldots,x_{n}$ --- $n$ {\bf independent variables},

\item $\mathbf{A}\vec{x} \leq \vec{b}$ --- $m$ {\bf restrictions},

\item $\vec{x} \geq \vec{0}$ ---
$n$ {\bf non-negativity constraints}.

\end{itemize}

\noindent
\underline{\bf Remark:} In an analogous fashion one may also formulate a {\bf standard minimum problem of linear programming}, which can be cast into the form
\[
\text{min}\left\{z=\vec{c}^{T}\cdot\vec{x}
+d\left|\mathbf{A}\vec{x}
\geq \vec{b}, \vec{x} \geq \vec{0}\right.\right\} \ .
\]
In this case, the components of the vector $\vec{b}$ need to be 
interpreted as lower limits on certain capacities.

\medskip
\noindent
For given linear objective function $z(x_{1},\ldots,x_{n})$, the set of points $\vec{x} = (x_{1},\ldots,x_{n})^{T}$ satisfying the condition
\be
\fbox{$\displaystyle
z(x_{1},\ldots,x_{n}) = C = \text{constant} \in \mathbb{R} \ ,
$}
\ee
for fixed value of $C$, is referred to as an {\bf isoquant} of 
$z$. {\bf Isoquants} of linear objective functions of $n=2$ 
independent variables constitute straight lines, of $n=3$ 
independent variables Euclidian planes, of $n=4$ independent 
variables Euclidian 3-spaces (or hyperplanes), and of $n\geq 5$ 
independent variables Euclidian $(n-1)$-spaces (or hyperplanes).

\medskip
\noindent
In the simplest cases of {\bf linear programming}, the linear {\bf 
objective function}~$z$ depends on $n=2$ {\bf variables}~$x_{1}$ 
and $x_{2}$ only. An illustrative and efficient method of solving 
problems of this kind will be looked at in the following section.

\section[Graphical solution method]%
{Graphical method for solving problems with two independent 
variables}
\lb{sec:lopgraf}
The systematic graphical solution method of standard maximum 
problems of {\bf linear programming} with $n=2$~independent 
variables comprises the following steps:
\begin{enumerate}

\item Derivation of the {\bf linear objective function}
$$
z(x_{1},x_{2})=c_{1}x_{1}+c_{2}x_{2}+d
$$
in dependence on the {\bf variables}~$x_{1}$ and $x_{2}$.

\item Identification in the $x_{1},x_{2}$--plane of the {\bf feasible region} $D$ of $z$ which is determined by the $m$ restrictions imposed on $x_{1}$ and $x_{2}$. Specifically, $D$ constitutes the domain of $z$ (cf. Ch.~\ref{ch7}).

\item Plotting in the $x_{1},x_{2}$--plane of the projection of the {\bf isoquant} of the linear objective function~$z$ which intersects the origin ($0=x_{1}=x_{2}$). When $c_{2} \neq 0$, this projection is described by the equation
\[
x_{2} = -(c_{1}/c_{2})x_{1} \ .
\]

\item Erecting in the origin of the $x_{1},x_{2}$--plane the {\bf direction of optimisation} for $z$ which is determined by the constant $z$-gradient
$$
(\boldsymbol{\nabla} z)^{T}
=\left(\begin{array}{c}\frac{\ptl z}{\ptl x_{1}} \\
\frac{\ptl z}{\ptl x_{2}}\end{array}\right)
=\left(\begin{array}{c} c_{1} \\ c_{2}\end{array}\right) \ .
$$

\item {\bf Parallel displacement} in the $x_{1},x_{2}$--plane of the projection of the $(0,0)$-isoquant of~$z$ along the direction of optimisation~$(\boldsymbol{\nabla} z)^{T}$ across the feasible region~$D$ out to a distance where the projected isoquant just about touches~$D$.

\item Determination of the {\bf optimal solution} $(x_{1{\rm O}},
x_{2{\rm O}})$ as the point resp.~set of points of intersection between the displaced projection of the $(0,0)$-isoquant of~$z$ and the \emph{far} boundary of~$D$.

\item Computation of the {\bf optimal value} of the linear objective function~$z_{{\rm O}}=z(x_{1{\rm O}},x_{2{\rm O}})$ from the optimal solution~$(x_{1{\rm O}}, x_{2{\rm O}})$.

\item Specification of potential {\bf remaining resources} by substitution of the optimal solution $(x_{1{\rm O}}, x_{2{\rm O}})$ into the $m$~restrictions.

\end{enumerate}
In general one finds that for a linear {\bf objective 
function}~$z$ with $n=2$ {\bf independent variables}~$x_{1}$ and 
$x_{2}$, the feasible region~$D$, when \emph{non-empty and 
bounded}, constitutes an area in the $x_{1},x_{2}$--plane with 
straight edges and a certain number of vertices. In these cases, 
the {\bf optimal values} of the linear objective function~$z$ are 
always to be found either at the vertices or on the edges of the 
feasible region~$D$. When $D$ is an empty set, then there exists 
no solution to the corresponding linear programming problem. When 
$D$ is unbounded, again there may not exist a solution to the 
linear programming problem, but this then depends on the specific 
circumstances that apply.

\medskip
\noindent
\underline{\bf Remark:} To solve a {\bf standard minimum problem 
of linear programming} with $n=2$ independent variables by means 
of the graphical method, one needs to parallelly displace in the 
$x_{1},x_{2}$--plane the projection of the $(0,0)$-isoquant of~$z$ 
along the direction of optimisation~$(\boldsymbol{\nabla} z)^{T}$ 
until contact is made with the feasible region~$D$ for the first 
time. The optimal solution is then given by the point resp.~set of 
points of intersection between the displaced projection of the 
$(0,0)$-isoquant of~$z$ and the \emph{near} boundary of~$D$.

\section[Dantzig's simplex algorithm]%
{Dantzig's simplex algorithm}
\lb{sec:lopsimplex}
The main disadvantage of the graphical solution method is its 
limitation to problems with only $n=2$~independent variables. In 
actual practice, however, one is often concerned with {\bf linear 
programming problems} that depend on \emph{more} than two {\bf 
independent variables}. To deal with these more complex problems 
in a systematic fashion, the US-American mathematician 
\href{http://www-groups.dcs.st-and.ac.uk/~history/Biographies/Dantzig_George.html}{George
Bernard Dantzig (1914--2005)} has devised during the 1940ies an 
efficient algorithm which can be programmed on a computer in a 
fairly straightforward fashion; cf. Dantzig 
(1949,1955)~\ct{dan1949,dan1955}.

\medskip
\noindent
In mathematics, {\bf simplex} is an alternative name used to refer 
to a convex polyhedron, i.e.,~a body of finite (hyper-)volume in 
two or more dimensions bounded by linear (hyper-)surfaces which 
intersect in linear edges and vertices. In general the feasible 
regions of linear programming problems constitute such simplexes. 
Since the {\bf optimal solutions} for the {\bf independent 
variables} of {\bf linear programming problems},
when they exist, are always to be found at a vertex or along an 
edge of simplex feasible regions, Dantzig developed his so-called 
{\bf simplex algorithm} such that it systematically scans the 
edges and vertices of a feasible region to identify the {\bf 
optimal solution} (when it exists) in as few steps as possible.

\medskip
\noindent
The starting point be a {\bf standard maximum problem of linear 
programming} with $n$~{\bf independent variables} in the form of 
relations~(\ref{lops1})--(\ref{lopsm1}). First, by introducing $m$ 
non-negative {\bf slack variables} $s_{1}, \ldots, s_{m}$, one 
transforms the $m$ linear {\bf restrictions} (inequalities) into 
an equivalent set of $m$ linear equations. In this way, potential 
differences between the left-hand and the right-hand sides of the 
$m$ inequalities are represented by the slack variables. In 
combination with the defining equation of the linear {\bf objective
function}~$z$, one thus is confronted with a system of $1+m$ 
linear algebraic equations for the $1+n+m$ variables $z, x_{1},
\ldots, x_{n}, s_{1}, \ldots, s_{m}$, given by

\medskip
\noindent
{\bf Maximum problem of linear programming in canonical form}
\nopagebreak
\begin{eqnarray}
\lb{lopk1}
z - c_{1}x_{1} - c_{2}x_{2} - \ldots - c_{n}x_{n} & = & d \\
a_{11}x_{1} + a_{12}x_{2} + \ldots + a_{1n}x_{n} + s_{1} & = & b_{1} \\
a_{21}x_{1} + a_{22}x_{2} + \ldots + a_{2n}x_{n} + s_{2} & = & b_{2} \\
 & \vdots & \nonumber \\
\lb{lopkm1}
a_{m1}x_{1} + a_{m2}x_{2} + \ldots + a_{mn}x_{n} + s_{m} & = & b_{m} \ .
\end{eqnarray}
As discussed previously in Ch.~\ref{ch3}, a system of linear 
algebraic equations of format $(1+m) \times (1+n+m)$ is 
\emph{under-determined} and so, at most, allows for \emph{multiple 
solutions}. The general $(1+n+m)$-dimensional solution vector
\be
\vec{x}_{L} = \left(z_{L}, x_{1, L}, \ldots, x_{n, L},
s_{1, L}, \ldots, s_{m, L}\right)^{T}
\ee
thus contains $n$ variables the values of which can be chosen 
\emph{arbitrarily}. It is very important to be aware of this fact. 
It implies that, given the linear system is solvable in the first 
place, one has a \emph{choice} amongst different solutions, and so 
one can pick the solution which proves {\bf optimal} for the given 
problem at hand. {\bf Dantzig's simplex algorithm} constitues a 
tool for determining such an {\bf optimal solution} in a 
systematic way.

\medskip
\noindent
Let us begin by transferring the coefficients and right-hand sides 
(RHS) of the under-determined linear system introduced above into 
a particular kind of {\bf simplex tableau}.

\medskip
\noindent
{\bf Initial simplex tableau}
\be
\lb{tableau1}
\begin{array}{c|rrcr|cccc|r}
z & x_{1} & x_{2} & \ldots & x_{n} & s_{1} & s_{2} & \ldots
& s_{m} & \text{RHS} \\
\hline
1 & -c_{1} & -c_{2} & \ldots & -c_{n} & 0 & 0 & \ldots & 0 & d \\
\hline
0 & a_{11} & a_{12} & \ldots & a_{1n} & 1 & 0 & \ldots & 0 & b_{1} \\
0 & a_{21} & a_{22} & \ldots & a_{2n} & 0 & 1 & \ldots & 0 & b_{2} \\
\vdots & \vdots & \vdots & \ddots & \vdots & \vdots & \vdots &
\ddots & \vdots & \vdots \\
0 & a_{m1} & a_{m2} & \ldots & a_{mn} & 0 & 0 & \ldots & 1 & b_{m}
\end{array}
\ee
In such a {\bf simplex tableau} one distinguishes so-called {\bf 
basis variables} from {\bf non-basis variables}. Basis variables 
are those that contain in their respective columns in the number 
tableau a $(1+m)$-component canonical unit vector [cf.\ 
Eq.~(\ref{kanbasis})]; in total the {\bf simplex tableau} contains 
$1+m$~of these. Non-basis variables are the remaining ones that do 
\emph{not} contain a canonical basis vector in their respective 
columns; there exist $n$ of this kind. The complete basis can thus 
be perceived as spanning a $(1+m)$-dimensional Euclidian space 
${\mathbb R}^{1+m}$. Initially, always $z$ and the $m$ slack 
variables $s_{1}, \ldots, s_{m}$ constitute the basis variables, 
while the $n$ independent variables $x_{1}, \ldots, x_{n}$ 
classify as non-basis variables [cf.\ the initial 
tableau~(\ref{tableau1})]. The corresponding so-called (first) 
{\bf basis solution} has the general appearance
\[
\vec{x}_{B_{1}} = \left(z_{B_{1}},
x_{1,B_{1}}, \ldots, x_{n, B_{1}},
s_{1, B_{1}}, \ldots, s_{m, B_{1}}\right)^{T}
= \left(d, 0, \ldots, 0,
b_{1}, \ldots, b_{m}\right)^{T} \ ,
\]
since, for simplicity, each of the $n$ arbitrarily specifiable 
non-basis variables may be assigned the special value zero. In 
this respect basis solutions will always be \emph{special 
solutions} (as opposed to general ones) of the under-determined 
system~(\ref{lopk1})--(\ref{lopkm1}) --- the maximum problem of 
linear programming in canonical form.

\medskip
\noindent
Central aim of the {\bf simplex algorithm} is to bring as many of 
the $n$~{\bf independent variables}~$x_{1}, \ldots, x_{n}$ as 
possible into the $(1+m)$-dimensional basis, at the expense of one 
of the $m$~{\bf slack variables}~$s_{1}, \ldots, s_{m}$, one at a 
time, in order to construct successively more favourable special 
vector-valued solutions to the optimisation problem at hand. 
Ultimately, the {\bf simplex algorithm} needs to be viewed as a 
special variant of Gau\ss ian elimination as discussed in 
Ch.~\ref{ch3}, with a set of systematic instructions concerning 
allowable equivalence transformations of the underlying 
under-determined linear system~(\ref{lopk1})--(\ref{lopkm1}), 
resp.\ the initial {\bf simplex tableau}~(\ref{tableau1}). This 
set of systematic algebraic simplex operations can be summarised 
as follows:

\medskip
\noindent
{\bf Simplex operations}
\nopagebreak
\begin{itemize}
\item[S1:] Does the current simplex tableau show $-c_{j} \geq 0$
for all $j \in \{1,\ldots,n\}$? If so, then the corresponding basis solution is {\bf optimal}. {\it END}. Otherwise goto S2.

\item[S2:] Choose a {\bf pivot column index} $j^{*} \in
\{1,\ldots,n\}$ such that $-c_{j^{*}}:= \text{min}\{-c_{j}|j
\in \{1,\ldots,n\}\} < 0$.

\item[S3:] Is there a row index $i^{*} \in \{1,\ldots,m\}$ such that $a_{i^{*}j^{*}} > 0$? If not, the objective function $z$ is unbounded from above. {\it END}. Otherwise goto S4.

\item[S4:] Choose a {\bf pivot row index} $i^{*}$ such that
$a_{i^{*}j^{*}} > 0$ and $b_{i^{*}}/a_{i^{*}j^{*}}
:=\text{min}\{b_{i}/a_{i^{*}j^{*}}|a_{i^{*}j^{*}} > 0,
i \in \{1,\ldots,m\}\}$. Perform a {\bf pivot operation} with the 
{\bf pivot element} $a_{i^{*}j^{*}}$. Goto S1.
\end{itemize}

\medskip
\noindent
When the final {\bf simplex tableau} has been arrived at, one 
again assigns the non-basis variables the value zero. The values 
of the final basis variables corresponding to the {\bf optimal 
solution} of the given {\bf linear programming problem} are then 
to be determined from the final {\bf simplex tableau} by backward 
substitution, beginning at the bottom row. Note that slack 
variables with positive values belonging to the basis variables in 
the {\bf optimal solution} provide immediate information on 
existing remaining capacities in the problem at hand.


\chapter[Elementary financial mathematics]%
{Elementary financial mathematics}
\lb{ch6}

\vspace{10mm}
\noindent
In this chapter we want to provide a brief introduction into some 
basic concepts of {\bf financial mathematics}. As we will try to 
emphasise, many applications of these concepts (that have 
immediate practical relevance) are founded on only two simple and 
easily accessible mathematical structures: the so-called 
arithmetical and geometrical real-valued sequences and their 
associated finite series.

\section[Arithmetical and geometrical sequences and series]%
{Arithmetical and geometrical sequences and series}
\lb{sec:folgreih}
\subsection{Arithmetical sequence and series}
\lb{subsec:arithseq}
An {\bf arithmetical sequence} of $n \in \mathbb{N}$ real 
numbers~$a_{n} \in \mathbb{R}$,
\[
(a_{n})_{n \in \mathbb{N}} \ ,
\]
is defined by the property that the {\bf difference} $d$ between 
neighbouring elements in the sequence be \emph{constant}, i.e., 
for  $n>1$
\be
\lb{arifolrek}
\fbox{$\displaystyle
a_{n}-a_{n-1}=:d=\text{constant}\neq 0 \ ,
$}
\ee
with $a_{n}, a_{n-1}, d \in \mathbb{R}$. Given this recursive 
formation rule, one may infer the {\bf explicit representation} of 
an {\bf arithmetical sequence} as
\be
\lb{arifolex}
a_{n} = a_{1} + (n-1)d
\quad\text{with}\quad
n \in \mathbb{N} \ .
\ee
Note that any {\bf arithmetical sequence} is \emph{uniquely 
determined} by the two free parameters $a_{1}$ and $d$, the 
starting value of the sequence and the constant difference between 
neighbours in the sequence, respectively. 
Equation~(\ref{arifolex}) shows that the elements $a_{n}$ in a 
non-trivial {\bf arithmetical sequence} exhibit either {\bf 
linear} growth or {\bf linear} decay with $n$.

\medskip
\noindent
When one calculates for an {\bf arithmetical sequence} of $n+1$ 
real numbers the {\bf arithmetical mean} of the immediate 
neighbours of any particular element $a_{n}$ (with $n \geq 2$), 
one finds that
\be
\frac{1}{2}\,(a_{n-1}+a_{n+1})
= \frac{1}{2}\left(a_{1}+(n-2)d+a_{1}+nd\right)
= a_{1}+(n-1)d
= a_{n} \ .
\ee

\medskip
\noindent
Summation of the first $n$ elements of an arbitrary {\bf 
arithmetical sequence} of real numbers leads to a {\bf finite 
arithmetical series},
\be
S_{n} := a_{1}+a_{2}+\ldots+a_{n} = \sum_{k=1}^{n}a_{k}
= \sum_{k=1}^{n}\left[a_{1} + (k-1)d\right]
= na_{1}+ \frac{d}{2}\,(n-1)n \ .
\ee
In the last algebraic step use was made of the Gau\ss ian {\bf 
identity}\footnote{Analogously, the modified Gau\ss ian identity 
$\displaystyle \sum_{k=1}^{n}(2k-1) \equiv n^{2}$ applies.} (cf., 
e.g., Bosch (2003)~\ct[p~21]{bos2003})
\be
\lb{id1}
\fbox{$\displaystyle
\sum_{k=1}^{n-1}k \equiv \frac{1}{2}\,(n-1)n \ .
$}
\ee
%

\subsection{Geometrical sequence and series}
\lb{subsec:geomseq}
A {\bf geometrical sequence} of $n \in \mathbb{N}$ real 
numbers~$a_{n} \in \mathbb{R}$,
\[
(a_{n})_{n \in \mathbb{N}} \ ,
\]
is defined by the property that the {\bf quotient} $q$ between 
neighbouring elements in the sequence be \emph{constant}, i.e., 
for $n>1$
\be
\lb{geofolrek}
\fbox{$\displaystyle
\frac{a_{n}}{a_{n-1}}=:q=\text{constant}\neq 0 \ ,
$}
\ee
with $a_{n}, a_{n-1} \in \mathbb{R}$ and $q \in 
\mathbb{R}\backslash\{0,1\}$. Given this recursive formation rule, 
one may infer the {\bf explicit representation} of a {\bf 
geometrical sequence} as
\be
\lb{geofolex}
a_{n} = a_{1}q^{n-1}
\quad\text{with}\quad
n \in \mathbb{N}.
\ee
Note that any {\bf geometrical sequence} is \emph{uniquely 
determined} by the two free parameters $a_{1}$ and $q$, the 
starting value of the sequence and the constant quotient between 
neighbours in the sequence, respectively. 
Equation~(\ref{geofolex}) shows that the elements $a_{n}$ in a 
non-trivial {\bf geometrical sequence} exhibit either {\bf 
exponential} growth or {\bf exponential} decay with $n$ (cf. 
Sec.~\ref{subsec:exponentials}).

\medskip
\noindent
When one calculates for a {\bf geometrical sequence} of $n+1$ real 
numbers the {\bf geometrical mean} of the immediate neighbours of 
any particular element $a_{n}$ (with $n \geq 2$), one finds that
\be
\sqrt{a_{n-1}\cdot a_{n+1}}
= \sqrt{a_{1}q^{n-2}\cdot a_{1}q^{n}}
= a_{1}q^{n-1}
= a_{n} \ .
\ee

\medskip
\noindent
Summation of the first $n$ elements of an arbitrary {\bf 
geometrical sequence} of real numbers leads to a {\bf finite 
geometrical series},
\be
S_{n} := a_{1}+a_{2}+\ldots+a_{n} = \sum_{k=1}^{n}a_{k}
= \sum_{k=1}^{n}\left[a_{1}q^{k-1}\right]
= a_{1}\sum_{k=0}^{n-1}q^{k}
= a_{1}\,\frac{q^{n}-1}{q-1} \ .
\ee
In the last algebraic step use was made of the {\bf identity} (cf., e.g., Bosch (2003)~\ct[p~27]{bos2003})
\be
\lb{id2}
\fbox{$\displaystyle
\sum_{k=0}^{n-1}q^{k} \equiv \frac{q^{n}-1}{q-1}
\quad\text{for}\quad
q \in \mathbb{R}\backslash\{0,1\} \ .
$}
\ee
%

\section[Interest and compound interest]%
{Interest and compound interest}
\lb{sec:zins}
Let us consider a first rather simple interest model. Suppose 
given an {\bf initial capital} of positive value $K_{0} > 0~{\rm 
CU}$ paid into a bank account at some initial instant, and a time 
interval consisting of $n \in \mathbb{N}$ {\bf periods} of equal 
lengths. At the end of each period, the money in this bank account 
shall earn a service fee corresponding to an {\bf interest rate} 
of $p>0$ percent. Introducing the dimensionless {\bf interest 
factor}\footnote{Inverting this defining relation for $q$ leads to 
$p=100\cdot(q-1)$.}
\be
q:=1+\frac{p}{100} > 1 \ ,
\ee
one finds that by the end of the first interest period a total 
capital of value (in CU)
\[
K_{1} = K_{0} + K_{0}\cdot\frac{p}{100}
= K_{0}\left(1+\frac{p}{100}\right) = K_{0}q
\]
will have accumulated. When the entire time interval of $n$ 
interest periods has ended, a {\bf final capital} worth of (in CU)
\be
\lb{kap1}
\fbox{$\displaystyle
\text{recursively:}\ K_{n}=K_{n-1}q \ , \quad
n\in\mathbb{N} \ ,
$}
\ee
will have accumulated, where $K_{n-1}$ denotes the capital (in CU) 
accumulated by the end of $n-1$ interest periods. This recursive 
representation of the growth of the initial capital $K_{0}$ due to 
a total of $n$ interest payments and the effect of {\bf compound 
interest} makes explicit the direct link with the mathematical 
structure of a {\bf geometrical sequence} of real 
numbers~(\ref{geofolrek}).

\medskip
\noindent
It is a straightforward exercise to show that in this simple 
interest model the final capital $K_{n}$ is related to the initial 
capital $K_{0}$ by
\be
\lb{kap2}
\fbox{$\displaystyle
\text{explicitly:}\ K_{n}=K_{0}q^{n} \ , \quad
n\in\mathbb{N} \ .
$}
\ee
Note that this equation links the four non-negative quantities 
$K_{n}$, $K_{0}$, $q$ and $n$ to one another. Hence, knowing the 
values of three of these quantities, one may solve 
Eq.~(\ref{kap2}) to obtain the value of the fourth. For example, 
solving Eq.~(\ref{kap2}) for $K_{0}$ yields
\be
K_{0} = \frac{K_{n}}{q^{n}} =: B_{0} \ .
\ee
In this particular variant, $K_{0}$ is referred to as the {\bf 
present value} $B_{0}$ of the final capital $K_{n}$; this is 
obtained from $K_{n}$ by an $n$-fold division with the interest 
factor $q$. 

\medskip
\noindent
Further possibilities of re-arranging Eq.~(\ref{kap2}) are:
\begin{itemize}
\item[(i)] Solving for the {\bf interest factor} $q$:
\be
q = \sqrt[n]{\frac{K_{n}}{K_{0}}} \ ,
\ee
\item[(ii)] Solving for the {\bf contract period} $n$:
\be
n = \frac{\ln\left(K_{n}/K_{0}\right)}{\ln(q)} \ .
\ee
\end{itemize}
{}From now on, $n \in \mathbb{N}$ shall denote the number of full years that have passed in a specific interest model.

\medskip
\noindent
Now we turn to discuss a second, more refined interest model. Let 
us suppose that an {\bf initial capital} $K_{0}>0~{\rm CU}$ earns 
interest during one full year $m \in \mathbb{N}$ times at the 
$m$th part of a {\bf nominal annual interest rate} $p_{\rm 
nom}>0$. At the end of the first out of $m$ periods of equal 
length $1/m$, the initial capital $K_{0}$ will thus have increased 
to an amount
\[
K_{1/m} = K_{0} + K_{0}\cdot\frac{p_{\rm nom}}{m\cdot 100}
= K_{0}\left(1+\frac{p_{\rm nom}}{m\cdot 100}\right) \ .
\]
By the end of the $k$th ($k \leq m$) out of $m$ periods the {\bf 
account balance} will have become
\[
K_{k/m} = K_{0}\left(1+\frac{p_{\rm nom}}{m\cdot 100}\right)^{k}
\ ;
\]
the interest factor $\displaystyle
\left(1+\frac{p_{\rm nom}}{m\cdot 100}\right)$ will then have been 
applied $k$ times to $K_{0}$. At the end of the full year, $K_{0}$ 
in this interest model will have increased to
\[
K_{1} = K_{m/m} = K_{0}\left(1+\frac{p_{\rm nom}}{m\cdot 100}
\right)^{m} \ , \quad
m \in \mathbb{N} \ .
\]
This relation defines an {\bf effective interest factor}
\be
\lb{qeff}
q_{\rm eff} := \left(1+\frac{p_{\rm nom}}{m\cdot 100}
\right)^{m} \ ,
\ee
with associated {\bf effective annual interest rate}
\be
\lb{peff}
\fbox{$\displaystyle
p_{\rm eff} = 100\cdot\left[\left(1+\frac{p_{\rm nom}}{m\cdot 100}
\right)^{m}-1\right] \ , \quad m\in\mathbb{N} \ ,
$
}
\ee
obtained from re-arranging $\displaystyle q_{\rm 
eff}=1+\frac{p_{\rm eff}}{100}$.

\medskip
\noindent
When, ultimately, $n \in \mathbb{N}$ full years will have passed 
in the second interest model, the initial capital~$K_{0}$ will 
have been transformed into a final capital of value
\be
\lb{kap3}
\fbox{$\displaystyle
K_{n}=K_{0}\left(1+\frac{p_{\rm nom}}{m\cdot 100}\right)^{n\cdot m}
= K_{0}q_{\rm eff}^{n} \ , \quad n,m\in\mathbb{N} \ .
$}
\ee
The {\bf present value} $B_{0}$ of $K_{n}$ is thus given by
\be
B_{0} = \frac{K_{n}}{q_{\rm eff}^{n}} = K_{0} \ .
\ee

\medskip
\noindent
Finally, as a third interest model relevant to applications in 
{\bf Finance}, we turn to consider the concept of {\bf installment 
savings}. For simplicity, let us restrict our discussion to the 
case when $n \in \mathbb{N}$ equal {\bf installments} of 
\emph{constant} value $E>0~{\rm CU}$ are paid into an account that 
earns $p>0$ percent annual interest (i.e., $q>1$) at the beginning 
of each of $n$~full years. The initial account balance be 
$K_{0}=0~{\rm CU}$. At the end of a first full year in this 
interest model, the account balance will have increased to
\[
K_{1} = E + E\cdot\frac{p}{100}
= E\left(1+\frac{p}{100}\right)
= Eq \ .
\]
At the end of two full years one finds, substituting for $K_{1}$,
\[
K_{2} = (K_{1}+E)q = (Eq+E)q = E(q^{2}+q) = Eq(q+1) \ .
\]
At the end of $n$ full years we have, recursively substituting for $K_{n-1}$, $K_{n-2}$, etc.,
\[
K_{n} = (K_{n-1}+E)q
= \cdots
=E(q^{n}+\ldots+q^{2}+q)
=Eq(q^{n-1}+\ldots+q+1)
= Eq\sum_{k=0}^{n-1}q^{k} \ .
\]
Using the identity~(\ref{id2}), since presently $q>1$, the {\bf 
account balance} at the end of $n$ full years can be reduced to 
the expression
\be
\lb{kap4}
\fbox{$\displaystyle
K_{n}=Eq\,\frac{q^{n}-1}{q-1} \ , \quad
q \in \mathbb{R}_{>1} \ , \quad
n\in\mathbb{N} \ .
$}
\ee
The {\bf present value} $B_{0}$ associated with $K_{n}$ is 
obtained by $n$-fold division of $K_{n}$ with the interest 
factor~$q$:
\be
B_{0} := \frac{K_{n}}{q^{n}}
\overbrace{=}^{\text{Eq.}~\ref{kap4}}
 = \frac{E(q^{n}-1)}{q^{n-1}(q-1)} \ .
\ee
This gives the value of an initial capital $B_{0}$ which will grow 
to the \emph{same} final value $K_{n}$ after $n$ annual interest 
periods with constant interest factor $q>1$.

\medskip
\noindent
Lastly, re-arranging Eq.~(\ref{kap4}) to solve for the {\bf 
contract period} $n$ yields.
\be
n = \frac{\ln\left[1+(q-1)(K_{n}/Eq)\right]}{\ln(q)} \ .
\ee
%

\section[Redemption payments in constant annuities]%
{Redemption payments in constant annuities}
\lb{sec:tilg}
The starting point of the next discussion be a {\bf mortgage loan} 
of amount $R_{0}>0~{\rm CU}$ that an {\bf economic agent} borrowed 
from a bank at the obligation of annual service payments of $p>0$ 
percent (i.e., $q>1$) on the {\bf remaining debt}. We suppose that 
the contract between the agent and the bank fixes the following 
conditions:
\begin{itemize}
\item[(i)]~the first {\bf redemption payment} $T_{1}$ amount to 
$t>0$ percent of the mortgage $R_{0}$,

\item[(ii)]~the remaining debt shall be paid back to the bank in 
\emph{constant} {\bf annuities} of value $A>0~{\rm CU}$ at the end 
of each full year that has passed.
\end{itemize}
The {\bf annuity} $A$ is defined as the \emph{sum} of the variable 
$n$th {\bf interest payment} $Z_{n} > 0~{\rm CU}$ and the variable 
$n$th {\bf redemption payment} $T_{n} > 0~{\rm CU}$. In the 
present model we impose on the annuity the condition that it be 
\emph{constant} across full years,
\be
\lb{annuity1}
A = Z_{n} + T_{n}
\stackrel{!}{=} \text{constant} \ .
\ee
For $n=1$, for example, we thus obtain
\be
\lb{annuity2}
A = Z_{1} + T_{1}
= R_{0}\cdot\frac{p}{100} + R_{0}\cdot\frac{t}{100}
= R_{0}\left(\frac{p+t}{100}\right)
= R_{0}\left[(q-1)+\frac{t}{100}\right]
\stackrel{!}{=} \text{constant} \ .
\ee
For the first full year of a running mortgage contract, the 
interest payment, the redemption payment, and, following the 
payment of a first annuity, the remaining debt take the values
\begin{eqnarray*}
Z_{1} & = & R_{0}\cdot\frac{p}{100} \ = \ R_{0}(q-1) \\
T_{1} & = & A - Z_{1} \\
R_{1} & = & R_{0} + Z_{1} - A
\ \overbrace{=}^{\text{substitute for}\ Z_{1}}
\ R_{0} + R_{0}\cdot\frac{p}{100} - A
\ = \ R_{0}q - A \ .
\end{eqnarray*}
By the end of a second full year, these become
\begin{eqnarray*}
Z_{2} & = & R_{1}(q-1) \\
T_{2} & = & A - Z_{2} \\
R_{2} & = & R_{1} + Z_{2} - A
\ \overbrace{=}^{\text{substitute for}\ Z_{2}}
\ R_{1}q - A
\ \overbrace{=}^{\text{substitute for}\ R_{1}}
\ R_{0}q^{2} - A(q+1) \ .
\end{eqnarray*}
At this stage, it has become clear according to which patterns the 
different quantities involved in the redemption payment model need 
to be formed. The {\bf interest payment} for the $n$th full year 
in a mortgage contract of constant anuities amounts to 
(recursively)
\be
\lb{zinsn}
Z_{n}=R_{n-1}(q-1) \ , \quad n\in\mathbb{N} \ ,
\ee
where $R_{n-1}$ denotes the remaining debt at the end of the 
previous full year. The {\bf redemption payment} for full year $n$ 
is then given by (recursively)
\be
\lb{tilgn}
T_{n} = A - Z_{n} \ , \quad n\in\mathbb{N} \ .
\ee
The {\bf remaining debt} at the end of the $n$th full year then is 
(in CU)
\be
\lb{remdebtrek}
\fbox{$\displaystyle
\text{recursively:} \quad
R_{n} = R_{n-1}+Z_{n}-A
\ = \ R_{n-1}q-A \ , \quad n\in\mathbb{N} \ .
$}
\ee
By successive backward substitution for $R_{n-1}$, $R_{n-2}$, etc.,
$R_{n}$ can be re-expressed as
\[
R_{n} = R_{0}q^{n}-A(q^{n-1}+\ldots+q+1)
= R_{0}q^{n}-A\sum_{k=0}^{n-1}q^{k} \ .
\]
Now employing the identity~(\ref{id2}), we finally obtain (since $q>1$)
\be
\lb{remdebtex}
\fbox{$\displaystyle
\text{explicitly:} \quad
R_{n} = R_{0}q^{n}-A\,\frac{q^{n}-1}{q-1} \ , \quad
n\in\mathbb{N} \ .
$}
\ee

\medskip
\noindent
All the formulae we have now derived
for computing the values of the quantities $\{n, Z_{n}, T_{n}, 
R_{n}\}$ form the basis of a formal {\bf redemption payment plan}, 
given by
\begin{center}
		\begin{tabular}{c||c|c|c}
		$n$ & $Z_{n}$ [CU] & $T_{n}$ [CU] & $R_{n}$ [CU] \\
		\hline\hline
		$0$ & -- & -- & $R_{0}$ \\
		$1$ & $Z_{1}$ & $T_{1}$ & $R_{1}$ \\
		$2$ & $Z_{2}$ & $T_{2}$ & $R_{2}$ \\
		\vdots & \vdots & \vdots & \vdots
		\end{tabular} \ ,
\end{center}
a standard scheme that banks must make available to their mortgage 
customers for the purpose of financial orientation.

\medskip
\noindent
\underline{\bf Remark:} For known values of the free parameters 
$R_{0}>0~{\rm CU}$, $q>1$ and $A>0~{\rm CU}$, the simple recursive 
formulae (\ref{zinsn}), (\ref{tilgn}) and (\ref{remdebtrek}) can 
be used to implement a redemption payment plan in a modern 
spreadsheet programme such as EXCEL or OpenOffice.

\medskip
\noindent
We emphasise the following observation concerning 
Eq.~(\ref{remdebtex}): since the constant annuity $A$ contains 
implicitly a factor $(q-1)$ [cf. Eq.~(\ref{annuity2})], the two 
competing terms in this relation each grow exponentially with $n$. 
For the redemption payments to eventually terminate, it is thus 
essential to fix the free parameter $t$ (for known $p>0 
\Leftrightarrow q>1$) in such a way that the second term on the 
right-hand side of Eq.~(\ref{remdebtex}) is given the possibility 
to catch up with the first as $n$ progresses (the latter of which 
has a head start of $R_{0}>0~{\rm CU}$ at $n=0$). The necessary 
condition following from the requirement that $R_{n} 
\stackrel{!}{\leq} R_{n-1}$ is thus $t>0$.

\medskip
\noindent
Equation~(\ref{remdebtex}) links the five non-negative quantities 
$R_{n}$, $R_{0}$, $q$, $n$ and $A$ to one another. Given one knows 
the values of four of these, one can solve for the fifth. For 
example:
\begin{itemize}
\item[(i)] Calculation of the {\bf contract period} $n$ of a 
mortgage contract, knowing the mortgage $R_{0}$, the interest 
factor $q$ and the annuity $A$. Solving the condition  $R_{n} 
\stackrel{!}{=} 0$ imposed on $R_{n}$ for $n$ yields (after a few 
algebraic steps)
\be
n = \frac{\ln\left(1+\frac{p}{t}\right)}{\ln(q)} \ ;
\ee
the contract period is thus independent of the value of the 
mortgage loan, $R_{0}$.

\item[(ii)] Evaluation of the {\bf annuity} $A$, knowing the 
contract period $n$, the mortgage loan~$R_{0}$, and the interest 
factor $q$. Solving the condition $R_{n} \stackrel{!}{=} 0$ 
imposed on $R_{n}$ for $A$ immediately yields
\be
\lb{annuity3}
A = \frac{q^{n}(q-1)}{q^{n}-1}\,R_{0} \ .
\ee
Now equating the two expressions (\ref{annuity3}) and
(\ref{annuity2}) for the annuity $A$, one finds in addition that
\be
\frac{t}{100} = \frac{q-1}{q^{n}-1} \ .
\ee
\end{itemize}
%

\section[Pension calculations]{Pension calculations}
\lb{sec:rente}
Quantitative models for {\bf pension calculations} assume given an 
{\bf initial capital} $K_{0}>0~{\rm CU}$ that was paid into a bank 
account at a particular moment in time. The issue is to monitor 
the subsequent evolution in {\bf discrete time} $n$ of the {\bf 
account balance} $K_{n}$ (in CU), which is subjected to two 
competing influences: on the one-hand side, the bank account earns 
interest at an {\bf annual interest rate} of $p>0$ percent (i.e., 
$q>1$), on the other, it is supposed that throughout one full year 
a total of~$m \in \mathbb{N}$ pension payments of the 
\emph{constant} {\bf amount}~$a$ are made from this bank account, 
always at the beginning of each of $m$ intervals of equal duration 
per year.

\medskip
\noindent
Let us begin by evaluating the amount of interest earned per year 
by the bank account. An important point in this respect is the 
fact that throughout one full year there is a total of $m$ 
deductions of value $a$ from the bank account, i.e., in general 
the account balance does \emph{not} stay constant throughout that 
year but rather decreases in discrete steps. For this reason, the 
account is credited by the bank with interest only at the $m$th 
part of $p>0$ percent for each interval (out of the total of $m$) 
that has passed, with \emph{no} compound interest effect. Hence, 
at the end of the first out of $m$ intervals per year the bank 
account has earned interest worth of (in CU)
\[
Z_{1/m} = (K_{0}-a)\cdot\frac{p}{m\cdot 100}
= (K_{0}-a)\,\frac{(q-1)}{m} \ .
\]
The interest earned for the $k$th interval (out of $m$; $k \leq m$) is then given by
\[
Z_{k/m} = (K_{0}-ka)\,\frac{(q-1)}{m} \ .
\]
Summation over the contributions of each of the $m$ intervals to 
the interest earned then yields for the entire interest earned 
during the first full year (in CU)
\[
Z_{1} = \sum_{k=1}^{m}Z_{k/m}
= \sum_{k=1}^{m}(K_{0}-ka)\,\frac{(q-1)}{m}
= \frac{(q-1)}{m}\left[mK_{0}-a\sum_{k=1}^{m}k\right] \ .
\]
By means of substitution from the identity~(\ref{id1}), this 
result can be recast into the equivalent form
\be
\lb{rentezins1}
Z_{1} = \left[K_{0}-\frac{1}{2}\,(m+1)a\right](q-1) \ .
\ee
Note that this quantity decreases linearly with the number of 
deductions $m$ made per year resp.~with the pension payment 
amount~$a$.

\medskip
\noindent
One now finds that the account balance at the end of the first 
full year that has passed is given by
\[
K_{1} = K_{0} - ma + Z_{1}
\overbrace{=}^{\text{Eq.}~(\ref{rentezins1})} K_{0}q
- \left[m+\frac{1}{2}\,(m+1)(q-1)\right]a \ .
\]
At the end of a second full year of the pension payment contract the interest earned is
\[
Z_{2} = \left[K_{1}-\frac{1}{2}\,(m+1)a\right](q-1) \ ,
\]
while the account balance amounts to
\[
K_{2} = K_{1} - ma + Z_{2}
\overbrace{=}^{\text{substitute for}\ K_{1}
\ \text{and}\ Z_{2}} K_{0}q^{2}
- \left[m+\frac{1}{2}\,(m+1)(q-1)\right]a(q+1) \ .
\]
At this stage, certain fairly simple patterns for the {\bf 
interest earned} during full year $n$, and the {\bf account 
balance} after $n$ full years, reveal themselves. For $Z_{n}$ we 
have
\be
Z_{n} = \left[R_{n-1}-\frac{1}{2}\,(m+1)a\right](q-1) \ ,
\ee
and for $K_{n}$ one obtains

\[
K_{n} = K_{n-1} - ma + Z_{n}
\overbrace{=}^{\text{substitute for}\ K_{n-1}
\ \text{and}\ Z_{n}} K_{0}q^{n}
- \left[m+\frac{1}{2}\,(m+1)(q-1)\right]a\sum_{k=0}^{n-1}q^{k} \ .
\]
The latter result can be re-expressed upon substitution from the 
identity~(\ref{id2}). Thus, $K_{n}$ can finally be given by
\be
\lb{pensionex}
\fbox{$\displaystyle
\text{explicitly:} \quad
K_{n} = K_{0}q^{n} - \left[m+\frac{1}{2}\,(m+1)(q-1)\right]
a\,\frac{q^{n}-1}{q-1} \ , \quad
n, m \in \mathbb{N} \ .
$
}
\ee
In a fashion practically identical to our discussion of the 
redemption payment model in Sec.~\ref{sec:tilg}, the two competing 
terms on the right-hand side of Eq.~(\ref{pensionex}) likewise 
exhibit exponential growth with the number $n$ of full years 
passed. Specifically, it depends on the values of the parameters 
$K_{0} > 0~{\rm CU}$, $q>1$, $a>0~{\rm CU}$, as well as $m \geq 1$,
whether the second term eventually manages to catch up with the 
first as $n$ progresses (the latter of which, in this model, is 
given a head start of value $K_{0} > 0~{\rm CU}$ at $n=0$).

\medskip
\noindent
We remark that Eq.~(\ref{pensionex}), again, may be algebraically 
re-arranged at one's convenience (as long as division by zero is 
avoided). For example:
\begin{itemize}

\item[(i)] The {\bf duration} $n$ (in full years) of a particular 
pension contract is obtained from solving the condition 
$K_{n}\stackrel{!}{=}0$ accordingly. Given that 
$[\ldots]a-K_{0}(q-1)>0$, one thus finds\footnote{To avoid 
notational overload, the brackets $[\ldots]$ here represent the 
term $\left[m+\frac{1}{2}\,(m+1)(q-1)\right]$.}
\be
\lb{rentelaufzeit}
n = \frac{\ln\left(\frac{[\ldots]a}{
[\ldots]a-K_{0}(q-1)}\right)}{\ln(q)} \ .
\ee

\item[(ii)] The {\bf present value} $B_{0}$ of a pension scheme 
results from the following consideration: for fixed interest 
factor $q>1$, which initial capital $K_{0} > 0~{\rm CU}$ must be 
paid into a bank account such that for a duration of $n$ full 
years one can receive payments of constant amount $a$ at the 
beginning of each of $m$ intervals (of equal length) per year? The 
value of $B_{0}=K_{0}$ is again obtained from imposing on 
Eq.~(\ref{pensionex}) the condition $K_{n}\stackrel{!}{=}0$ and 
solving for $K_{0}$. This yields
\be
\lb{rentebarwert}
B_{0} = K_{0} = \left[m+\frac{1}{2}\,(m+1)(q-1)\right]a\,
\frac{q^{n}-1}{q^{n}(q-1)} \ .
\ee

\item[(iii)] The idea of so-called {\bf everlasting pension 
payments} of amount $a_{\rm ever} >0~{\rm CU}$ is based on the 
strategy to consume only the annual interest earned by an initial 
capital $K_{0}>0~{\rm CU}$ residing in a bank account with 
interest factor $q>1$. Imposing now on Eq.~(\ref{pensionex}) the 
condition $K_{n}\stackrel{!}{=}K_{0}$ to hold for all values of 
$n$, and then solving for $a$, yields the result
\be
a_{\rm ever} = \frac{q-1}{m+\frac{1}{2}\,(m+1)(q+1)}\,K_{0} \ ;
\ee
Note that, naturally, $a_{\rm ever}$ is directly proportional to 
the initial capital $K_{0}$!

\end{itemize}
%

\section[Linear and declining-balance depreciation methods]%
{Linear and declining-balance depreciation methods}
\lb{sec:abschr}
Attempts at the quantitative description of the process of 
declining material value of industrial goods, properties or other 
assets, of {\bf initial value} $K_{0}>0~{\rm CU}$, are referred to 
as {\bf depreciation}. International tax laws generally provide 
investors with a choice between two particular mathematical 
methods of calculating {\bf depreciation}. We will discuss 
these options in turn.

\subsection{Linear depreciation method}
When the {\bf initial value} $K_{0}>0~{\rm CU}$ is supposed to 
decline to $0~{\rm CU}$ in the space of $N$ full years by equal 
annual amounts, the {\bf remaining value} $R_{n}$ (in CU) at the 
end of $n$ full years is described by
\be
\fbox{$\displaystyle
R_{n} = K_{0} - n\left(\frac{K_{0}}{N}\right) \ ,
\quad n=1,\ldots,N \ .
$}
\ee
Note that for the difference of remaining values for years 
adjacent one obtains $\displaystyle R_{n}-R_{n-1}
= -\left(\frac{K_{0}}{N}\right) =: d < 0$. The underlying 
mathematical structure of the {\bf straight line depreciation 
method} is thus an {\bf arithmetical sequence} of real numbers, 
with constant \emph{negative} difference $d$ between neighbouring 
elements (cf. Sec.~\ref{subsec:arithseq}).

\subsection{Declining-balance depreciation method}
The foundation of the second depreciation method to be described 
here, for an industrial good of {\bf initial value} $K_{0}>0~{\rm 
CU}$, is the idea that per year the value declines by a certain 
{\bf percentage rate} $p>0$ of the value of the good during the 
previous year. Introducing a dimensionless {\bf depreciation 
factor} by
\be
q := 1-\frac{p}{100} < 1 \ ,
\ee
the {\bf remaining value} $R_{n}$ (in CU) after $n$ full years 
amounts to
\be
\lb{deprrec}
\fbox{$\displaystyle
\text{recursively:} \quad
R_{n} = R_{n-1}q \ , \quad R_{0} \equiv K_{0} \ ,\quad
n\in\mathbb{N} \ .
$}
\ee
The underlying mathematical structure of the {\bf declining 
balance depreciation method} is thus a {\bf geo\-metrical 
sequence} of real numbers, with constant ratio $0<q<1$ between 
neighbouring elements (cf. Sec.~\ref{subsec:geomseq}). With 
increasing $n$ the values of these elements become ever smaller. 
By means of successive backward substitution 
expression~(\ref{deprrec}) can be transformed to
\be
\lb{deprex}
\fbox{$\displaystyle
\text{explicitly:} \quad
R_{n} = K_{0}q^{n} \ , \quad 0 < q < 1 \ , \quad
n\in\mathbb{N} \ .
$}
\ee

\medskip
\noindent
{}From Eq.~(\ref{deprex}), one may derive results concerning the 
following questions of a quantitative nature:

\begin{itemize}

\item[(i)] Suppose given a depreciation factor $q$ and a projected 
remaining value $R_{n}$ for some industrial good. After which {\bf 
depreciation period} $n$ will this value be attained? One finds
\be
n = \frac{\ln\left(R_{n}/K_{0}\right)}{\ln(q)} \ .
\ee

\item[(ii)] Knowing a projected depreciation period $n$ and 
corresponding remaining value $R_{n}$, at which {\bf percentage 
rate} $p>0$ must the depreciation method be operated? This yields
\be
q = \sqrt[n]{\frac{R_{n}}{K_{0}}}
\quad\Rightarrow\quad
p = 100 \cdot \left(1-\sqrt[n]{\frac{R_{n}}{K_{0}}}\right) \ .
\ee

\end{itemize}
%

\section[Summarising formula]{Summarising formula}
\lb{sec:zus}
To conclude this chapter, let us summarise the results on {\bf 
elementary financial mathematics} that we derived along the way. 
Remarkably, these can be condensed in a single formula which 
contains the different concepts discussed as special cases. This 
formula, in which $n$ represents the number of full years that 
have passed, is given by (cf. Zeh--Marschke (2010) \ct{zeh2010}):
\be
\lb{fincmastereq}
\fbox{$\displaystyle
K_{n} = K_{0}q^{n} + R\,\frac{q^{n}-1}{q-1} \ , \quad
q \in \mathbb{R}_{>0}\backslash\{1\} \ , \quad
n\in\mathbb{N} \ .
$}
\ee

\medskip
\noindent
The different special cases contained therein are:
\begin{itemize}
\item[(i)] {\bf Compound interest} for an initial capital 
$K_{0}>0~{\rm CU}$: with $R=0$ and $q>1$, Eq.~(\ref{fincmastereq}) 
reduces to Eq.~(\ref{kap2}).

\item[(ii)] {\bf Installment savings} with constant installments 
$E>0~{\rm CU}$: with $K_{0}=0~{\rm CU}$, $q>1$ and $R=Eq$,
Eq.~(\ref{fincmastereq}) reduces to Eq.~(\ref{kap3}).

\item[(iii)] {\bf Redemption payments in constant annuities}:
with $K_{0}=-R_{0}<0~{\rm CU}$, $q>1$ and $R=A>0~{\rm CU}$, 
Eq.~(\ref{fincmastereq}) reduces to the \emph{negative (!)} of 
Eq.~(\ref{remdebtex}). In this dual formulation, remaining debt 
$K_{n}=-R_{n}$ is (meaningfully) expressed as a negative account 
balance.

\item[(iv)] {\bf Pension payments} of constant amount $a>0~{\rm 
CU}$: with $q>1$ and $\displaystyle R=-\left[m
+\frac{1}{2}\,(m+1)(q-1)\right]a$, Eq.~(\ref{fincmastereq}) 
transforms to Eq.~(\ref{pensionex}).

\item[(v)] {\bf Declining balance depreciation} of an asset of 
initial value $K_{0}>0~{\rm CU}$: with $R=0$ and $0<q<1$, 
Eq.~(\ref{fincmastereq}) converts to Eq.~(\ref{deprex}) for the 
remaining value $K_{n}=R_{n}$.

\end{itemize}
%


\chapter[Differential calculus of real-valued functions]%
{Differential calculus of real-valued functions of one real 
variable}
\lb{ch7}

\vspace{10mm}
\noindent
In Chs.~\ref{ch1} to \ref{ch5} of these lecture notes, we confined 
our considerations to functional relationships between {\bf INPUT 
quantities} and {\bf OUTPUT quantities} of a \emph{linear} nature. 
In this chapter now, we turn to discuss characteristic properties 
of truly {\bf non-linear functional relationships} between one 
{\bf INPUT quantity} and one {\bf OUTPUT quantity}.

\section[Real-valued functions]%
{Real-valued functions of one real variable}
\lb{sec:fkt}
Let us begin by introducing the concept of a {\bf real-valued 
function of one real variable}. This constitutes a special 
kind of a {\bf mapping}\footnote{Cf. our introduction in 
Ch.~\ref{ch2} of matrices as a particular class of mathematical 
objects.} that needs to satisfy the following simple but very 
strict rule:

\medskip
a mapping $f$ that assigns to \emph{every element} $x$ from a 
subset $D$ of the real numbers $\mathbb{R}$ (i.e., $D \subseteq 
\mathbb{R}$) \emph{one and only one element} $y$ from a second 
subset $W$ of the real numbers $\mathbb{R}$ (i.e., $W\subseteq 
\mathbb{R}$).

\medskip
\noindent
\underline{\bf Def.:}
A {\bf unique} mapping $f$ of a subset $D\subseteq\mathbb{R}$ of 
the real numbers onto a subset $W\subseteq\mathbb{R}$ of the real 
numbers,
\be
\fbox{$\displaystyle
f\!: D \rightarrow W \ ,
\hspace{10mm} x \mapsto y=f(x)
$}
\ee
is referred to as a {\bf real-valued function of one real 
variable}.

\medskip
\noindent
We now fix some terminology concerning the concept of a 
real-valued function of one real variable:
%
\begin{itemize}
\item $D$: {\bf domain} of $f$,
\item $W$: {\bf target space} of $f$,
\item $x \in D$: {\bf independent variable} of $f$, also referred 
to as the \emph{argument} of $f$,
\item $y \in W$: {\bf dependent variable} of $f$,
\item $f(x)$: {\bf mapping prescription},
\item {\bf graph} of $f$: the set of pairs of values 
$G=\{(x,f(x))|x \in D\} \subseteq \mathbb{R}^{2}$.
\end{itemize}

\medskip
\noindent
For later analysis of the mathematical properties of real-valued 
functions of one real variable, we need to address a few more 
technical issues.

\medskip
\noindent
\underline{\bf Def.:}
Given a mapping $f$ that is {\bf one-to-one and onto}, with domain 
$D(f) \subseteq \mathbb{R}$ and target space $W(f) \subseteq 
\mathbb{R}$, not only is every $x \in D(f)$ assigned to one and 
only one $y \in W(f)$, but also every $y \in W(f)$ is assigned to 
one and only one $x \in D(f)$. In this case, there exists an 
associated mapping $f^{-1}$, with $D(f^{-1})=W(f)$ and 
$W(f^{-1})=D(f)$, which is referred to as the {\bf inverse 
function} of $f$.

\medskip
\noindent
\underline{\bf Def.:}
A real-valued function $f$ of one real variable $x$ is {\bf continuous} at some value $x \in D(f)$ when for $\Delta x \in \mathbb{R}_{>0}$ the condition
\be
\lim_{\Delta x \to 0}f(x-\Delta x)
= \lim_{\Delta x \to 0}f(x+\Delta x)
= f(x)
\ee
obtains, i.e., when at $x$ the left and right limits of the 
function $f$ coincide and are equal to the value $f(x)$. A 
real-valued function $f$ as such is {\bf continuous} when $f$ is 
continuous \emph{for all} $x \in D(f)$.

\medskip
\noindent
\underline{\bf Def.:}
When a real-valued function $f$ of one real variable $x$ satisfies 
the condition
\be
f(a) < f(b)
\quad\quad\text{\emph{for all}}\quad
a,b \in D(f)\ \text{with}\ a < b \ ,
\ee
then $f$ is called {\bf strictly monotonously increasing}. When, 
however, $f$ satisfies the condition
\be
f(a) > f(b)
\quad\quad\text{\emph{for all}}\quad
a,b \in D(f)\ \text{with}\ a < b \ ,
\ee
then $f$ is called {\bf strictly monotonously decreasing}.

\medskip
\noindent
Note, in particular, that real-valued functions of one real 
variable that are strictly monotonous and continous are always 
one-to-one and onto and, therefore, are invertible.

\medskip
\noindent
In the following, we briefly review five elementary classes of 
real-valued functions of one real variable that find frequent 
application in the modelling of quantitative problems in {\bf 
economic theory}.

\subsection{Polynomials of degree $n$}
\lb{subsec:polynomials}
Polynomials of degree $n$ are real-valued functions of one real variable of the form
\be
\lb{npol}
\fbox{$\displaystyle\begin{array}{c}
y = f(x) = a_{n}x^{n}+a_{n-1}x^{n-1}+\ldots
+a_{i}x^{i}+\ldots+a_{2}x^{2}+a_{1}x+a_{0}
\\[5mm]
\text{with}\ a_{i}\in\mathbb{R},\ i=1,\ldots,n,
\ \ n\in\mathbb{N},\ a_{n}\neq 0 \ .
\end{array}
$}
\ee
Their domain comprises the entire set of real numbers, i.e., 
$D(f)=\mathbb{R}$. The extent of their target space depends 
specifically on the values of the real constant {\bf coefficients} 
$a_{i}\in\mathbb{R}$. Functions in this class possess a maximum of 
$n$ real {\bf roots}.

\subsection{Rational functions}
Rational functions are constructed by forming the {\bf ratio of two polynomials} of degrees $m$ resp.\ $n$, i.e.,
\be
\fbox{$\displaystyle\begin{array}{c}
y = f(x) = \displaystyle{\frac{p_{m}(x)}{q_{n}(x)}
= \frac{a_{m}x^{m}+\ldots+a_{1}x+a_{0}}{
b_{n}x^{n}+\ldots+b_{1}x+b_{0}}}
\\[5mm]
\text{with}\ a_{i},b_{j}\in\mathbb{R},\ i=1,\ldots,m,
\ j=1,\ldots,n,
\ \ m,n\in\mathbb{N},\ a_{m},b_{n}\neq 0 \ .
\end{array}
$}
\ee
Their domain is given by $D(f) = \mathbb{R} \backslash 
\{x|q_{n}(x)=0\}$. When for the degrees $m$ and $n$ of the 
polynomials $p_{m}(x)$ and $q_{n}(x)$ we have
\begin{itemize}
\item[(i)] $m<n$, then $f$ is referred to as a {\bf proper 
rational function}, or
\item[(ii)] $m\geq n$, then $f$ is referred to as an {\bf improper 
rational function}.
\end{itemize}
In the latter case, application of \emph{polynomial division} 
leads to a separation of $f$ into a purely polynomial part and a 
proper rational part. The {\bf roots} of $f$ always correspond to 
those roots of the numerator polynomial $p_{m}(x)$ for which 
simultaneously $q_{n}(x) \neq 0$ applies. The roots of the 
denominator polynomial $q_{n}(x)$ constitute {\bf poles} of $f$. 
Proper rational functions always tend for very small (i.e., $x \to 
-\infty$) and for very large (i.e., $x \to +\infty$) values of 
their argument to zero.

\subsection{Power-law functions}
Power-law functions exhibit the specific structure given by
\be
\fbox{$\displaystyle
y = f(x) = x^{\alpha} \quad
\text{with}\ \alpha\in\mathbb{R} \ .
$}
\ee
We here confine ourselves to cases with domains $D(f) = 
\mathbb{R}_{>0}$, such that for the coresponding target spaces we 
have $W(f) = \mathbb{R}_{>0}$. Under these conditions, power-law 
functions are strictly monotonously increasing when $\alpha > 0$, 
and strictly monotonously decreasing when $\alpha < 0$. Hence, 
they are inverted by $y = \sqrt[\alpha]{x} = x^{1/\alpha}$. There 
do \emph{not} exist any roots under the conditions stated here.

\subsection{Exponential functions}
\lb{subsec:exponentials}
Exponential functions have the general form
\be
\fbox{$\displaystyle
y = f(x) = a^{x} \quad
\text{with}\ a\in\mathbb{R}_{>0}\backslash\{1\} \ .
$}
\ee
Their domain is $D(f)=\mathbb{R}$, while their target space is 
$W(f)=\mathbb{R}_{>0}$. They exhibit strict monotonous increase 
for $a>1$, and strict monotonous decrease for $0<a<1$. Hence, they 
too are invertible. Their $y$-intercept is generally located at 
$y=1$. For $a>1$, exponential functions are also known as {\bf 
growth functions}.

\medskip
\noindent
\underline{Special case:} When the \emph{constant (!)} base number 
is chosen to be $a=e$, where $e$ denotes the irrational {\bf 
Euler's number} (according to the Swiss mathematician 
\href{http://turnbull.mcs.st-and.ac.uk/history/Biographies/Euler.html}{Leonhard Euler, 1707--1783}) defined by the infinite series
\[
e := \sum_{k=0}^{\infty}\frac{1}{k!}
= \frac{1}{0!} + \frac{1}{1!} + \frac{1}{2!}
+ \frac{1}{3!} + \ldots \ ,
\]
one obtains the {\bf natural exponential function}
\be
y=f(x)=e^{x} =: \exp(x) \ .
\ee
In analogy to the definition of $e$, the relation
\[
e^{x} = \exp(x)
= \sum_{k=0}^{\infty}\frac{x^{k}}{k!}
= \frac{x^{0}}{0!} + \frac{x^{1}}{1!} + \frac{x^{2}}{2!}
+ \frac{x^{3}}{3!} + \ldots
\]
applies.

\subsection{Logarithmic functions}
Logarithmic functions, denoted by
\be
\fbox{$\displaystyle
y = f(x) = \log_{a}(x) \quad
\text{with}\ a\in\mathbb{R}_{>0}\backslash\{1\} \ ,
$}
\ee
are defined as \emph{inverse functions} of the strictly monotonous 
exponential functions $y=f(x)=a^{x}$ --- and vice versa. 
Correspondingly, $D(f)=\mathbb{R}_{>0}$ and $W(f)=\mathbb{R}$ 
apply. Strictly monotonously increasing behaviour is given when 
$a>1$, strictly monotonously decreasing behaviour when $0<a<1$. In 
general, the $x$-intercept is located at $x=1$.

\medskip
\noindent
\underline{Special case:} The {\bf natural logarithmic function} 
(lat.: logarithmus naturalis) obtains when the constant basis 
number is set to $a=e$. This yields
\be
y=f(x)=\log_{e}(x):=\ln(x) \ .
\ee
%

\subsection{Concatenations of real-valued functions}
\lb{subsec:kombfkt}
Real-valued functions from all five categories considered in the 
previous sections may be combined arbitrarily (respecting relevant 
computational rules), either via the four {\bf fundamental 
arithmetical operations}, or via {\bf concatenations}.

\medskip
\noindent
\underline{\bf Theorem:} Let real-valued functions $f$ and $g$ be continuous on domains $D(f)$ resp.\ $D(g)$. Then the combined real-valued functions
\begin{enumerate}
	\item {\bf sum/difference} $f \pm g$, where $(f \pm g)(x)
	:=f(x) \pm g(x)$ with $D(f) \cap D(g)$,
	
	\item {\bf product} $f \cdot g$, where  $(f\cdot g)(x):=f(x)g(x)$
	with $D(f) \cap D(g)$,
	
	\item {\bf quotient} ${\displaystyle\frac{f}{g}}$, where 
	${\displaystyle\left(\frac{f}{g}\right)(x):=\frac{f(x)}{g(x)}}$ 
	with $g(x) \neq 0$ and $D(f) \cap D(g)
	\backslash\{x|g(x)=0\}$,
	
	\item {\bf concatenation} $f \circ g$, where
	$(f\circ g)(x):=f(g(x))$ mit $\{x \in D(g)|g(x) \in D(f)\}$,
\end{enumerate}
are also continuous on the respective domains.

\section[Derivation of differentiable real-valued functions]%
{Derivation of differentiable real-valued functions}
\lb{sec:ablt}
The central theme of this chapter is the mathematical description 
of the {\bf local variability} of \emph{continuous} real-valued 
function of one real variable, $f:D\subseteq\mathbb{R} \rightarrow 
W\subseteq\mathbb{R}$. To this end, let us consider the effect on 
$f$ of a small change of its argument $x$. Supposing we affect a 
change $x \rightarrow x+\Delta x$, with $\Delta x \in \mathbb{R}$, 
what are the resultant consequences for $f$? We immediately find 
that $y \rightarrow y+\Delta y = f(x+\Delta x)$, with $\Delta y
\in \mathbb{R}$, obtains. Hence, a prescribed change of the 
argument~$x$ by a (small) value $\Delta x$ triggers in $f$ a 
change by the amount $\Delta y = f(x+\Delta x)-f(x)$. It is of 
general quantitative interest to compare the {\bf sizes} of these 
two changes. This is accomplished by forming the respective {\bf 
difference quotient}
\[
\frac{\Delta y}{\Delta x} = \frac{f(x+\Delta x)-f(x)}{\Delta x}
\ .
\]
It is then natural, for given $f$, to investigate the limit 
behaviour of this difference quotient as the change $\Delta x$ of 
the argument of $f$ is made successively smaller.

\medskip
\noindent
\underline{\bf Def.:}
A continuous real-valued function $f$ of one real variable is 
called {\bf differentiable at} $\boldsymbol{x \in D(f)}$, when for 
arbitrary $\Delta x \in \mathbb{R}$ the limit
\be
\fbox{$\displaystyle
f^{\prime}(x) := \lim_{\Delta x \to 0}
\frac{\Delta y}{\Delta x}
= \lim_{\Delta x \to 0}
\frac{f(x+\Delta x)-f(x)}{\Delta x}
$}
\ee
exists and is unique. When $f$ is differentiable \emph{for all} $x 
\in D(f)$, then $f$ as such is referred to as being {\bf 
differentiable}.

\medskip
\noindent
The existence of this limit in a point $(x,f(x))$ for a 
real-valued function $f$ requires that the latter exhibits neither 
``jumps'' nor ``kinks,'' i.e., that at $(x,f(x))$ the function is 
sufficiently ``smooth.'' The quantity $f^{\prime}(x)$ is referred 
to as the {\bf first derivative} of the (differentiable) 
function~$f$ at position~$x$. It provides a quantitative measure 
for the {\bf local rate of change} of the function~$f$ in the 
point~$(x,f(x))$. In general one interprets the first 
derivative~$f^{\prime}(x)$ as follows: an increase of the 
argument~$x$ of a differentiable real-valued function~$f$ by $1$ 
(one) unit leads to a change in the value of~$f$ by approximately 
$f^{\prime}(x)\cdot 1$ units.

\medskip
\noindent
Alternative notation for the first derivative of $f$:
$$
f^{\prime}(x) \equiv \frac{{\rm d}f(x)}{{\rm d}x} \ .
$$

\medskip
\noindent
The differential calculus was developed in parallel with the 
integral calculus (see Ch.~\ref{ch7}) during the second half of 
the $17^{\rm th}$ Century, independent of one another by the 
English physicist, mathematiccian, astronomer and philosopher
\href{http://turnbull.mcs.st-and.ac.uk/history/Biographies/Newton.html}{Sir Isaac Newton (1643--1727)} and the German philosopher,
mathematician and physicist
\href{http://turnbull.mcs.st-and.ac.uk/history/Biographies/Leibniz.html}{Gottfried Wilhelm Leibniz (1646--1716)}.

\medskip
\noindent
Via the first derivative of a differentiable function~$f$ at an 
argument $x_{0} \in D(f)$, i.e., $f^{\prime}(x_{0})$, one defines 
the so-called {\bf linearisation} of~$f$ in a neighbourhood 
of~$x_{0}$. The equation describing the associated {\bf tangent} 
to~$f$ in the point~$(x_{0},f(x_{0}))$ is given by
\be
\lb{ftangente}
\fbox{$\displaystyle
y = f(x_{0}) + f^{\prime}\left(x_{0})(x-x_{0}\right) \ .
$}
\ee

\medskip
\noindent
\underline{\bf GDC:} Local values~$f^{\prime}(x_{0})$ of first 
derivatives can be computed for given function~$f$ in mode~{\tt 
CALC} using the interactive routine~{\tt dy/dx}.

\medskip
\noindent
The following rules of differentiation apply for the five families 
of elementary real-valued functions discussed in 
Sec.~\ref{sec:fkt}, as well as concatenations thereof:

\medskip
\noindent
{\bf Rules of differentiation}
\begin{enumerate}
\item $(c)^{\prime} = 0$ for $c = \text{constant} \in \mathbb{R}$
\hfill ({\bf constants})
\item $(x)^{\prime} = 1$ \hfill ({\bf linear function})
\item $(x^{n})^{\prime} = nx^{n-1}$ for $n \in \mathbb{N}$
\hfill ({\bf natural power-law functions})
\item $(x^{\alpha})^{\prime} = \alpha x^{\alpha-1}$ for
$\alpha \in \mathbb{R}$ and $x \in \mathbb{R}_{> 0}$
\hfill ({\bf general power-law functions})
\item $(a^{x})^{\prime} = \ln(a)a^{x}$ for
$a \in \mathbb{R}_{> 0}\backslash\{1\}$
\hfill ({\bf exponential functions})
\item $(e^{ax})^{\prime} = ae^{ax}$ for $a \in \mathbb{R}$
\hfill ({\bf natural exponential functions})
\item $\displaystyle (\log_{a}(x))^{\prime} = \frac{1}{x\ln(a)}$
for $a \in \mathbb{R}_{> 0}\backslash\{1\}$ and $x \in
\mathbb{R}_{> 0}$ \hfill ({\bf logarithmic functions})
\item $(\displaystyle \ln(x))^{\prime} = \frac{1}{x}$
for $x \in \mathbb{R}_{> 0}$
\hfill ({\bf natural logarithmic function}).
\end{enumerate}
For differentiable real-valued functions~$f$ and~$g$ it holds that:
\begin{enumerate}
\item $(cf(x))^{\prime} = cf^{\prime}(x)$ for
$c = \text{constant} \in \mathbb{R}$
\item $(f(x) \pm g(x))^{\prime} = f^{\prime}(x) \pm g^{\prime}(x)$
\hfill ({\bf summation rule})
\item $(f(x)g(x))^{\prime} = f^{\prime}(x)g(x) + f(x)g^{\prime}(x)$
\hfill ({\bf product rule})
\item $\displaystyle \left(\frac{f(x)}{g(x)}\right)^{\prime}
= \frac{f^{\prime}(x)g(x) - f(x)g^{\prime}(x)}{(g(x))^{2}}$
\hfill ({\bf quotient rule})
\item $((f \circ g)(x))^{\prime}
= \left.f^{\prime}(g)\right|_{g=g(x)}\cdot
g^{\prime}(x)$ \hfill ({\bf chain rule})
\item $\displaystyle (\ln(f(x)))^{\prime}
= \frac{f^{\prime}(x)}{f(x)}$ for $f(x) > 0$
\hfill ({\bf logarithmic differentiation})
\item $\displaystyle (f^{-1}(x))^{\prime}
= \left.\frac{1}{f^{\prime}(y)}
\right|_{y=f^{-1}(x)}$, if $f$ is one-to-one and onto.

\hfill ({\bf differentiation of inverse functions}).
\end{enumerate}

\medskip
\noindent
The methods of differential calculus just introduced shall now be 
employed to describe the local change behaviour of a few simple 
examples of functions in {\bf economic theory}, and also to 
determine their local extremal values. The following section 
provides an overview of such frequently occurring {\bf economic 
functions}.

\section[Common functions in economic theory]%
{Common functions in economic theory}
\lb{sec:oekfkt}
%
\begin{enumerate}

\item {\bf total cost function} $K(x) \geq 
0$ \hfill (dim: CU) \\
argument: level of physical output $x 
\geq 0$ (dim: units)

\item {\bf marginal cost function}
$K^{\prime}(x) > 0$ \hfill (dim: CU/unit) \\
argument: level of physical output $x \geq 0$ (dim: units)

\item {\bf average cost function} $K(x)/x > 0$
\hfill (dim: CU/unit) \\
argument: level of physical output $x > 0$ (dim: units)

\item {\bf unit price function}
$p(x) \geq 0$ \hfill (dim: CU/unit) \\
argument: level of physical output $x > 0$ (dim: units)

\item {\bf total revenue function} $E(x) 
:= xp(x) \geq 0$ \hfill (dim: CU) \\
argument: level of physical output $x > 0$ (dim: units)

\item {\bf marginal revenue function}
$E^{\prime}(x) = xp^{\prime}(x)+p(x)$
\hfill (dim: CU/unit) \\
argument: level of physical output $x > 0$ (dim: units)

\item {\bf profit function}
$G(x) := E(x)-K(x)$
\hfill (dim: CU) \\
argument: level of physical output $x > 0$ (dim: units)

\item {\bf marginal profit function}
$G^{\prime}(x) := E^{\prime}(x)-K^{\prime}(x)
= xp^{\prime}(x)+p(x)-K^{\prime}(x)$
\hfill (dim: CU/unit) \\
argument: level of physical output $x > 0$ (dim: units)

\item {\bf utility function}
$U(x)$ \hfill (dim: case dependent) \\
argument: material wealth, opportunity, action $x$ 
(dim: case dependent)

The fundamental concept of a utility function as a means to 
capture in quantitative terms the psychological value (happiness) 
assigned by an economic agent to a certain amount of money, or to 
owning a specific good, was introduced to {\bf economic theory} in 
1738 by the Swiss mathematician and physicist
\href{http://www-history.mcs.st-and.ac.uk/Biographies/Bernoulli_Daniel.html}{Daniel Bernoulli FRS (1700--1782)}; cf. Bernoulli 
(1738)~\ct{ber1738}. The utility function is part of the folklore 
of the theory, and often taken to be a piecewise differentiable, 
right-handedly curved (concave) function, i.e., 
$U^{\prime\prime}(x) < 0$, on the grounds of the \emph{assumption} 
of diminishing marginal utility (happiness) with increasing 
material wealth.

\item {\bf economic efficiency}
$W(x) := E(x)/K(x) \geq 0$
\hfill (dim: 1) \\
argument: level of physical output $x > 0$ (dim: units)

\item {\bf demand function} $N(p) \geq 
0$, monotonously decreasing \hfill (dim: units) \\
argument: unit price $p$, ($0 \leq p \leq p_{\rm max}$)
(dim: CU/unit)

\item {\bf supply function} $A(p) \geq 
0$, monotonously increasing \hfill (dim: units) \\
argument: unit price $p$, ($p_{\rm min} \leq p$)
(dim: CU/units).

\end{enumerate}

\medskip
\noindent
A particularly prominent example of a real-valued economic 
function of one real variable constitutes the {\bf psychological 
value function}, devised by the Israeli--US-American experimental 
psychologists Daniel Kahneman and Amos Tversky (1937--1996) in the 
context of their {\bf Prospect Theory} (a pillar of {\bf 
Behavioural Economics}), which was later awarded a 
\href{http://www.nobelprize.org/nobel_prizes/economics/laureates/2002/}{Sveriges Riksbank Prize in Economic Sciences in Memory of 
Alfred Nobel} in 2002 (cf. Kahneman and Tversky 
(1979)~\ct[p~279]{kahtve1979}, and Kahneman 
(2011)~\ct[p~282f]{kah2011}). A possible representation of this 
function is given by the piecewise description
\be
\lb{psychvaluefct}
v(x)=
\begin{cases}
a\log_{10}\left(1+x\right) & \text{for}\quad
x \in \mathbb{R}_{\geq 0} \\
\\
-2a\log_{10}\left(1-x\right) & \text{for}\quad
x \in \mathbb{R}_{< 0}
\end{cases} \ ,
\ee
with parameter $a \in \mathbb{R}_{>0}$. Overcoming a conceptual 
problem of Bernoulli's utility function, here, in contrast, the 
argument~$x$ quantifies a \emph{change in wealth (or welfare)}
with respect to some given reference point (rather than a specific 
value of wealth itself).

\section{Curve sketching}
\lb{sec:kurvdisk}
Before we turn to discuss applications of {\bf differential 
calculus} to simple quantitative problems in {\bf economic 
theory}, we briefly summarize the main steps of {\bf curve 
sketching} for a real-valued function of one real variable, also 
referred to as {\bf analysis} of the properties of 
differentiability of a real-valued function.
\begin{enumerate}

\item {\bf domain}: $D(f)=\{x \in 
\mathbb{R}|f(x) \ \text{is regular}\}$

\item {\bf symmetries}: for all $x \in D(f)$, is
\begin{itemize}
\item[(i)] $f(-x) = f(x)$, i.e., is $f$ {\bf even}, 
or
\item[(ii)] $f(-x) = -f(x)$, i.e., is $f$ {\bf odd}, 
or
\item[(iii)] $f(-x) \neq f(x) \neq -f(x)$, i.e., $f$ exhibits {\bf 
no symmetries}?
\end{itemize}

\item {\bf roots}: identify all $x_{N} \in D(f)$ that satisfy the 
condition $f(x) \stackrel{!}{=} 0$.

\item {\bf local extremal values}:
\begin{itemize}
\item[(i)] {\bf local minima} of $f$ exist at all $x_{E} \in 
D(f)$, for which the

necessary condition $f^{\prime}(x) \stackrel{!}{=} 0$, and the

sufficient condition $f^{\prime\prime}(x) \stackrel{!}{>} 0$ are 
satisfied simultaneously.

\item[(ii)] {\bf local maxima} of $f$ exist at all $x_{E} \in 
D(f)$, for which the

necessary condition $f^{\prime}(x) \stackrel{!}{=} 0$, and the

sufficient condition $f^{\prime\prime}(x) \stackrel{!}{<} 0$ are 
satisfied simultaneously.
\end{itemize}

\item {\bf points of inflection}: find all $x_{W} 
\in D(f)$, for which the

necessary condition $f^{\prime\prime}(x) \stackrel{!}{=} 0$, and 
the

sufficient condition $f^{\prime\prime\prime}(x) 
\stackrel{!}{\neq} 0$ are satisfied simultaneously.

\item {\bf monotonous behaviour}:
\begin{itemize}
\item[(i)] $f$ is {\bf monotonously 
increasing} for all $x \in D(f)$ with $f^{\prime}(x) > 0$
\item[(ii)] $f$ is {\bf monotonously 
decreasing} for all $x \in D(f)$ with $f^{\prime}(x) < 0$
\end{itemize}

\item {\bf local curvature}:
\begin{itemize}
\item[(i)] $f$ behaves {\bf left-handedly curved}
for $x \in D(f)$ with $f^{\prime\prime}(x) > 0$
\item[(ii)] $f$ behaves {\bf right-handedly 
curved} for $x \in D(f)$ with $f^{\prime\prime}(x) < 0$
\end{itemize}

\item {\bf asymptotic behaviour}:

asymptotes to $f$ are constituted by
\begin{itemize}
\item[(i)] straight lines $y=ax+b$ with the property
$\lim_{x \to +\infty}[f(x)-ax-b]=0$
or $\lim_{x \to -\infty}[f(x)-ax-b]=0$
\item[(ii)] straight lines $x=x_{0}$ at poles
$x_{0} \notin D(f)$
\end{itemize}

\item {\bf range}: $W(f)=\{y \in 
\mathbb{R}|y=f(x)\}$.

\end{enumerate}
%

\section[Analytic investigations of economic functions]%
{Analytic investigations of economic functions}
\lb{sec:extroekfkt}
\subsection{Total cost functions according to Turgot and von 
Th\"unen}
According to the {\bf law of diminishing returns}, which was 
introduced to {\bf economic theory} by the French economist and 
statesman 
\href{http://en.wikipedia.org/wiki/Anne-Robert-Jacques_Turgot,_Baron_de_Laune}{Anne Robert Jacques Turgot (1727--1781)} and also by 
the German economist 
\href{http://en.wikipedia.org/wiki/Johann_Heinrich_von_Thünen}{Johann Heinrich von Th\"unen (1783--1850)}, it is meaningful to 
model non-negative {\bf total cost functions}~$K(x)$ (in CU) 
relating to typical production processes, with argument {\bf 
level of physical output}~$x \geq 0~\text{units}$, as 
a mathematical mapping in terms of a special \emph{polynomial of 
degree 3} [cf.\ Eq.~(\ref{npol})], which is given by
\be
\lb{totalcostfct}
\fbox{$\displaystyle\begin{array}{c}
K(x) = \underbrace{a_{3}x^{3}
+ a_{2}x^{2} + a_{1}x}_{=K_{v}(x)}
+ \underbrace{a_{0}}_{=K_{f}} \\[10mm]
\text{with}\ a_{3},a_{1} > 0,\, a_{2} < 0,
\, a_{0} \geq 0,
\, a_{2}^{2}-3a_{3}a_{1} < 0 \ .
\end{array}
$}
\ee
The model thus contains a total of four free parameters. It is 
the outcome of a systematic {\bf regression analysis} of 
agricultural quantitative--empirical data with the aim to describe 
an inherently {\bf non-linear functional relationship} between a 
few economic variables. As such, the functional relationship 
for~$K(x)$ expressed in Eq.~(\ref{totalcostfct}) was derived from a
practical consideration. It is a reflection of the following 
observed features:
\begin{itemize}
\item[(i)]~for levels of physical output~$x \geq 
0~\text{units}$, the total costs relating to typical production 
processes exhibit strictly monotonously increasing behaviour; thus 

\item[(ii)]~for the total costs there do \emph{not} exist neither 
roots nor local extremal values;\footnote{The last condition in 
Eq.~(\ref{totalcostfct}) ensures a first derivative of~$K(x)$ that 
does \emph{not} possess any roots; cf. the case of a quadratic 
algebraic equation $0\stackrel{!}{=}ax^{2}+bx+c$, with 
discriminant $b^{2}-4ac<0$.} however,

\item[(iii)]~the total costs display \emph{exactly one} point of 
inflection.
\end{itemize}
The continuous curve for~$K(x)$ resulting from these 
considerations exhibits the characteristic shape of an inverted 
capital letter ``S'': beginning at a positive value corresponding 
to fixed costs, the total costs  first increase degressively up to 
a point of inflection, whereafter they continue to increase, but 
in a progressive fashion.

\medskip
\noindent
In broad terms, the functional expression given in 
Eq.~(\ref{totalcostfct}) to model totals costs in dependence of 
the level of physical output is the sum of two contributions, 
the {\bf variable costs}~$K_{v}(x)$ and the {\bf fixed 
costs}~$K_{f}=a_{0}$, viz.
\be
K(x)=K_{v}(x)+K_{f} \ .
\ee

\medskip
\noindent
In {\bf economic theory}, it is commonplace to partition {\bf 
total cost functions} in the diminishing returns picture into 
\emph{four phases}, the boundaries of which are designated by 
special values of the level of physical output of a production 
process: 
\begin{itemize}

\item {\bf phase I} (interval $0~\text{units} \leq x \leq 
x_{W}$):

the total costs~$K(x)$ possess at a level of physical 
output~$x_{W} = -a_{2}/(3a_{3}) > 0~\text{units}$ a {\bf point of 
inflection}. For values of $x$ smaller than $x_{W}$, one 
obtains~$K^{\prime\prime}(x) < 0~\text{CU}/\text{unit}^{2}$, 
i.e., $K(x)$ increases in a degressive fashion. For values of $x$ 
larger than $x_{W}$, the opposite applies, $K^{\prime\prime}(x) > 
0~\text{CU}/\text{unit}^{2}$, i.e., $K(x)$ increases in a 
progressive fashion. The {\bf marginal costs}, given by
\be
K^{\prime}(x) = 3a_{3}x^{2}+2a_{2}x+a_{1}
> 0~\text{CU}/\text{unit}
\quad\quad \text{for\ all} \quad x \geq 0~\text{units} \ ,
\ee
attain a {\bf minimum} at the same level of physical output, 
$x_{W} = -a_{2}/(3a_{3})$.

\item {\bf phase II} (interval $x_{W} < x \leq x_{g_{1}}$):

the {\bf variable average costs}
\be
\frac{K_{v}(x)}{x} = a_{3}x^{2} + a_{2}x + a_{1}
\ ,\quad\quad x > 0~\text{units}
\ee
become {\bf minimal} at a level of physical 
output~$x_{g_{1}}=-a_{2}/(2a_{3}) > 0~\text{units}$. At this value 
of~$x$, \emph{equality of variable average costs and marginal 
costs} applies, i.e.,
\be
\lb{betrmin}
\frac{K_{v}(x)}{x} = K^{\prime}(x) \ ,
\ee
which follows by the quotient rule of differentiation from the 
necessary condition for an extremum of the variable average costs,
\[
0 \stackrel{!}{=} \left(\frac{K_{v}(x)}{x}\right)^{\prime}
= \frac{(K(x)-K_{f})^{\prime}\cdot x - K_{v}(x)\cdot 1}{x^{2}}
\ ,
\]
and the fact that $K_{f}^{\prime}=0~\text{CU/unit}$. Taking care 
of the equality~(\ref{betrmin}), one finds for the tangent 
to~$K(x)$ in the point~$(x_{g_{1}},K(x_{g_{1}}))$ the equation 
[cf. Eq.~(\ref{ftangente})] 
\[
T(x) = K(x_{g_{1}}) + K^{\prime}(x_{g_{1}})(x-x_{g_{1}})
= K_{v}(x_{g_{1}}) + K_{f} + \frac{K_{v}(x_{g_{1}})}{x_{g_{1}}}\,
(x-x_{g_{1}})
= K_{f} + \frac{K_{v}(x_{g_{1}})}{x_{g_{1}}}\,x \ .
\]
Its intercept with the $K$-axis is at~$K_{f}$.

\item {\bf phase III} (interval $x_{g_{1}} < x \leq x_{g_{2}}$):

The {\bf average costs}
\be
\frac{K(x)}{x} = a_{3}x^{2} + a_{2}x + a_{1}
+ \frac{a_{0}}{x} \ ,\quad\quad x > 0~\text{units}
\ee
attain a {\bf minimum} at a level of physical output~$x_{g_{2}} > 
0~\text{units}$, the defining equation of which is given by
$0~\text{CU}\stackrel{!}{=}2a_{3}x_{g_{2}}^{3}+a_{2}x_{g_{2}}^{2}
-a_{0}$. At this value of~$x$, \emph{equality of average costs and marginal costs} obtains, viz.
\be
\lb{betropt1}
\frac{K(x)}{x} = K^{\prime}(x) \ ,
\ee
which follows by the quotient rule of differentiation from the necessary condition for an extremum of the average costs,
\[
0 \stackrel{!}{=} \left(\frac{K(x)}{x}\right)^{\prime}
= \frac{K^{\prime}(x)\cdot x - K(x)\cdot 1}{x^{2}} \ .
\]
Since a quotient can be zero only when its numerator vanishes (and 
its denominator remains non-zero), one finds from re-arranging the 
numerator expression equated to zero the property
\be
\lb{betropt2}
\frac{K^{\prime}(x)}{K(x)/x}
=x\,\frac{K^{\prime}(x)}{K(x)}=1
\quad\quad\text{for}\quad x=x_{g_{2}} \ .
\ee
The corresponding extremal value pair~$(x_{g_{2}},K(x_{g_{2}}))$ 
is referred to in {\bf economic theory} as the {\bf minimum 
efficient scale (MES)}. From a business economics perspective, at 
a level of physical output~$x=x_{g_{2}}$ the (compared to our 
remarks in the Introduction inverted) ratio ``INPUT over OUTPUT,'' 
i.e., $\displaystyle\frac{K(x)}{x}$, becomes most favourable. By 
respecting the property~(\ref{betropt1}), the equation for the 
tangent to~$K(x)$ in this point [cf. Eq.~(\ref{ftangente})] 
becomes 
\[
T(x) = K(x_{g_{2}}) + K^{\prime}(x_{g_{2}})(x-x_{g_{2}})
= K(x_{g_{2}}) + \frac{K(x_{g_{2}})}{x_{g_{2}}}\,(x-x_{g_{2}})
= \frac{K(x_{g_{2}})}{x_{g_{2}}}\,x \ .
\]
Its intercept with the $K$-axis is thus at~$0~\text{CU}$.

\item {\bf phase IV} (half-interval $x > x_{g_{2}}$):

In this phase $K^{\prime}(x)/K(x)/x>1$ obtains; the costs 
associated with the production of one additional unit of a good, 
approximately the marginal costs~$K^{\prime}(x)$, now exceed the 
average costs, $K(x)/x$. This situation is considered unfavourable 
from a business economics perspective.
\end{itemize}
%

\subsection{Profit functions in the diminishing returns picture}
In this section, we confine our considerations, for reasons of 
\emph{simplicity}, to a market sitution with only a single 
supplier of a good in demand. The price policy that this single 
supplier may thus inact defines a state of {\bf monopoly}. 
Moreover, in addition we want to \emph{assume} that for the market 
situation considered {\bf economic equilibrium} obtains. This
manifests itself in equality of {\bf supply} and {\bf demand}, viz.
\be
\lb{eq:ecoequil}
x(p) = N(p) \ ,
\ee
wherein~$x$ denotes a non-negative {\bf supply function} (in 
$\text{units}$) (which is synonymous with the supplier's level of 
physical output) and $N$ a non-negative {\bf demand function} (in 
$\text{units}$), both of which are taken to depend on the positive 
{\bf unit price}~$p$ (in $\text{CU}/\text{unit}$) of the good in 
question. The {\bf supply function}, and with it the {\bf unit 
price}, can, of course, be prescribed by the monopolistic supplier 
in an arbitrary fashion. In a specific quantitative economic 
model, for instance, the {\bf demand function}~$x(p)$ (recall that 
by Eq.~(\ref{eq:ecoequil}) $x(p) = N(p)$ obtains) could be assumed 
to be either a linear or a quadratic function of~$p$. In any case, 
in order for~$x(p)$ to realistically describe an actual 
demand--unit price relationship, it should be chosen as a 
strictly monotonously decreasing function, and as such it is 
\emph{invertible}. The non-negative {\bf demand function}~$x(p)$ 
features two characteristic points, signified by its intercepts 
with the $x$- and the $p$-axes. The {\bf prohibitive 
price}~$p_{\rm proh}$ is to be determined from the condition 
$x(p_{\rm proh}) \stackrel{!}{=} 0~\text{units}$; therefore, it 
constitutes a root of~$x(p)$. The {\bf saturation 
quantity}~$x_{\rm sat}$, on the other hand, is defined by $x_{\rm 
sat}:=x(0~\text{CU/unit})$.

\medskip
\noindent
The inverse function associated with the strictly monotonously 
decreasing non-negative {\bf demand function}~$x(p)$, the {\bf 
unit price function}~$p(x)$ (in $\text{CU}/\text{unit}$), is 
likewise strictly monotonously decreasing. Via $p(x)$, one 
calculates, in dependence on a known amount~$x$ of units 
supplied/demanded (i.e., sold), the {\bf total revenue} (in 
$\text{CU}$) made by a monopolist according to (cf. 
Sec.~\ref{sec:oekfkt})
\be
\lb{eq:ertrag}
E(x) = xp(x) \ .
\ee
Under the \emph{assumption} that the non-negative {\bf total 
costs}~$K(x)$ (in $\text{CU}$) underlying the production process 
of the good in demand can be modelled according to the diminishing 
returns picture of Turgot and von Th\"unen, the {\bf profit 
function} (in $\text{CU}$) of the monopolist in dependence on the 
level of physical output takes the form
\be
\lb{eq:gewinn}
G(x) = E(x) - K(x)
= \underbrace{x\overbrace{p(x)}^{\text{unit\ price}}}_{
\text{total\ revenue}}
- \underbrace{\left[a_{3}x^{3}+a_{2}x^{2}
+a_{1}x+a_{0}\right]}_{\text{total\ costs}} \ .
\ee

\medskip
\noindent
The first two derivatives of $G(x)$ with respect to its 
argument~$x$ are given by
\begin{eqnarray}
G^{\prime}(x) = E^{\prime}(x) - K^{\prime}(x)
& = & xp^{\prime}(x)+p(x)
-\left[3a_{3}x^{2}+2a_{2}x+a_{1}\right] \\
G^{\prime\prime}(x) = E^{\prime\prime}(x) - K^{\prime\prime}(x)
& = & xp^{\prime\prime}(x)
+2p^{\prime}(x)-\left[6a_{3}x+2a_{2}\right] \ .
\end{eqnarray}
Employing the principles of curve sketching set out in 
Sec.~\ref{sec:kurvdisk}, the following characteristic values 
of~$G(x)$ can thus be identified:
\begin{itemize}

\item {\bf break-even point}

$x_{S} > 0~\text{units}$, as the unique solution to the conditions
\be
G(x) \stackrel{!}{=} 0~\text{CU} \quad\quad\text{(necessary 
condition)}
\ee
and
\be
G^{\prime}(x) \stackrel{!}{>} 0~\text{CU}/\text{unit}
\quad\quad\text{(sufficient condition)} \ ,
\ee

\item {\bf end of the profitable zone}

$x_{G} > 0~\text{units}$, as the unique solution to the conditions
\be
G(x) \stackrel{!}{=} 0~\text{CU} \quad\quad\text{(necessary 
condition)}
\ee
and
\be
G^{\prime}(x) \stackrel{!}{<} 0~\text{CU}/\text{unit}
\quad\quad\text{(sufficient condition)} \ ,
\ee

\item {\bf maximum profit}

$x_{M} > 0~\text{CU}$, as the unique solution to the conditions
\be
G^{\prime}(x) \stackrel{!}{=} 0~\text{CU}/\text{unit}
\quad\quad\text{(necessary condition)}
\ee
and
\be
G^{\prime\prime}(x) \stackrel{!}{<} 0~\text{CU}/\text{unit}^{2}
\quad\quad\text{(sufficient condition)} \ .
\ee
\end{itemize}
At this point, we like to draw the reader's attention to a special 
geometric property of the quantitative model for {\bf profit} that 
we just have outlined: at maximum profit, the {\bf total revenue 
function}~$E(x)$ and the {\bf total cost function}~$K(x)$ always 
possess \emph{parallel tangents}. This is due to the fact that by 
the necessary condition for an extremum to exist, one finds that
\be
0~\text{CU}/\text{unit} \stackrel{!}{=} G^{\prime}(x)
= E^{\prime}(x) - K^{\prime}(x)
\qquad\Leftrightarrow\qquad
E^{\prime}(x) \stackrel{!}{=} K^{\prime}(x) \ .
\ee

\medskip
\noindent
\underline{\bf GDC:} Roots and local maxima resp.\ minima can be 
easily determined for a given stored function in mode {\tt CALC} 
by employing the interactive routines {\tt zero} and
{\tt maximum} resp.\ {\tt minimum}.

\leftout{
\medskip
\noindent
Von besonderem mathematischen Interesse ist in diesem Zusammenhang
die folgende Betrachtung. Bekannt sei eine Gewinnfunktion $G(x)$
der polynomialen Form
\be
G(x) = -a_{3}(x+a)(x-b)(x-c)
= a_{3}\left[-x^{3}+(b+c-a)x^{2}+(ab-bc+ca)x-abc\right] \ ,
\ee
mit $a,b,c > 0$, $b < c$ und $x \geq 0~\text{ME}$, der nach Gln. 
(\ref{eq:ertrag}) und (\ref{eq:gewinn}) eine {\em lineare\/} 
Preis--Absatz--Funktion $p(x)$ zugrunde liegt. Dann
\"uberf\"uhren die simultanen {\bf Skalentransformationen} (engl.: 
scale transformation)
\be
x \mapsto \lambda x \ , \qquad
G(x) \mapsto \frac{1}{\lambda^{3}}\,G(\lambda x) \ ,
\qquad
\lambda > 0
\ee
von unabh\"angiger und abh\"angiger Variable die
Gewinnfunktion $G(x)$ in eine zweite, zu $G(x)$
{\bf selbst\"ahnliche} (engl.: self-similar) Gewinnfunktion
\bea
\tilde{G}(x) & = & -a_{3}\left(x+\frac{a}{\lambda}\right)
\left(x-\frac{b}{\lambda}\right)\left(x-\frac{c}{\lambda}\right)
\nonumber \\
& = & a_{3}\left[-x^{3}+\frac{(b+c-a)}{\lambda}\,x^{2}
+\frac{(ab-bc+ca)}{\lambda^{2}}\,x
-\frac{abc}{\lambda^{3}}\right] \ .
\eea
Analog l\"asst sich mit polynomialen \"okonomischen 
Funktionen $p(x)$, $E(x)$ und $K(x)$ verfahren.
}

\medskip
\noindent
To conclude these considerations, we briefly turn to elucidate the 
technical term {\bf Cournot's point}, which frequently arises in 
quantitative discussions in {\bf economic theory}; this is named 
after the French mathematician and economist 
\href{http://turnbull.mcs.st-and.ac.uk/history/Biographies/Cournot.html}{Antoine--Augustin Cournot (1801--1877)}. {\bf Cournot's 
point} simply labels the profit-optimal combination of the level 
of physical output and the associated unit price, $(x_{M}, 
p(x_{M}))$, for the {\bf unit price function}~$p(x)$ of a good in 
a monopolistic market situation. Note that for this specific 
combination of optimal values the {\bf Amoroso--Robinson formula} 
applies, which was developed by the Italian mathematician and 
economist \href{http://en.wikipedia.org/wiki/Luigi_Amoroso}{Luigi 
Amoroso (1886--1965)} and the British economist
\href{http://en.wikipedia.org/wiki/Joan_Robinson}{Joan Violet
Robinson (1903--1983)}. This states that
\be
p(x_{M}) = \frac{K^{\prime}(x_{M})}{1+\varepsilon_{p}(x_{M})} \ ,
\ee
with $K^{\prime}(x_{M})$ the value of the marginal costs 
at~$x_{M}$, and $\varepsilon_{p}(x_{M})$ the value of the {\bf  
elasticity} of the unit price function at~$x_{M}$ (see the 
following Sec.~\ref{sec:elast}). Starting from the defining 
equation of the {\bf total revenue}~$E(x)=xp(x)$, the {\bf 
Amoroso--Robinson formula} is derived by evaluating the first 
derivative of~$E(x)$ at $x_{M}$, so
\[
E^{\prime}(x_{M}) = p(x_{M}) + x_{M}p^{\prime}(x_{M})
= p(x_{M})\left[1
+ x_{M}\,\frac{p^{\prime}(x_{M})}{p(x_{M})}\right]
\overbrace{=}^{\text{Sec.~\ref{sec:elast}}}= p(x_{M})\left[1 + 
\varepsilon_{p}(x_{M})\right] \ ,
\]
and then re-arranging to solve for $p(x_{M})$, using the fact that 
$E^{\prime}(x_{M})=K^{\prime}(x_{M})$.

\medskip
\noindent
\underline{\bf Remark:} In a market situation where {\bf perfect 
competition} applies, one \emph{assumes} that the {\bf unit price 
function} has settled to a \emph{constant} value~$p(x) = p = 
\text{constant}>0~\text{CU/unit}$ (and, hence, $p^{\prime}(x) = 
0~\text{CU/unit}^{2}$ obtains).

\subsection{Extremal values of rational economic functions}
Now we want to address the determination of extremal values of 
economic functions that constitute ratios in the sense of the 
construction
\[
\frac{\text{OUTPUT}}{\text{INPUT}} \ ,
\]
a topic raised in the Introduction.

\medskip
\noindent
Let us consider two examples for determining {\bf local maxima} of 
ratios of this kind.
\begin{itemize}
\item[(i)]
We begin with the {\bf average profit} in dependence on the level 
of physical output~$x \geq 0~\text{units}$,
\be
\frac{G(x)}{x} \ .
\ee
The conditions that determine a local maximum are 
$[G(x)/x]^{\prime} \stackrel{!}{=} 0~\text{CU}/\text{unit}^{2}$ 
and $[G(x)/x]^{\prime\prime} \stackrel{!}{<} 
0~\text{CU}/\text{unit}^{3}$. Respecting the quotient rule of 
differentiation (cf. Sec.~\ref{sec:ablt}), the first condition 
yields
\be
\frac{G^{\prime}(x)x-G(x)}{x^{2}} = 0~\text{GE}/\text{ME}^{2} \ .
\ee
Since a quotient can only be zero when its numerator vanishes 
while its denominator remains non-zero, it immediately follows that
\be
\lb{dgextr2}
G^{\prime}(x)x-G(x) = 0~\text{CU}
\quad\Rightarrow\quad
x\,\frac{G^{\prime}(x)}{G(x)} = 1 \ .
\ee
The task at hand now is to find a (unique) value of the level of 
physical output~$x$ which satisfies this last condition, and for 
which the second derivative of the average profit becomes negative.

\item[(ii)]
To compare the performance of two companies over a given period of 
time in a meaningful way, it is recommended to adhere only to 
measures that are \emph{dimensionless ratios}, and so independent 
of {\bf scale}. An example of such a dimensionless ratio is the 
measure referred to as {\bf economic efficiency},
\be
W(x)=\frac{E(x)}{K(x)} \ ,
\ee
which expresses the {\bf total revenue} (in $\text{CU}$) of a 
company for a given period as a multiple of the {\bf total costs} 
(in $\text{CU}$) it had to endure during this period, both as 
functions of the {\bf level of physical output}. In analogy to our 
discussion in~(i), the conditions for the existence of a local 
maximum amount to $[E(x)/K(x)]^{\prime} \stackrel{!}{=} 0 \times 
1/\text{unit}$ and $[E(x)/K(x)]^{\prime\prime} \stackrel{!}{<} 0 
\times 1/\text{unit}^{2}$. By the quotient rule of differentiation 
(see Sec.~\ref{sec:ablt}), the first condition leads to
\be
\frac{E^{\prime}(x)K(x)-E(x)K^{\prime}(x)}{K^{2}(x)}
= 0\times 1/\text{unit} \ ,
\ee
i.e., for $K(x) > 0~\text{CU}$,
\be
\lb{wirtextr1}
E^{\prime}(x)K(x)-E(x)K^{\prime}(x) = 0~\text{CU}^{2}/\text{unit}
\ .
\ee
By re-arranging and multiplication with~$x > 0~\text{unit}$, this 
can be cast into the particular form
\be
\lb{wirtextr2}
x\,\frac{E^{\prime}(x)}{E(x)} = x\,\frac{K^{\prime}(x)}{K(x)} \ .
\ee
The reason for this special kind of representation of the 
necessary condition for a local maximum to exist [and also for 
Eq.~(\ref{dgextr2})] will be clarified in the subsequent section.
Again, a value of the level of physical output which satisfies 
Eq.~(\ref{wirtextr2}) must in addition lead to a negative second 
derivative of the {\bf economic efficiency} in order to satisfy 
the sufficient condition for a local maximum to exist.
\end{itemize}
%

\section[Elasticities]{Elasticities}
\lb{sec:elast}
Finally, we pick up once more the discussion on quantifying the 
{\bf local variability} of differentiable real-valued functions of 
one real variable, $f:D\subseteq\mathbb{R} \rightarrow 
W\subseteq\mathbb{R}$, though from a slightly different 
perspective. For reasons to be elucidated shortly, we confine 
ourselves to considerations of 
regimes of $f$ with \emph{positive} values of the argument~$x$ and 
also \emph{positive} values $y=f(x) > 0$ of the function itself.

\medskip
\noindent 
As before in Sec.~\ref{sec:ablt}, we want to assume a small change 
of the value of the argument~$x$ and evaluate its resultant effect 
on the value~$y=f(x)$. This yields
\be
x \stackrel{\Delta x \in \mathbb{R}}{\longrightarrow} x+\Delta x
\qquad \Longrightarrow \qquad
y=f(x) \stackrel{\Delta y \in \mathbb{R}}{\longrightarrow}
y+\Delta y=f(x+\Delta x) \ .
\ee
We remark in passing that {\bf relative changes} of non-negative 
quantities are defined by the quotient
\[
\displaystyle
\frac{\text{new\ value}-\text{old\ value}}{\text{old\ value}}
\]
under the prerequisite that ``$\text{old\ value}>0$'' applies. It 
follows from this specific construction that the minimum value a 
relative change can possibly attain amounts to ``$-1$'' 
(corresponding to a decrease of the ``$\text{old\ value}$'' by 
100\%).

\medskip
\noindent
Related to this consideration we identify the following terms:
\begin{itemize}
\item a prescribed {\bf absolute change} of the independent 
variable~$x$: \qquad $\Delta x$\ ,
\item the resultant {\bf absolute change} of the function $f$:
\qquad $\Delta y=f(x+\Delta x)-f(x)$\ ,
\item the associated {\bf relative change} of the independent 
variable~$x$:
\qquad $\displaystyle \frac{\Delta x}{x}$\ ,
\item the associated resultant {\bf relative change} of the 
function~$f$:
\qquad $\displaystyle \frac{\Delta y}{y}
=\frac{f(x+\Delta x)-f(x)}{f(x)}$\ .
\end{itemize}

\medskip
\noindent
Now let us compare the {\bf order-of-magnitudes} of the two 
relative changes just envisaged, $\displaystyle \frac{\Delta 
x}{x}$ and $\displaystyle \frac{\Delta y}{y}$. This is realised by 
considering the value of their quotient, ``resultant relative 
change of~$f$ divided by the prescribed relative change of~$x$"':
\[
\frac{\displaystyle\frac{\Delta y}{y}}{\displaystyle\frac{\Delta 
x}{x}}
=\frac{\displaystyle\frac{f(x+\Delta 
x)-f(x)}{f(x)}}{\displaystyle\frac{\Delta x}{x}} \ .
\]
Since we assumed $f$ to be differentiable, it is possible to 
investigate the behaviour of this quotient of relative changes in 
the limit of increasingly smaller prescribed relative 
changes~$\displaystyle \frac{\Delta x}{x} \to 0 \Rightarrow \Delta 
x \to 0$ near some $x > 0$. One thus defines:

\medskip
\noindent
\underline{\bf Def.:}
For a differentiable real-valued function~$f$ of one real 
variable~$x$, the \emph{dimensionless} (i.e., units-independent) 
quantity
\be
\fbox{$\displaystyle
\varepsilon_{f}(x)
:=\lim_{\Delta x\to 0}\frac{\displaystyle\frac{\Delta 
y}{y}}{\displaystyle\frac{\Delta x}{x}}
= \lim_{\Delta x\to 0}\frac{\displaystyle\frac{f(x+\Delta 
x)-f(x)}{f(x)}}{\displaystyle\frac{\Delta x}{x}}
= x\,\frac{f^{\prime}(x)}{f(x)}
$}
\ee
is referred to as the {\bf elasticity} of the function~$f$ at position~$x$.

\medskip
\noindent
The elasticity of~$f$ quantifies its resultant relative change in 
response to a prescribed infinitesimally small relative change of 
its argument~$x$, starting from some positive initial value $x>0$. 
As such it constitutes a measure for the {\bf relative local rate 
of change} of a function~$f$ in a point~$(x,f(x))$. In {\bf 
economic theory}, in particular, one adheres to the following 
interpretation of the elasticity~$\varepsilon_{f}(x)$: when the 
postive argument~$x$ of some positive differentiable real-valued 
function~$f$ is increased by $1~\%$, then in consequence $f$ will 
change approximately by~$\varepsilon_{f}(x)\times
1~\%$.

\medskip
\noindent
In the scientific literature one often finds the elasticity of a 
positive differentiable function~$f$ of a positive argument~$x$ 
expressed in terms of logarithmic differentiation. That is,
$$
\varepsilon_{f}(x)
:= \frac{{\rm d}\ln[f(x)]}{{\rm d}\ln(x)}
\qquad\text{for}\ x>0 \ \text{and} \ f(x)>0 \ ,
$$
since by the chain rule of differentiation it holds that
$$
\frac{{\rm d}\ln[f(x)]}{{\rm d}\ln(x)}
= \frac{\displaystyle\frac{{\rm 
d}f(x)}{f(x)}}{\displaystyle\frac{{\rm d}x}{x}}
= x\,\frac{\displaystyle\frac{{\rm d}f(x)}{{\rm d}x}}{f(x)}
= x\,\frac{f^{\prime}(x)}{f(x)} \ .
$$
The logarithmic representation of the elasticity of a 
differentiable function~$f$ immediately explains why, at the 
beginning, we confined our considerations to positive 
differentiable functions of positive arguments only.\footnote{To 
extend the regime of applicability of the 
measure~$\varepsilon_{f}$, one may consider working in terms of 
absolute values~$|x|$ and $|f(x)|$. Then one has to distinguish 
between four cases, which need to be looked at separately: 
(i)~$x>0$, $f(x)>0$, (ii)~$x<0$, $f(x)>0$, (iii)~$x<0$,
$f(x)<0$ and (iv) $x>0$, $f(x)<0$.} A brief look at the list of 
standard economic functions provided in Sec.~\ref{sec:oekfkt} 
reveals that most of these (though not all) are positive functions 
of non-negative arguments.

\medskip
\noindent
For the elementary classes of real-valued functions of one real 
variable discussed in Sec.~\ref{sec:fkt} one finds:

\medskip
\noindent
{\bf Standard elasticities}
\begin{enumerate}
\item $f(x)=x^{n}$ for $n \in \mathbb{N}$
and $x \in \mathbb{R}_{> 0}
\ \Rightarrow\ \varepsilon_{f}(x)=n$
\hfill ({\bf natural power-law functions})
\item $f(x)=x^{\alpha}$ for $\alpha \in \mathbb{R}$
and $x \in \mathbb{R}_{> 0}
\ \Rightarrow\ \varepsilon_{f}(x)=\alpha$
\hfill ({\bf general power-law functions})
\item $f(x)=a^{x}$ for $a \in \mathbb{R}_{> 0}\backslash\{1\}$
and $x \in \mathbb{R}_{> 0}
\ \Rightarrow\ \varepsilon_{f}(x)=\ln(a)x$
\hfill ({\bf exponential functions})
\item $f(x)=e^{ax}$ for $a \in \mathbb{R}$
and $x \in \mathbb{R}_{> 0}
\ \Rightarrow\ \varepsilon_{f}(x)=ax$
\hfill ({\bf natural exponential functions})
\item $f(x)=\log_{a}(x)$ for $a \in \mathbb{R}_{> 0}
\backslash\{1\}$ and $x \in \mathbb{R}_{> 0}$

\hfill\hfill$\displaystyle \quad\Rightarrow\quad
\varepsilon_{f}(x)=\frac{1}{\ln(a)\log_{a}(x)}$
\hfill ({\bf logarithmic functions})
\item $f(x)=\ln(x)$ for $\displaystyle
x \in \mathbb{R}_{> 0}
\quad\Rightarrow\quad \varepsilon_{f}(x)=\frac{1}{\ln(x)}$
\hfill ({\bf natural logarithmic function}).
\end{enumerate}

\medskip
\noindent
In view of these results, we would like to emphasise the fact that 
for the entire family of {\bf general power-law functions} the 
elasticity~$\varepsilon_{f}(x)$ has a \emph{constant value}, 
independent of the value of the argument~$x$. It is this very 
property which classifies {\bf general power-law functions} as 
{\bf scale-invariant}. When {\bf scale-invariance} obtains, 
dimensionless ratios, i.e., quotients of variables of the same 
physical dimension, reduce to \emph{constants}. In this context, 
we would like to remark that scale-invariant (fractal) power-law 
functions of the form $f(x)=Kx^{\alpha}$, with $K > 0$ and $\alpha 
\in \mathbb{R}_{<0}\backslash \{-1\}$, are frequently employed in 
{\bf Economics} and the {\bf Social Sciences} for modelling {\bf 
uncertainty} of {\bf economic agents} in {\bf decision-making 
processes}, or for describing probability distributions of {\bf 
rare event phenomena}; see, e.g., Taleb 
(2007)~\ct[p~326ff]{tal2007} or Gleick 
(1987)~\ct[Chs.~5~and~6]{gle1987}. This is due, in part, to 
the curious property that for certain values of the 
exponent~$\alpha$ general power-law probability distributions 
attain unbounded variance; cf. Ref.~\ct[Sec.~8.9]{hve2015}.

\medskip
\noindent
Practical applications in {\bf economic theory} of the concept of 
an elasticty as a measure of relative change of a differentiable 
real-valued function~$f$ of one real variable~$x$ are generally 
based on the following \emph{linear (!)} approximation: beginning 
at~$x_{0}>0$, for small prescribed percentage changes of the 
argument~$x$ in the interval $0~\% < {\displaystyle\frac{\Delta 
x}{x_{0}}} \leq 5~\%$, the resultant percentage changes of~$f$ 
amount approximately to
\be
(\text{percentage\ change\ of}\ f)
\approx (\text{elasticity\ of}\ f\ \text{at}\ x_{0}) \times
(\text{percentage\ change\ of}\ x) \ ,
\ee
or, in terms of a mathematical formula, to
\be
\frac{f(x_{0}+\Delta x)-f(x_{0})}{f(x_{0})}
\approx \varepsilon_{f}(x_{0})
\frac{\Delta x}{x_{0}} \ .
\ee

\medskip
\noindent
We now draw the reader's attention to a special kind of 
terminology developed in {\bf economic theory} to describe the 
{\bf relative local change behaviour} of economic functions in 
qualitative terms. For $x \in D(f)$, the relative local change 
behaviour of a function~$f$ is called
\begin{itemize}
	\item {\bf inelastic}, whenever $|\varepsilon_{f}(x)|<1$,
	\item {\bf unit elastic}, when $|\varepsilon_{f}(x)|=1$, and
	\item {\bf elastic}, whenever $|\varepsilon_{f}(x)|>1$.
\end{itemize}
For example, a total cost function~$K(x)$ in the diminishing 
returns picture exhibits unit elastic behaviour at the minimum 
efficient scale~$x=x_{g_{2}}$ where, by Eq.~(\ref{betropt2}), 
$\varepsilon_{K}(x_{g_{2}})=1$. Also, at the local maximum of an 
average profit function~$G(x)/x$, the property
$\varepsilon_{G}(x)=1$ applies; cf. Eq.~(\ref{dgextr2}).

\medskip
\noindent
Next, we review the computational rules one needs to adhere to 
when calculating elasticities for combinations of two real-valued 
functions of one real variable in the sense of 
Sec.~\ref{subsec:kombfkt}:

\medskip
\noindent
{\bf Computational rules for elasticities}

\noindent
If $f$ and $g$ are differentiable real-valued functions of one 
real variable, with elasticities~$\varepsilon_{f}$ and 
$\varepsilon_{g}$, it holds that:
\begin{enumerate}
	\item {\bf product} $f \cdot g$: \qquad\qquad
	$\varepsilon_{f \cdot g}(x)
	= \varepsilon_{f}(x) + \varepsilon_{g}(x)$,
	\item {\bf quotient} ${\displaystyle\frac{f}{g}}$, $g \neq 0$: 
	\qquad\qquad 	$\varepsilon_{f/g}(x)
	= \varepsilon_{f}(x) - \varepsilon_{g}(x)$,
	\item {\bf concatenation} $f \circ g$: \qquad\qquad
	$\varepsilon_{f \circ g}(x)
	= \varepsilon_{f}(g(x))\cdot\varepsilon_{g}(x)$,
	\item {\bf inverse function} $f^{-1}$: \qquad\qquad
	$\displaystyle \varepsilon_{f^{-1}}(x)
	= \left.\frac{1}{\varepsilon_{f}(y)}\right|_{y=f^{-1}(x)}$.
\end{enumerate}

\medskip
\noindent
To end this chapter, we remark that for a positive differentiable 
real-valued function~$f$ of one positive real variable~$x$, a 
second elasticity may be defined according to
\be
\lb{eq:secondelasticity}
\varepsilon_{f}\left[\varepsilon_{f}(x)\right]
:= x\,\frac{{\rm d}}{{\rm d}x}\left[\frac{x}{f(x)}\,\frac{{\rm 
d}f(x)}{{\rm d}x}\right] \ .
\ee
Of course, by analogy this procedure may be generalised to higher 
derivatives of~$f$ still.


\chapter[Integral calculus of real-valued functions]%
{Integral calculus of real-valued functions of one real variable}
\lb{ch8}

\vspace{10mm}
\noindent
In the final chapter of these lecture notes we give a brief 
overview of the main definitions and laws of the {\bf integral 
calculus} of real-valued functions of one variable. Subsequently 
we consider a simple application of this tool in {\bf economic 
theory}.

\section[Indefinite integrals]{Indefinite integrals}
\lb{sec:unbint}
\underline{\bf Def.:} Let $f$ be a continuous real-valued function 
of one real variable and $F$ a differentiable real-valued function 
of the same real variable, with $D(f)=D(F)$. Given that $f$ and 
$F$ are related according to
\be
\fbox{$\displaystyle
F^{\prime}(x) = f(x)
\quad\quad\text{\emph{for all}}\quad x\in D(f) \ ,
$}
\ee
then $F$ is referred to as a {\bf primitive} of $f$.

\medskip
\noindent
\underline{\bf Remark:} The primitive of a given continuous 
real-valued function $f$ \emph{cannot} be unique. By the rules of 
differentiation discussed in Sec.~\ref{sec:ablt}, besides $F$ 
also $F+c$, with $c \in \mathbb{R}$ a real-valued constant, 
constitutes a primitive of $f$ since $(c)^{\prime} = 0$.

\medskip
\noindent
\underline{\bf Def.:} If $F$ is a primitive of a continuous 
real-valued function $f$ of one real variable, then
\be
\fbox{$\displaystyle
\int f(x)\,{\rm d}x = F(x) + c \ , \quad
c~=~\text{constant}~\in~\mathbb{R} \ ,
\quad \text{with}\ F^{\prime}(x) = f(x)
$}
\ee
defines the {\bf indefinite integral} of the function $f$. The 
following names are used to refer to the different ingredients in 
this expression:
\begin{itemize}
\item $x$ --- the {\bf integration variable},
\item $f(x)$ --- the {\bf integrand},
\item ${\rm d}x$ --- the {\bf differential}, and, lastly,
\item $c$ --- the {\bf constant of integration}.
\end{itemize}

\medskip
\noindent
For the elementary, continuous real-valued functions of one 
variable introduced in Sec.~\ref{sec:fkt}, the following rules of 
indefinite integration apply:

\medskip
\noindent
{\bf Rules of indefinite integration}
\begin{enumerate}
\item $\int \alpha\,{\rm d}x = \alpha x + c$ with $\alpha
= \text{constant} \in \mathbb{R}$ \hfill ({\bf constants})
\item $\displaystyle \int x\,{\rm d}x = \frac{x^{2}}{2} + c$
 \hfill ({\bf linear functions})
\item $\displaystyle \int x^{n}\,{\rm d}x = \frac{x^{n+1}}{n+1} + 
c$ for $n \in \mathbb{N}$ \hfill ({\bf natural power-law 
functions})
\item $\displaystyle \int x^{\alpha}\,{\rm d}x = 
\frac{x^{\alpha+1}}{\alpha+1}
+ c$ for $\alpha \in \mathbb{R}\backslash\{-1\}$ and
$x \in \mathbb{R}_{> 0}$ \hfill ({\bf general power-law functions})
\item $\displaystyle \int a^{x}\,{\rm d}x = \frac{a^{x}}{\ln(a)} + c$ for $a \in \mathbb{R}_{> 0}\backslash\{1\}$
\hfill ({\bf exponential functions})
\item $\displaystyle \int e^{ax}\,{\rm d}x = \frac{e^{ax}}{a} + c$
for $a \in \mathbb{R}\backslash\{0\}$
\hfill ({\bf natural exponential functions})
\item $\int x^{-1}\,{\rm d}x = \ln|x| + c$ for $x \in \mathbb{R}\backslash\{0\}$.
\end{enumerate}
Special methods of integration need to be employed when the 
integrand consists of a concatanation of elementary real-valued 
functions. Here we provide a list with the main tools for this 
purpose. For differentiable real-valued functions $f$ and $g$, it 
holds that
\begin{enumerate}
\item $\int(\alpha f(x) \pm \beta g(x))\,{\rm d}x
= \alpha\int f(x)\,{\rm d}x \pm \beta\int g(x)\,{\rm d}x$ \\
with $\alpha,\beta = \text{constant} \in \mathbb{R}$
\hfill ({\bf summation rule})
\item $\int f(x)g^{\prime}(x)\,{\rm d}x = f(x)g(x)
- \int f^{\prime}(x)g(x)\,{\rm d}x$
\hfill ({\bf integration by parts})
\item $\int f(g(x))g^{\prime}(x)\,{\rm d}x
\overbrace{=}^{u=g(x)\ \text{and}\ {\rm d}u=g^{\prime}(x){\rm d}x}
\int f(u)\,{\rm d}u = F(g(x)) + c$
\hfill ({\bf substitution method})
\item $\displaystyle \int\frac{f^{\prime}(x)}{f(x)}\,{\rm d}x
= \ln|f(x)| + c$ for $f(x) \neq 0$\hfill ({\bf logarithmic integration}).
\end{enumerate}
%

\section[Definite integrals]{Definite integrals}
\lb{sec:bint}
\medskip
\noindent
\underline{\bf Def.:} Let $f$ be a real-valued function of one variable which is continuous on an interval $[a,b] \subset D(f)$, and let $F$ be a primitive of $f$. Then the expression
\be
\fbox{$\displaystyle
\int_{a}^{b}f(x)\,{\rm d}x
= {\displaystyle\left.F(x)\right|_{x=a}^{x=b}}
= F(b) - F(a)
$}
\ee
defines the {\bf definite integral} of $f$ in the 
{\bf limits of integration} $a$ and $b$.

\medskip
\noindent
For definite integrals the following general rules apply:
\begin{enumerate}
\item $\displaystyle\int_{a}^{a}f(x)\,{\rm d}x = 0$
\hfill ({\bf identical limits of integration})
\item $\displaystyle\int_{b}^{a}f(x)\,{\rm d}x = 
-\int_{a}^{b}f(x)\,{\rm d}x$
\hfill ({\bf interchange of limits of integration})
\item $\displaystyle\int_{a}^{b}f(x)\,{\rm d}x = 
\int_{a}^{c}f(x)\,{\rm d}x
+ \int_{c}^{b}f(x)\,{\rm d}x$ for $c \in [a,b]$
\hfill ({\bf split of integration interval}).
\end{enumerate}

\medskip
\noindent
\underline{\bf Remark:} The main qualitative difference between an 
(i)~indefinite integral and a (ii)~definite integral of a 
continuous real-valued function of one variable reveals intself in 
the different kinds of outcome: while (i)~yields as a result a 
real-valued (primitive) \emph{function}, (ii)~simply yields a 
single real \emph{number}.

\medskip
\noindent
\underline{\bf GDC:} For a stored real-valued function, the 
evaluation of a definite integral can be performed in mode {\tt 
CALC} with the pre-programmed function $\int${\tt f(x)dx}. The 
corresponding limits of integration need to be specified 
interactively.

\medskip
\noindent
As indicated in Sec.~\ref{sec:elast}, the scale-invariant 
power-law functions $f(x)=x^{\alpha}$ for $\alpha \in \mathbb{R}$
and $x \in \mathbb{R}_{> 0}$ play a special role in practical 
applications. For $x \in \left[a,b\right] \subset \mathbb{R}_{> 
0}$ and $\alpha \neq -1$ it holds that
\be
\int_{a}^{b}x^{\alpha}\,{\rm d}x
= \left.\frac{x^{\alpha+1}}{\alpha+1}\right|_{x=a}^{x=b}
= \frac{1}{\alpha+1}\left(b^{\alpha+1}-a^{\alpha+1}\right) \ .
\ee
Problematic in this context can be considerations of taking limits 
of the form $a \to 0$ resp. $b \to \infty$, since for either of 
the two cases
\begin{itemize}
\item[(i)] case $\alpha < -1$:
\be
\lim_{a \to 0}\int_{a}^{b}x^{\alpha}\,{\rm d}x \to \infty \ ,
\ee
\item[(ii)] case $\alpha > -1$:
\be
\lim_{b \to \infty}\int_{a}^{b}x^{\alpha}\,{\rm d}x \to \infty \ ,
\ee
\end{itemize}
one ends up with {\bf divergent} mathematical expressions.

\section[Applications in economic theory]{Applications in economic 
theory}
\lb{sec:intanw}
The starting point shall be a simple market situation for a single 
product. For this product, on the one-hand side, there be a {\bf 
demand function}~$N(p)$ (in $\text{units}$) which is monotonously 
decreasing on the price interval $[p_{u},p_{o}]$; the limit values 
$p_{u}$ and $p_{o}$ denote the minimum and maximum prices per unit 
(in CU/u) acceptable for the product. On the other hand, the 
market situation be described by a {\bf supply function}~$A(p)$ 
(in $\text{units}$) which is monotonously increasing on 
$[p_{u},p_{o}]$.

\medskip
\noindent
The {\bf equilibrium unit price} $p_{M}$ (in 
$\text{CU}/\text{unit}$) for this product is defined by assuming a 
state of {\bf economic equilibrium} of the market, quantitatively 
expressed by the condition
\be
A(p_{M}) = N(p_{M}) \ .
\ee
Geometrically, this condition defines common points of 
intersection for the functions $A(p)$ and $N(p)$ (when they exist).

\medskip
\noindent
\underline{\bf GDC:} Common points of intersection for stored 
functions $f$ and $g$ can be easily determined interactively in 
mode {\tt CALC} employing the routine {\tt intersect}.

\medskip
\noindent
In a drastically simplified fashion, we now turn to compute the 
revenue made on the market by the suppliers of a new product for 
either of three possible {\bf strategies of market entry}.
\begin{enumerate}

\item {\bf Strategy~1:} The revenue obtained by the suppliers when 
the new product is being sold straight at the equilibrium unit 
price $p_{M}$, in an amount $N(p_{M})$, is simply given by
\be
U_{1} = U(p_{M}) = p_{M}N(p_{M}) \qquad
(\text{in\ CU}) \ .
\ee

\item {\bf Strategy~2:} Some consumers would be willing to 
purchase the product intially also at a unit price which is higher 
than $p_{M}$. If, hence, the suppliers decide to offer the product 
initially at a unit price $p_{o} > p_{M}$, and then, in order to 
generate further demand, to continuously\footnote{This is a strong 
mathematical assumption aimed at facilitating the actual 
calculation to follow.} (!) \emph{reduce} the unit price to the 
lower $p_{M}$, the revenue obtained yields the larger value
\be
U_{2} = U(p_{M}) + \int_{p_{M}}^{p_{o}}N(p)\,{\rm d}p \ .
\ee
Since the amount of money
\be
K := \int_{p_{M}}^{p_{o}}N(p)\,{\rm d}p \qquad
(\text{in\ CU})
\ee
is (theoretically) safed by the consumers when the product is 
introduced to the market according to strategy~1, this amount is 
referred to in the economic literature as {\bf consumer surplus}.

\item {\bf Strategy~3:} Some suppliers would be willing to 
introduce the product to the market initially at a unit price 
which is lower than $p_{M}$. If, hence, the suppliers decide to 
offer the product initially at a unit price $p_{u} < p_{M}$, and 
then to continuously\footnote{See previous footnote.} (!) 
\emph{raise} it to the higher $p_{M}$, the revenue obtaines 
amounts to the smaller value
\be
U_{3} = U(p_{M}) - \int_{p_{u}}^{p_{M}}A(p)\,{\rm d}p \ .
\ee
Since the suppliers (theoretically) earn the extra amount
\be
P := \int_{p_{u}}^{p_{M}}A(p)\,{\rm d}p \qquad
(\text{in\ CU})
\ee
when the product is introduced to the market according to 
strategy~1, this amount is referred to in the economic literature 
as {\bf producer surplus}.

\end{enumerate}
%

\appendix
\chapter[Glossary of technical terms (GB -- D)]{Glossary of 
technical terms (GB -- D)}
\lb{app1}

\noindent
{\bf A}\\
absolute change: absolute \"Anderung\\
absolute value: Betrag\\
account balance: Kontostand\\
addition: Addition\\
analysis: Analysis, Untersuchung auf 
Differenzierbarkeitseigenschaften\\
arithmetical mean: arithmetischer Mittelwert\\
arithmetical sequence: arithmetische Zahlenfolge\\
arithmetical series: arithmetische Reihe\\
augmented coefficient matrix: erweiterte Koeffizientenmatrix\\
average costs: St\"uckkosten\\
average profit: Durchschnittsgewinn, Gewinn pro St\"uck

\medskip
\noindent
{\bf B}\\
backward substitution: r\"uckwertige Substitution\\
balance equation: Bilanzgleichung\\
basis: Basis\\
basis solution: Basisl\"osung\\
basis variable: Basisvariable\\
Behavioural Economics: Verhaltens\"{o}konomik\\
boundary condition: Randbedingung\\
break-even point: Gewinnschwelle

\medskip
\noindent
{\bf C}\\
chain rule: Kettenregel\\
characteristic equation: charakteristische Gleichung\\
coefficient matrix: Koeffizientenmatrix\\
column: Spalte\\
column vector: Spaltenvektor\\
component: Komponente\\
compound interest: Zinseszins\\
concatenation: Verschachtelung, Verkn\"{u}pfung\\
conservation law: Erhaltungssatz\\
constant of integration: Integrationskonstante\\
constraint: Zwangsbedingung\\
cost function: Kostenfunktion\\
consumer surplus: Konsumentenrente\\
continuity: Stetigkeit\\
contract period: Laufzeit\\
Cournot's point: Cournotscher Punkt\\
curve sketching: Kurvendiskussion

\medskip
\noindent
{\bf D}\\
decision-making: Entscheidungsfindung\\
declining-balance depreciation method: geometrisch--degressive 
Abschreibung\\
definite integral: bestimmtes Integral\\
demand function: Nachfragefunktion\\
dependent variable: abh\"{a}ngige Variable\\
depreciation: Abschreibung\\
depreciation factor: Abschreibungsfaktor\\
derivative: Ableitung\\
determinant: Determinante\\
difference: Differenz\\
difference quotient: Differenzenquotient\\
differentiable: differenzierbar\\
differential: Integrationsdifferenzial\\
differential calculus: Differenzialrechnung\\
dimension: Dimension\\
direction of optimisation: Optimierungsrichtung\\
divergent: divergent, unbeschr"ankt\\
domain: Definitionsbereich

\medskip
\noindent
{\bf E}\\
economic agent: Wirtschaftstreibende(r) (meistens ein \emph{homo 
oeconomicus})\\
economic efficiency: Wirtschaftlichkeit\\
economic equilibrium: \"okonomisches Gleichgewicht\\
economic principle: \"okonomisches Prinzip\\
economic theory: Wirtschaftstheorie\\
Econophysics: \"{O}konophysik\\
eigenvalue: Eigenwert\\
eigenvector: Eigenvektor\\
elastic: elastisch\\
elasticity: Elastizit\"at\\
element: Element\\
end of profitable zone: Gewinngrenze\\
endogenous: endogen\\
equilibrium price: Marktpreis\\
equivalence transformation: \"{A}quivalenztransformation\\
exogenous: exogen\\
exponential function: Exponentialfunktion\\
extrapolation: Extrapolation, \"uber bekannten G\"utigkeitsbereich 
hinaus verallgemeinern

\medskip
\noindent
{\bf F}\\
feasible region: zul\"{a}ssiger Bereich\\
final capital: Endkapital\\
fixed costs: Fixkosten\\
forecasting: Vorhersagen erstellen\\
function: Funktion

\medskip
\noindent
{\bf G}\\
Gau\ss ian elimination: Gau\ss'scher Algorithmus\\
GDC: GTR, grafikf\"ahiger Taschenrechner\\
geometrical mean: geometrischer Mittelwert\\
geometrical sequence: geometrische Zahlenfolge\\
geometrical series: geometrische Reihe\\
growth function: Wachstumsfunktion

\medskip
\noindent
{\bf H}\\

\medskip
\noindent
{\bf I}\\
identity: Identit\"{a}t\\
image vector: Absolutgliedvektor\\
indefinite integral: unbestimmtes Integral\\
independent variable: unabh\"{a}ngige Variable\\
inelastic: unelastisch\\
initial capital: Anfangskapital\\
installment: Ratenzahlung\\
installment savings: Ratensparen\\
integral calculus: Integralrechnung\\
integrand: Integrand\\
integration variable: Integrationsvariable\\
interest factor: Aufzinsfaktor\\
interest rate: Zinsfu\ss\\
inverse function: Inversfunktion, Umkehrfunktion\\
inverse matrix: inverse Matrix, Umkehrmatrix\\
isoquant: Isoquante

\medskip
\noindent
{\bf J}\\

\medskip
\noindent
{\bf K}\\

\medskip
\noindent
{\bf L}\\
law of diminishing returns: Ertragsgesetz\\
length: L\"{an}ge\\
level of physical output: Ausbringungsmenge\\
limits of integration: Integrationsgrenzen\\
linear combination: Linearkombination\\
linearisation: Linearisierung\\
linear programming: lineare Optimierung\\
local rate of change: lokale \"{A}nderungsrate\\
logarithmic function: Logarithmusfunktion

\medskip
\noindent
{\bf M}\\
mapping: Abbildung\\
marginal costs: Grenzkosten\\
maximisation: Maximierung\\
minimisation: Minimierung\\
minimum efficient scale: Betriebsoptimum\\
monetary value: Geldwert\\
monopoly: Monopol\\
monotonicity: Monotonie\\
mortgage loan: Darlehen

\medskip
\noindent
{\bf N}\\
non-basis variable: Nichtbasisvariable\\
non-negativity constraints: Nichtnegativit\"atsbedingungen\\
non-linear functional relationship: nichtlineare 
Funktionalbeziehung

\medskip
\noindent
{\bf O}\\
objective function: Zielfunktion\\
one-to-one and onto: eineindeutig\\
optimal solution: optimalen L\"osung\\
optimal value: optimaler Wert\\
optimisation: Optimierung\\
order-of-magnitude: Gr\"o\ss enordnung\\
orthogonal: orthogonal\\
over-determined: \"{u}berbestimmt

\medskip
\noindent
{\bf P}\\
parallel displacement: Parallelverschiebung\\
pension calculations: Rentenrechnung\\
percentage rate: Prozentsatz\\
perfect competition: totale Konkurrenz\\
period: Periode\\
pivot column index: Pivotspaltenindex\\
pivot element: Pivotelement\\
pivot operation: Pivotschritt\\
pivot row index: Pivotzeilenindex\\
pole: Polstelle, Singularit\"at\\
polynomial division: Polynomdivision\\
polynomial of degree $n$: Polynom vom Grad $n$\\
power-law function: Potenzfunktion\\
present value: Barwert\\
primitive: Stammfunktion\\
principal component analysis: Hauptkomponentenanalyse\\
producer surplus: Produzentenrente\\
product rule: Produktregel\\
profit function: Gewinnfunktion\\
prohibitive price: Pohibitivpreis\\
Prospect Theory: Neue Erwartungstheorie\\
psychological value function: psychologische Wertfunktion

\medskip
\noindent
{\bf Q}\\
quadratic matrix: quadratische Matrix\\
quotient: Quotient\\
quotient rule: Quotientenregel

\medskip
\noindent
{\bf R}\\
range: Wertespektrum\\
rank: Rang\\
rare event: seltenes Ereignis\\
rational function: gebrochen rationale Funktion\\
real-valued function: reellwertige Funktion\\
reference period: Referenzzeitraum\\
regression analysis: Regressionsanalyse\\
regular: regul\"ar\\
relative change: relative \"Anderung\\
remaining debt: Restschuld\\
remaining resources: Restkapazit\"aten\\
remaining value: Restwert\\
rescaling: Skalierung\\
resources: Rohstoffe\\
resource consumption matrix: Rohstoffverbrauchsmatrix\\
restrictions: Restriktionen\\
root: Nullstelle\\
row: Reihe\\
row vector: Zeilenvektor

\medskip
\noindent
{\bf S}\\
saturation quantity: S\"attigungsmenge\\
scale: Skala, Gr\"o\ss enordnung\\
scale-invariant: skaleninvariant\\
simplex: Simplex, konvexer Polyeder\\
simplex tableau: Simplextabelle\\
singular: singul\"ar\\
sink: Senke\\
slack variable: Schlupfvariable\\
source: Quelle\\
stationary: station\"ar, konstant in der Zeit\\
straight line depreciation method: lineare Abschreibung\\
strictly monotonously decreasing: streng monoton fallend\\
strictly monotonously increasing: streng monoton steigend\\
summation rule: Summationsregel\\
supply function: Angebotsfunktion

\medskip
\noindent
{\bf T}\\
tangent: Tangente\\
target space: Wertebereich\\
technology matrix: Technologiematrix\\
total demand matrix: Gesamtbedarfsmatrix\\
total revenue: Ertrag\\
transpose: Transponierte

\medskip
\noindent
{\bf U}\\
uncertainty: Unsicherheit\\
under-determined: unterbestimmt\\
uniqueness: Eindeutigkeit\\
unit elastic: proportional elastisch\\
unit matrix: Einheitsmatrix\\
unit price: St\"uckpreis\\
unit vector: Einheitsvektor\\
utility function: Nutzenfunktion

\medskip
\noindent
{\bf V}\\
value chain: Wertsch\"{o}pfungskette\\
variability: \"Anderungsverhalten, Variabilit\"at\\
variable average costs: variable St\"uckkosten\\
variable costs: variable Kosten\\
variable vector: Variablenvektor\\
vector: Vektor\\
vector algebra: Vektoralgebra

\medskip
\noindent
{\bf W}\\
well-determined: wohlbestimmt

\medskip
\noindent
{\bf Z}\\
zero matrix: Nullmatrix\\
zero vector: Nullvektor


\addcontentsline{toc}{chapter}{Bibliography}


\end{document}